\documentclass[aps,twocolumn]{revtex4}
\makeatletter

\newcommand{\Rmnum}[1]{\expandafter\@slowromancap\romannumeral #1@}
\makeatother
\usepackage{graphicx}
\usepackage{dcolumn}
\usepackage{bm}
\usepackage{CJK}
\usepackage{amsfonts}
\usepackage{psfrag}
\usepackage{wrapfig}
\usepackage{subfigure}
\usepackage{makeidx}
\usepackage{multirow}
\usepackage{epsf}
\usepackage{amsmath}
\usepackage{cases}
\usepackage{amssymb}
\usepackage{soul}
\usepackage[version=3]{mhchem}

\usepackage[colorlinks,linkcolor=red,anchorcolor=red,citecolor=blue,urlcolor=blue]{hyperref}

\begin{document}
\title{Fundamental and second-order dark soliton solutions of 2- and 3-component Manakov equations in the defocusing regime}
\author{Wen-Juan Che$^{1}$}
\author{Chong Liu$^{1,2,3,4}$}\email{chongliu@nwu.edu.cn}
\author{Nail Akhmediev$^{2,5}$}
\address{$^1$School of Physics, Northwest University, Xi'an 710127, China}
\address{$^2$Department of Fundamental and Theoretical Physics, Research School of Physics, The Australian National University, Canberra, ACT 2600, Australia}
\address{$^3$Shaanxi Key Laboratory for Theoretical Physics Frontiers, Xi'an 710127, China}
\address{$^4$Peng Huanwu Center for Fundamental Theory, Xi'an 710127, China}
\address{$^5$Arts $\&$ Sciences Division, Texas A$\&$M University at Qatar, Doha, Qatar}
\begin{abstract}
We present exact multi-parameter families of soliton solutions for  two- and three-component Manakov equations in the \emph{defocusing} regime.
Existence diagrams for such solutions in the space of parameters are presented.
Fundamental soliton solutions exist only in finite areas on the plane of parameters. Within these areas, the solutions demonstrate rich spatio-temporal dynamics.
The complexity increases in the case of 3-component solutions.
The fundamental solutions are dark solitons with complex oscillating patterns in the individual wave components.
At the boundaries of existence, the solutions are transformed into plain (non-oscillating) vector dark solitons.
The superposition of two dark solitons in the solution adds more frequencies in the patterns of oscillating dynamics. These solutions admit degeneracy when the eigenvalues of fundamental solitons in the superposition coincide.
\end{abstract}

\maketitle
\section{INTRODUCTION}

Variety of oscillating localised structures associated with the scalar nonlinear Schr\"odinger equation (NLSE) is enormous \cite{Book97,PR2013,Dudley1,Dudley2}.
Oscillating nonlinear solutions are commonly dubbed as `breathers' \cite{KMH,BS1}.
They are multi-parameter families of solutions that are periodic either in space or time   with periods being the free parameters of the families. More general family
of the lowest order double-periodic solution is periodic both in space and in time \cite{TMP1987}. It
contains particular subsets such as Akhmediev breathers \cite{AB} and Kuznetsov-Ma solitons \cite{KM}.
Each of them is still a family of solutions with a free parameter.
Their limiting cases when each period is infinite leads to a special solution known as  Peregrine rogue wave \cite{PRW}.

Periodic breathers do exist in the \emph{focusing} regime of the NLSE. They describe a variety of physical phenomena such as modulation instability \cite{AB,Dudley3,MC,MC2}, rogue wave events \cite{RW}, Fermi-Pasta-Ulam recurrence \cite{FPU1,FPU2,FPU3,FPU4}, supercontinuum generation \cite{SCG}, and even turbulence \cite{Crespo}.
Exact multi-parameter families of solutions also exist in the case of defocusing NLSE \cite{AA93}.
These families also contain double-periodic solutions, although they describe different sets of physical phenomena. They involve dark solitons and their interactions.

Vector (two-component) generalisation of the NLSE
describes more complex systems such as nonlinear interaction of two wave components in optical fibres \cite{OF}, two-atom Bose-Einstein condensates (BECs) \cite{BEC,BEC1}, and the two-way wave propagation in the ocean (crossing seas) \cite{F}. The integrable version of this system is known as the set of Manakov equations \cite{MM}. As mentioned, oscillating structures do exist in the \textit{defocusing} NLSE case as well \cite{AA93}. Their investigation can be naturally extended to the case of Manakov equations \cite{DF0,DF1,DF2,DF3,DF4,DF5,Vobservation1,Vobservation2}.
For example, vector defocusing rogue waves have been predicted in \cite{DF2} and  observed experimentally in fiber optics \cite{Vobservation1,Vobservation2}.
Vector Akhmediev breathers also do exist in the defocusing regime \cite{DF4,DF5} and they can exhibit unique `hidden' dynamics in the nonlinear stage \cite{DF4}.
We can expect variety of other interesting phenomena when dealing with the whole family of exact solutions of Manakov equations in the defocusing regime.

Even fundamental (lowest-order) solutions of the Manakov model are not as simple as we would initially expect. Clearly, superposition of these solutions produces highly nontrivial structures especially, when the number of components in the model exceeds two.
Among these phenomena are multisoliton complexes \cite{SC0,SC1,SC2}, `beating solitons' \cite{BS1a,BS2,BS3}, non-degenerate solitons \cite{NDS1,NDS2,NDS3,NDS4} etc.
Another physical phenomenon is the multi-valley dark structure that exists in the defocusing regime when the number of components $N\geq3$ \cite{NDS4}.
Soliton on a background is one type of the structures that exist in these systems \cite{Wabnitz,KM1}.
In the defocusing media, solitons on a background are dark solitons.
The study of these objects for Manakov equations is still incomplete.
In this paper, we fill this gap in the knowledge.
In particular, we have found several new types of dark solitons in
the \emph{defocusing} regime of Manakov system and revealed their properties.

The paper is organised as follows.
Exact fundamental (lowest-order) soliton solutions in the defocusing regime of Manakov system and their symmetries are presented in Section \ref{Sec2}. Existence diagrams and characteristics of these solutions in the two- and three-component cases of the Manakov system are given in Sections \ref{Sec3} and \ref{Sec4}, respectively. A special case when all background amplitudes are equal $a_j=a$
is considered in Section \ref{Sec5}. Two other special cases when one ($a_1=a_2=a$, $a_3=0$ ) or two ($a_1=a$, $a_2=a_3=0$) background amplitudes are zero are considered in Sections \ref{Sec6} and \ref{Sec7}, respectively. Finally,
Section \ref{Sec8} contains our conclusions.

\section{Fundamental soliton solutions and their symmetries}\label{Sec2}

We consider here the set of Manakov equations generally consisting of $N$ wave components. In dimensionless form, they are given by
\begin{equation}\label{eq1}
\begin{split}
i\frac{\partial\psi^{(j)}}{\partial t}+\frac{1}{2}\frac{\partial^2\psi^{(j)}}{\partial x^2}+\sigma\sum_{j=1}^{j=N}(|\psi^{(j)}|^2)\psi^{(j)}&=0,
\end{split}
\end{equation}
where $\psi^{(j)}(t,x)$ are the nonlinearly coupled wave components of the vector wave field. The physical meaning of independent variables $x$ and $t$ depends on a particular physical problem of interest.
We have normalized Eqs. (\ref{eq1}) in a way such that $\sigma=\pm1$.
Note that in the case $\sigma=1$, Eqs. (\ref{eq1}) refer to either the focusing
(or anomalous dispersion) regime in optics or the attractive interaction between the atomic components of BEC; in the case $\sigma=-1$,
Eqs. (\ref{eq1}) refer to either the defocusing (or normal dispersion)
regime in optics or the repulsive interaction between the atomic components of BEC.

In our previous work \cite{VKMS-f}, we have demonstrated the dynamics of vector solitons
in the focusing regime $\sigma=1$ for the basic Manakov system, when $N=2$.
In contrast, we present here an exact multi-parameter family of fundamental soliton solutions in the defocusing regime $\sigma=-1$ of $N$-component Manakov equations when
$N=2$ and $N=3$. We reveal the existence conditions and the exact dynamics of solitons separately for $N=2$ and $N=3$.
This is different from the fundamental dark-dark and bright-dark soliton solutions of the defocusing Manakov equations reported recently \cite{DS2015,DS2022}.

The applicability of Eqs. (\ref{eq1}) with $N=2$ in physics has been verified experimentally in optics \cite{ME1,ME2,ME3,ME-MI} and for description of multicomponent BECs \cite{ME4,ME5}.
This task becomes significantly more difficult when the number of components in Eqs. (\ref{eq1}) increases. Nevertheless, recent experiments \cite{ME6} confirmed the physical relevance of Eqs. (\ref{eq1}) with $N=3$ by observing the bright-dark-bright solitons in BECs with repulsive forces between the atomic components.
Our present theoretical results may provide a basis for observing more complex wave patterns in such experiments.

\subsection{Fundamental soliton solutions in general form}

A fundamental (first-order) vector soliton solution of Eqs. (\ref{eq1}) can be obtained using a Darboux transformation scheme \cite{DT} with the seed in the form of a plane wave. In compact form, it is given by:
\begin{equation}
\psi_1^{(j)}(t,x)=\rho^{(j)}\psi_{0}^{(j)}(t,x) \psi_{fs}^{(j)}(t,x),\label{eqkmb}
\end{equation}
where $\psi_{0}^{(j)}(t,x)$ is the seed plane wave solution:
\begin{equation}
\psi_{0}^{(j)}= a_j \exp{\left\{i \left[\beta_jx +\sigma\left(\sum_{j=1}^{j=N}a_j^2+\frac{1}{2}\beta_j^2\right) t \right] \right\}}
\label{eqpw}
\end{equation}
with the real parameters $a_j$, and $\beta_j$ being the amplitudes and wave numbers, respectively, and
\begin{eqnarray}
\rho^{(j)}=\frac{\tilde{\bm\chi}^*+\beta_j}{\tilde{\bm\chi}+\beta_j}\sqrt{\frac{(\bm\chi^*+\beta_j)(\tilde{\bm\chi}+\beta_j)}{(\bm\chi+\beta_j)
(\tilde{\bm\chi}^*+\beta_j)}},
\end{eqnarray}
where $$\tilde{\bm\chi}=\bm\chi+i\alpha$$
with $\alpha(\neq0)$ being a real parameter. One can readily confirm that $|\rho^{(j)}|=1$.
Moreover,
\begin{eqnarray}
\psi_{fs}^{(j)}&=&\frac{\varpi\cosh({\bm{\Gamma}+i\delta_j})
+\cos{(\bm{\Omega}+i\gamma_j)}}{\varpi\cosh{\bm{\Gamma}}+\cos{\bm{\Omega}}},\label{eqkmbp}
\end{eqnarray}
where
\begin{eqnarray}
&&\bm{\Gamma}=\alpha (x+\bm\chi_{r}t)+\frac{1}{2}\ln\left(\frac{\alpha+\bm\chi_{i}}{\bm\chi_{i}}\right),\label{kmbcosh}\\
&&\bm{\Omega}=\Omega t=\alpha\left(\frac{\alpha}{2}+\bm\chi_{i}\right)t.
\end{eqnarray}
Subscripts $r$ and $i$ denote the real and imaginary parts of the complex parameter $\bm\chi$, respectively. The latter denotes the eigenvalue of the Manakov system (\ref{eq1})
which obeys the relation:
\begin{equation}
1+\sigma\sum_{j=1}^N\frac{a_j^2}{(\bm\chi-\beta_j)(\tilde{\bm\chi}-\beta_j)}=0.\label{eqchi}
\end{equation}
In principle, $N$-component model can admit $2N$ roots for $\bm\chi$.
The one-to-one correspondence between the eigenvalue and the spectral parameter of the associated Lax pair is given by:
\begin{equation}\label{Eqlambda}
\lambda=\bm\chi-\sigma\sum_{j=1}^N\frac{a_j^2}{\bm\chi+\beta_j}.
\end{equation}
The remaining notations in Eq. (\ref{eqkmbp}) are:
\begin{eqnarray}
\delta_j&=&\arg[{(\bm\chi^*+\beta_j)(\bm\chi+{i\alpha}+\beta_j)}],\nonumber\\
\gamma_j&=&-\frac{1}{2}\ln\left[\frac{(\bm\chi^*-{i\alpha}+\beta_j)(\bm\chi+{i\alpha}+\beta_j)}{(\bm\chi^*+\beta_j)(\bm\chi+\beta_j)}\right],\nonumber\\
\varpi&=&\frac{\alpha+2\bm\chi_{i}}{2\alpha+2\bm\chi_{i}}\sqrt{\frac{\alpha+\bm\chi_{i}}{\bm\chi_{i}}}.\nonumber
\end{eqnarray}

Clearly, the solution (\ref{eqkmb}) depends on the background wave parameters $a_j$, $\beta_j$, and the real parameter $\alpha$ $(\neq0)$.
For any $N$-component Manakov system, the solution (\ref{eqkmb}) describes fundamental dark vector soliton with the plane wave background (\ref{eqpw}) around it. The solution (\ref{eqkmb}) is the direct analog of the dark soliton of the single component NLSE in the defocusing ($\sigma=-1$) regime. At any $t$, the deviation of the soliton profile from the background is localised in $x$ with the width $\sim 1/\alpha$. These solitons can move with the group velocity $V_g=-\bm\chi_r$.

The new notable feature of the dark soliton of the Manakov system is that its components may exchange energy and therefore may oscillate in $t$. Period of these oscillations as we can see from (\ref{eqkmbp}) is $2\pi/\Omega$. Additional oscillations may appear when two dark solitons are superposed at the same location. The frequency of these oscillations will be equal to the beating frequency of two dark solitons. Such superpositions will be considered below.

The choice of parameters $a_j$, $\beta_j$ strongly influences the dynamics of solitons. As there are several of them, variety of possible dynamics is very large.
First, let us consider the case of identical background amplitudes $a_j=a$. Such condition ($a_j=a$) has been used in experimental observations of optical rogue waves in the two-component Manakov system \cite{Vobservation1,Vobservation2}. As particular cases, we consider the characteristics of dark solitons when one or two of the background amplitudes vanish.
As for the wave numbers $\beta_j$, we set them as follows:
\begin{eqnarray}
{\beta_1}=-{\beta_2}=\beta,~~~ \textrm{for}~~~ N=2,\\
{\beta_1}=-{\beta_3}=\beta,~{\beta_2}=0,~~~ \textrm{for} ~~~N=3.
\end{eqnarray}

\subsection{Symmetries of the solutions}

Before entering the details, let us consider the two main symmetries of the fundamental  soliton solution (\ref{eqkmb}). Taking them into account will simplify the analysis.
The first one is the symmetry of the solution (\ref{eqkmb}) relative
to the sign change of $\beta$ and simultaneous change of the wave component.
For the case of identical background amplitudes $a_j=a$, we have
\begin{eqnarray}\label{eqsymm1}
&\psi_1^{(1)}(\beta)=\psi_1^{(2)}(-\beta),~~~~\mbox{when}~N=2,\\
&\psi_1^{(1)}(\beta)=\psi_1^{(3)}(-\beta),~~~~\mbox{when}~N=3.
\end{eqnarray}

The second symmetry involving the eigenvalue $\bm\chi$, is not that simple. Namely, if $\bm\chi_i\Rightarrow-\bm\chi_i-\alpha$, we have
\begin{eqnarray}
\psi_1^{(j)}\{(x,t);\bm\chi_i\}=\psi_1^{(j)}\{(x',t');-\bm\chi_i-\alpha\}e^{(ir_j)},\label{eqsymm2}
\end{eqnarray}
where $r_j=2\arg(\rho_j)$ denotes a constant phase, and $x'=x+\Delta x$, $t'=t+\Delta t$, with $\Delta x$ and $\Delta t$ fixed
constant shifts along the $x$ and $t$ axes, respectively. They are given by
\begin{center}
\begin{eqnarray}
\begin{split}
&\Delta x=-\frac{1}{\alpha}\left[\frac{4\pi\bm\chi_r}{\alpha+2\bm\chi_i}+\ln \left( \frac{\alpha+\bm\chi_i}{\bm\chi_i} \right) \right],\\
&\Delta t=\frac{4\pi}{\alpha^2+2\alpha \bm\chi_{i}}.
\end{split}
\end{eqnarray}
\end{center}
Thus, the symmetry (\ref{eqsymm2}) defines the periods in the oscillating patterns of dark soliton.
The symmetries (\ref{eqsymm1})-(\ref{eqsymm2}) provide more insight in revealing the richness of soliton properties as it is demonstrated below. Let us start with the analysis of dark solitons in the two-component Manakov system.

\section{Dark solitons in the two-component Manakov system}\label{Sec3}

In the defocusing regime of the two-component Manakov system ($N=2$),
Eq. (\ref{eqchi}) admits four roots for the eigenvalue $\bm\chi$.
For the case of identical background amplitudes $a_j=a$, and for ${\beta_1}=-{\beta_2}=\beta$, the explicit expressions for them are given by
\begin{eqnarray}\label{Eqchi1234}
\begin{split}
\bm{\chi}_{1}=-\frac{i}{2}\alpha-\sqrt{\kappa-\sqrt{\eta}},~~\bm{\chi}_{2}=-\frac{i}{2}\alpha+\sqrt{\kappa-\sqrt{\eta}},\\
\bm{\chi}_{3}=-\frac{i}{2}\alpha-\sqrt{\kappa+\sqrt{\eta}},~~\bm{\chi}_{4}=-\frac{i}{2}\alpha+\sqrt{\kappa+\sqrt{\eta}},
\end{split}
\end{eqnarray}
where $\kappa=\beta^2+a^2-\alpha^2/4$, and $\eta=a^4+4a^2\beta^2-\alpha^2\beta^2$.

It follows, from (\ref{Eqchi1234}), that
\begin{eqnarray}\label{chi}
\bm\chi_{1i}+\bm\chi_{2i}=-\alpha,~~\bm\chi_{3i}+\bm\chi_{4i}=-\alpha.
\end{eqnarray}
Then, from (\ref{eqsymm2}), it also follows that the wave components $\{\psi^{(j)}_1(\bm\chi_{1})$, $\psi^{(j)}_1(\bm\chi_{2})\}$ or $\{\psi^{(j)}_1(\bm\chi_{3})$, $\psi^{(j)}_1(\bm\chi_{4})\}$ have the same amplitude profiles. The only difference between them is the shifts in $x$ and $t$ equal to $\Delta x$, $\Delta t$.

A direct analysis shows that $\bm\chi_{3i}=\bm\chi_{4i}\equiv-\alpha/2$, implying that $\bm{\Omega}\equiv0$. This indicates that the period of $\psi^{(j)}_1(\bm\chi_{3})$ in $t$ (i.e., $2\pi/\bm{\Omega}$) becomes infinite (no oscillations). Moreover, the solutions $\psi^{(j)}_1(\bm\chi_{3})$, $\psi^{(j)}_1(\bm\chi_{4})$ reduce to the background level everywhere on the ($x$,$t$)-plane.
Thus, these two eigenvalues describe trivial background wave solutions. They can be ignored in further analysis.

\begin{figure*}[htbp]
\centering
\includegraphics[width=130mm]{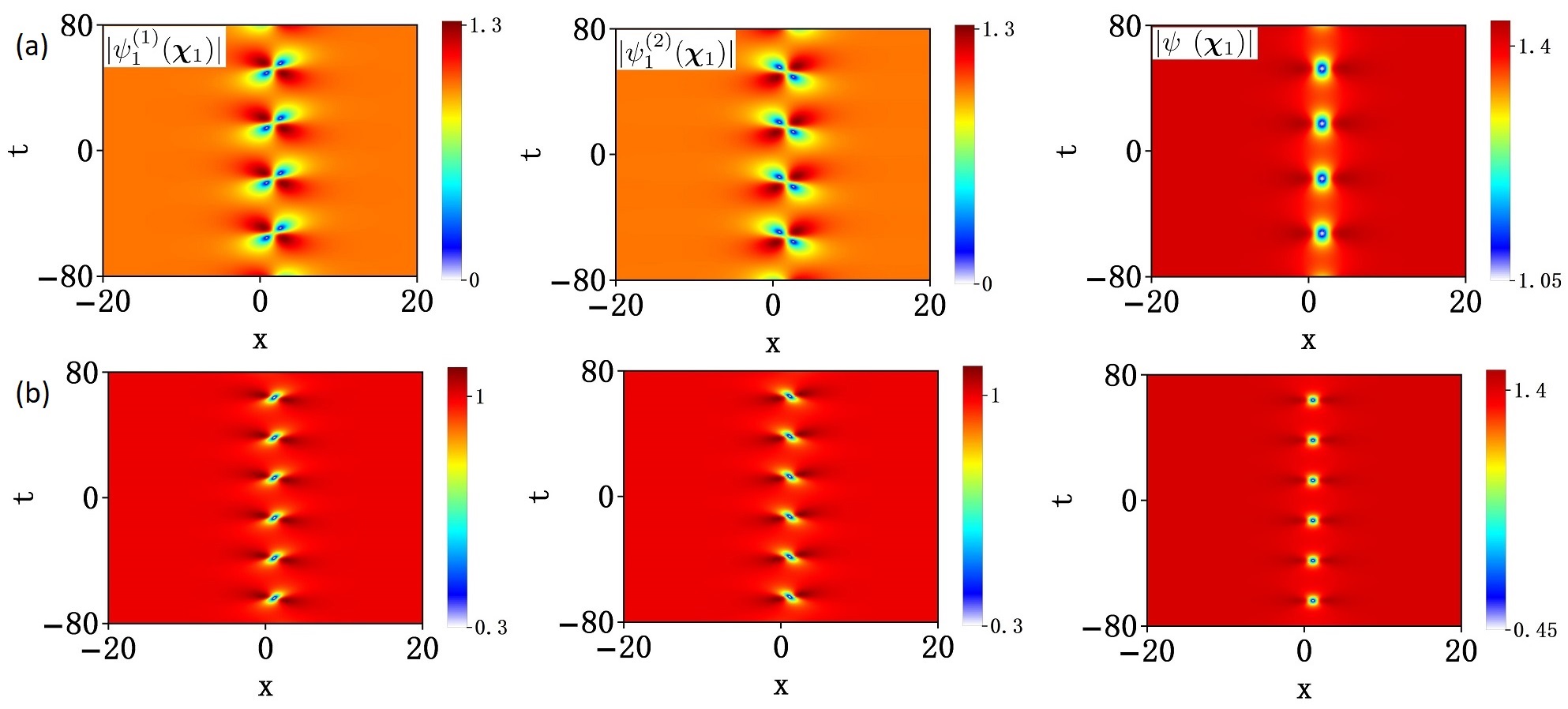}
\caption{Individual $|\psi^{(j)}_1(\bm{\chi}_1)|$ and the total $|\psi|=\sqrt{|\psi^{(1)}_1|^2+|\psi^{(2)}_1|^2}$ amplitude profiles of the two-component dark soliton (\ref{eqkmb}) on the ($x$,$t$)-plane for two  relative wavenumbers (a) $\beta = 0.3$, (b) $\beta = 1$. Parameters $a=1$, and $\alpha=0.5$.}
\label{fig1-f-kms2}
\end{figure*}

For illustration, Figures \ref{fig1-f-kms2}(a) and \ref{fig1-f-kms2}(b) show the individual and total component profiles of the fundamental dark soliton on the ($x$,$t$)-plane that corresponds to the eigenvalue $\bm{\chi}_1$. Two different relative wavenumbers $\beta=0.3$ and $\beta=1.0$ are used.
The individual components are periodic in $t$ due to the energy exchange between them.
Figure \ref{fig1-f-kms2} (a) shows a `four-petal' pattern in each period of oscillations with two
areas of depressed and two areas of elevated amplitudes diagonally located relative to the centre. The central point in this pattern is a saddle.
Fig. \ref{fig1-f-kms2}(b) displays a similar pattern but with the amplitude at the central point being transformed from a saddle to a minimum. The two areas with depressed amplitudes are now combined into a single one. With further increase of $\beta$, the oscillations disappear and each component is gradually transformed into a plain (non-periodic) dark soliton.

The total amplitudes of the dark soliton $|\psi|=\sqrt{|\psi^{(1)}_1|^2+|\psi^{(2)}_1|^2}$ shown in the r.h.s. columns of Fig. \ref{fig1-f-kms2} are also oscillating. The minima are located at the centres of each four-petal patterns in (a) or coincide with the minima of the two components in (b). Thus, the solution (\ref{eqkmb}) generally describes oscillating dark solitons.

Clearly, the choice of the parameters $\alpha$, $\beta$ is not arbitrary.
We need to analyse Hessian matrix for (\ref{eqkmb}) in order to find the regions of existence of these solutions. Using the technique presented in \cite{VKMS-f}, we constructed the existence diagrams for the solutions for each of the eigenvalues.

\begin{figure}[htbp]
\centering
\includegraphics[width=84mm]{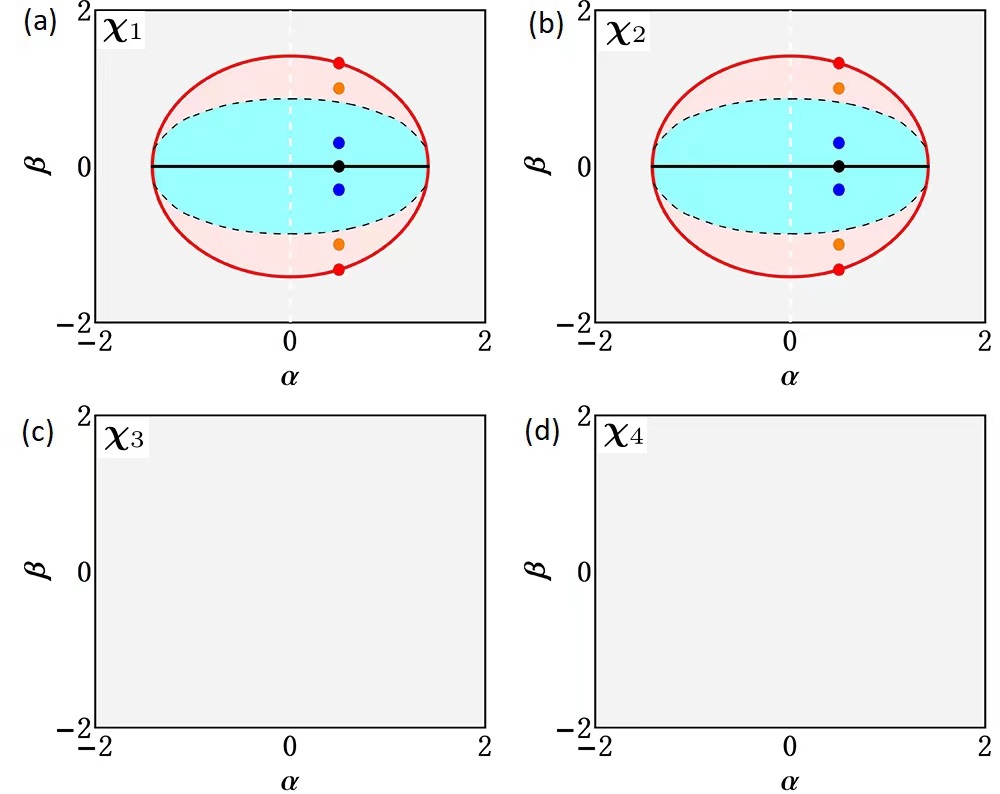}
\caption{Existence diagrams for the two-component dark solitons on the ($\alpha$, $\beta$) plane for four eigenvalues $\bm{\chi}_i$
given by Eqs. (\ref{Eqchi1234}). Cyan and pink areas correspond to dark solitons with
two types of patterns shown in Fig. \ref{fig1-f-kms2}. The blue and orange solid circles correspond to specific solutions shown in Figs. \ref{fig1-f-kms2}(a) and (b) respectively.
The red solid lines correspond to the plain dark solitons (no oscillations).
The black solid lines ($\beta=0$) correspond to dark solitons with the oscillating components but no oscillations in the total amplitude.
 The red and black solid circles correspond to specific solutions shown below in Figs. \ref{fig3-f-rd2}(a) and \ref{fig3-f-rd2}(b) respectively. Parameter $a=1$.}
\label{fig2-f-ed2}
\end{figure}

Figure \ref{fig2-f-ed2} shows these diagrams on the ($\alpha$, $\beta$) plane.
For the first two eigenvalues $\bm\chi_1$ and $\bm\chi_2$, the solutions are confined to the elliptical regions bounded by the red solid curves in Figs. \ref{fig2-f-ed2}(a) and \ref{fig2-f-ed2}(b).
Dark solitons do exist in the pink and cyan areas which correspond to the two types of structures shown in Fig. \ref{fig1-f-kms2}. The black dashed curve
corresponds to the transition from the saddle point at the centre of each periodic pattern to a minimum. Dark soliton solutions do not exist in grey areas.
The existence diagrams for $\bm\chi_1$ and $\bm\chi_2$ are identical.
In the limiting case of $\alpha=0$, dark solitons are transformed into vector rogue waves \cite{DF2}. As mentioned, the eigenvalues $\bm\chi_3$ and $\bm\chi_4$ describe only trivial solutions.
Thus, the diagrams corresponding to these eigenvalues are fully grey in Fig. \ref{fig2-f-ed2}.

The analytical expression for the boundary of the dark soliton existence in Fig. \ref{fig2-f-ed2} (the red solid curves) can be extracted from the conditions
\begin{eqnarray}
\bm\chi_{1i}=-\alpha~~\textmd{or}~~\bm\chi_{2i}=0.
\end{eqnarray}
Namely, from Eqs. (\ref{Eqchi1234}) we obtain
\begin{eqnarray}
\beta_c^2=2a^2-\alpha^2,\label{eqbc}
\end{eqnarray}
where $\beta_c$ denotes the critical wavenumber. Dark solitons do exist in the region confined by the condition $\beta^2<\beta_c^2$.
When this is the case, the two eigenvalues $\bm\chi_1$ and $\bm\chi_2$ are purely imaginary ($\bm\chi_r=0$).
This implies that these dark solitons have zero velocity ($v_g=0$).
Two examples are shown in Fig. \ref{fig1-f-kms2}.

\begin{figure*}[htbp]
\centering
\includegraphics[width=130mm]{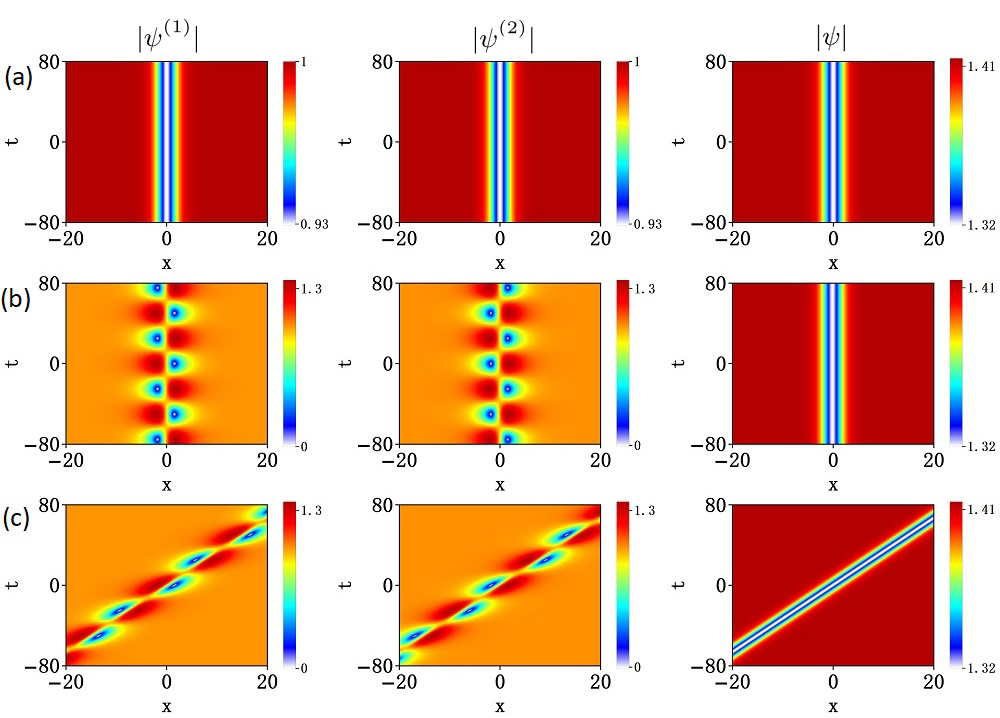}
\caption{Amplitude distributions of individual components $|\psi^{(j)}|$ and the total amplitude $|\psi|=\sqrt{|\psi^{(1)}_1|^2+|\psi^{(2)}_1|^2}$ of the two-component dark solitons. (a) Dark soliton given by (\ref{eqds1}) with $\beta=\beta_c=\sqrt{2a^2-\alpha^2}$ (red solid circle in Fig.\ref{fig2-f-ed2}(a)). (b) Dark soliton with oscillating components given by (\ref{eqvbs1}) with $\beta=0$ (black solid circle in Fig.\ref{fig2-f-ed2}(a)). (c) Moving dark soliton with oscillating components given by (\ref{eqmvbs2}) with $\beta_1=\beta_2=0.5$. Parameters $a=1$, and $\alpha=0.5$.}
\label{fig3-f-rd2}
\end{figure*}

When $\beta^2=\beta_c^2$, the solution is converted into a plain dark soiton (no oscillations).  The explicit expressions for the components of this dark soliton follow from Eq. (\ref{eqkmb}):
\begin{eqnarray}\label{eqds1}
\psi^{(j)}_{DS}=\psi^{(j)}_0\left[\frac{\beta_j}{\beta_j-i\alpha}+\frac{i\alpha}{\beta_j-i\alpha}\tanh{(\alpha x)}\right].
\end{eqnarray}
This dark soliton has zero velocity. The solution (\ref{eqds1}) is
reduced to the plane wave $\psi^{(j)}_0$ when $\alpha=0$. One example of the
amplitude profiles of the dark soliton (\ref{eqds1}) is shown in Fig. \ref{fig3-f-rd2}(a) with the choice of parameters $\alpha=0.5$, $\beta=\beta_c=\sqrt{7}/2$.
The amplitude profiles of the two components of the dark soliton are identical:
\begin{eqnarray}\label{eqta1}
|\psi^{(1)}_{DS}|=|\psi^{(2)}_{DS}|=
a\sqrt{\frac{\alpha ^2 \tanh ^2(\alpha  x)+\beta^2}{\alpha ^2+\beta^2}}.
\end{eqnarray}
Consequently, the total amplitude $\sqrt{|\psi_{DS}^{(1)}|^2+|\psi_{DS}^{(2)}|^2}$ of the dark soliton also has a dark-soliton shape.

When $\beta=0$, dark solitons are located on the black solid line in Figs. \ref{fig2-f-ed2}(a) and \ref{fig2-f-ed2}(b). They are confined by the condition $\alpha^2<2a^2$ ($\alpha\neq0$). Analytical expression for these dark solitons can be found from Eq. (\ref{eqkmb}):
\begin{eqnarray}
\psi^{(j)}_{VBS}(\bm\chi_1)&=&\psi^{(j)}_0(\psi_{DS}\mp\psi_{BS}),\label{eqvbs1}\\
\psi^{(j)}_{VBS}(\bm\chi_2)&=&\psi^{(j)}_0(\tilde{\psi}_{DS}\mp\tilde{\psi}_{BS}),\label{eqvbs2}
\end{eqnarray}
where
\begin{eqnarray}
&\psi_{DS}&=\frac{(4a^2-\alpha^2)\sinh{(\alpha x)}-\alpha^2\cosh{(\alpha x)}}{(4a^2-\alpha^2)\cosh{(\alpha x)}-\alpha^2\sinh{(\alpha x)}},\\
&\psi_{BS}&=\frac{-2(2a^2-\alpha^2)\exp{(\frac{1}{2}i\alpha^2t})}{(4a^2-\alpha^2)\cosh{(\alpha x)}-\alpha^2\sinh{(\alpha x)}}.
\label{BS}
\end{eqnarray}
and
\begin{eqnarray}
\tilde{\psi}_{DS}=\psi_{DS}(-x),~~~~\tilde{\psi}_{BS}=-\psi_{BS}(-x).
\end{eqnarray}
The solution (\ref{eqvbs2}) is the same as (\ref{eqvbs1}) but reversed in space.
The two components of the dark soliton are oscillating in $t$ with the frequency $\alpha ^2/2$. This follows from Eq. (\ref{BS}). The soliton profiles for this case are shown in Fig. \ref{fig3-f-rd2}(b). The two components are oscillating in the opposite phases. The elevations (depressions) in $\psi^{(1)}_{VBS}$ correspond to the depressions (elevations) in $\psi^{(2)}_{VBS}$. This allows the total amplitude profile $\sqrt{|\psi_1^{(1)}|^2+|\psi_1^{(2)}|^2}$ of the dark soliton to be constant in $t$. The explicit expression for it is given by:
\begin{eqnarray}
\label{eqta2}
\sqrt{|\psi_{VBS}^{(1)}|^2+|\psi_{VBS}^{(2)}|^2}=a\sqrt{2\frac{\alpha ^2 \tanh ^2(\alpha  x)+\beta^2}{\alpha ^2+\beta^2}}.
\end{eqnarray}
This is the same profile as for the dark soliton (\ref{eqds1}). Indeed, Eqs. (\ref{eqta1}) and (\ref{eqta2})] are the same. This can also be seen from the comparison of
Figs. \ref{fig3-f-rd2}(a) and \ref{fig3-f-rd2}(b).

When $\beta_1=\beta_2\neq0$, the dark soliton acquires nonzero velocity.
The solution can be derived either directly from Eq. (\ref{eqkmb}) or obtained from the expressions (\ref{eqvbs1}-\ref{eqvbs2}) using Galilean transformation.
For the absolute values of the components, we have
\begin{eqnarray}
\begin{split}\label{eqmvbs2}
\left| \psi^{(j)}(t,x) \right| &= \left| \psi^{(j)}_{VBS}(\bm\chi_1;~t, x-\beta_1t) \right|,\\
\left| \psi^{(j)}(t,x) \right| &= \left| \psi^{(j)}_{VBS}(\bm\chi_2;~t, x-\beta_1t) \right|.
\end{split}
\end{eqnarray}
The amplitude profiles of this solution is shown in Fig. \ref{fig3-f-rd2}(c). This dark soliton propagates with the velocity $\beta_1$. Two of its components remain oscillating. The shape of the total amplitude remains fixed in $t$. It coincides with the shape of dark solitons in Figs \ref{fig3-f-rd2}(a) and \ref{fig3-f-rd2}(b).

\begin{figure*}[htbp]
\centering
\includegraphics[width=160mm]{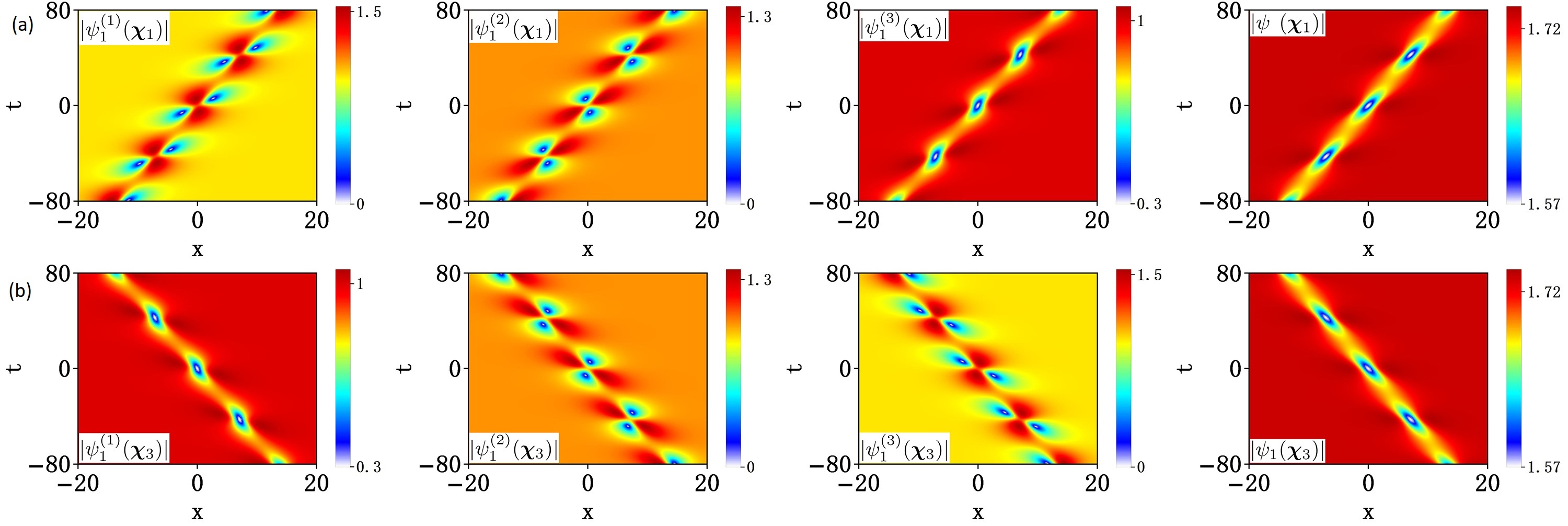}
\caption{Individual $|\psi^{(j)}_1|$ and the total amplitude distributions $|\psi|=\sqrt{|\psi^{(1)}_1|^2+|\psi^{(2)}_1|^2+|\psi^{(3)}_1|^2}$ of the three-component dark solitons (\ref{eqkmb}) corresponding to the eigenvalues $\bm{\chi}_1$, and $\bm{\chi}_3$, given by Eqs. (\ref{Eqchi1-6}). Parameters $a = 1$, $\beta=0.3$, and $\alpha=0.5$.}
\label{fig4-f-kms3}
\end{figure*}

\section{Dark solitons in the three - component Manakov system}\label{Sec4}

Now, we explore the properties of three-component ($N=3$) vector dark solitons in the defocusing regime. There are six eigenvalues $\bm{\chi}_j$ in this case.
The explicit expressions for them when $a_j=a$, ${\beta_1}=-{\beta_3}=\beta$, and ${\beta_2}=0$
are given by
\begin{eqnarray}\label{Eqchi1-6}
\begin{split}
\bm{\chi}_{1}&=\frac{i}{2}(-\alpha-\sqrt{\kappa_1}),~~~~
\bm{\chi}_{2}=\frac{i}{2}(-\alpha+\sqrt{\kappa_2}),\\
\bm{\chi}_{3}&=\frac{i}{2}(-\alpha-\sqrt{\kappa_2}),~~~~
\bm{\chi}_{4}=\frac{i}{2}(-\alpha+\sqrt{\kappa_1}),\\
\bm{\chi}_{5}&=\frac{i}{2}(-\alpha-\sqrt{\kappa_3}),~~~~
\bm{\chi}_{6}=\frac{i}{2}(-\alpha+\sqrt{\kappa_3}).
\end{split}
\end{eqnarray}
Here
\begin{eqnarray} \nonumber
\kappa_1&=&\alpha^2+ \frac{4}{3} i \left( \mathcal{A}+\frac{i 2^{1/3}\mathcal{C}}{\mathcal{B}}-\frac{i \mathcal{B}}{2^{1/3}} \right),\\   \nonumber
\kappa_2&=&\alpha^2+ \frac{4}{3} i \left( \mathcal{A}-\frac{(i+\sqrt{3})\mathcal{C}}{(2^{2/3})\mathcal{B}}+ \frac{(i-\sqrt{3})\mathcal{B}}{2^{4/3}} \right),\\   \nonumber
\kappa_3&=&\alpha^2+ \frac{4}{3} i(\mathcal{A}-\frac{(i-\sqrt{3})\mathcal{C}}{(2^{2/3})\mathcal{B}}+ \frac{(i+\sqrt{3})\mathcal{B}}{2^{4/3}}),
\end{eqnarray}
with
\begin{eqnarray} \nonumber
\mathcal{A}&=&i(3 a^2+2 \beta^2),~~~~~~~~~~~\mathcal{B}=\mathcal{D}+\sqrt{4\mathcal{C}^3+\mathcal{D}^2},\\   \nonumber
\mathcal{C}&=&-9 a^4-12 a^2 \beta^2+3 \alpha^2 \beta^2-\beta^4, \\   \nonumber
\mathcal{D}&=&-54 a^6-108 a^4 \beta^2-72 a^2 \beta^4+18 \alpha^2\beta^4+2\beta^6.
\end{eqnarray}

Similar to the case $N=2$ considered above, here, not all eigenvalues describe a soliton. We have found that $\bm{\chi}_{5}$ and $\bm{\chi}_{6}$ correspond to the trivial background solutions, while four other eigenvalues do correspond to dark solitons.
 They obey the relations:
\begin{eqnarray}
&&\bm\chi_{1i}+\bm\chi_{2i}=-\alpha,~~~~~~
\bm\chi_{3i}+\bm\chi_{4i}=-\alpha,\\
&&\bm\chi_{1r}=\bm\chi_{2r}=-\bm\chi_{3r}=-\bm\chi_{4r}.\label{eqchi-r}
\end{eqnarray}
The corresponding amplitude profiles satisfy the symmetry (\ref{eqsymm2}):
\begin{eqnarray}
\left|\psi^{(j)}_1[(x,t);\bm\chi_{1}]\right|&=&\left|\psi^{(j)}_1[(x',t');\bm\chi_{2}]\right|,\label{eqchi12}\\
\left|\psi^{(j)}_1[(x,t);\bm\chi_{3}]\right|&=&\left|\psi^{(j)}_1[(x',t');\bm\chi_{4}]\right|.\label{eqchi34}
\end{eqnarray}
This means that $\psi^{(j)}_1(\bm\chi_{1})$ and $\psi^{(j)}_1(\bm\chi_{2})$ [or $\psi^{(j)}_1(\bm\chi_{3})$ and $\psi^{(j)}_1(\bm\chi_{4})$] have the same amplitude distributions. However,
\begin{eqnarray}
\left|\psi^{(j)}_1[(x,t);\bm\chi_{1}]\right|\neq\left|\psi^{(j)}_1[(x,t);\bm\chi_{3}]\right|.\label{eqchi13}
\end{eqnarray}
This means that for given values of $a$, $\beta$ and $\alpha$, we have two different dark solitons  with opposite group velocities ($\bm\chi_{1r}=-\bm\chi_{3r}$).

The amplitude profiles of these two solitons, $|\psi_1^{(j)}(\bm\chi_1)|$, $|\psi_1^{(j)}(\bm\chi_3)|$ are shown in Figs. \ref{fig4-f-kms3}(a) and \ref{fig4-f-kms3}(b) respectively. Oscillations are now due to the energy exchange between the three wave components.
The first two components in Fig. \ref{fig4-f-kms3}(a) show the four-petal patterns in each period with a saddle point at the centre. The third component has a minimum at the centre.
The total amplitude (r.h.s. panel) is still an oscillating dark soliton.
The two solitons shown in Figs. \ref{fig4-f-kms3}(a) and \ref{fig4-f-kms3}(b) have the same oscillating period. The patterns in the second case are reversed as well as the direction of propagation.
Nonlinear superposition of these two dark vector solitons is a second-order `non-degenerate' dark soliton
(see Section \ref{Sec5}).

\begin{figure*}[htbp]
\centering
\includegraphics[width=130mm]{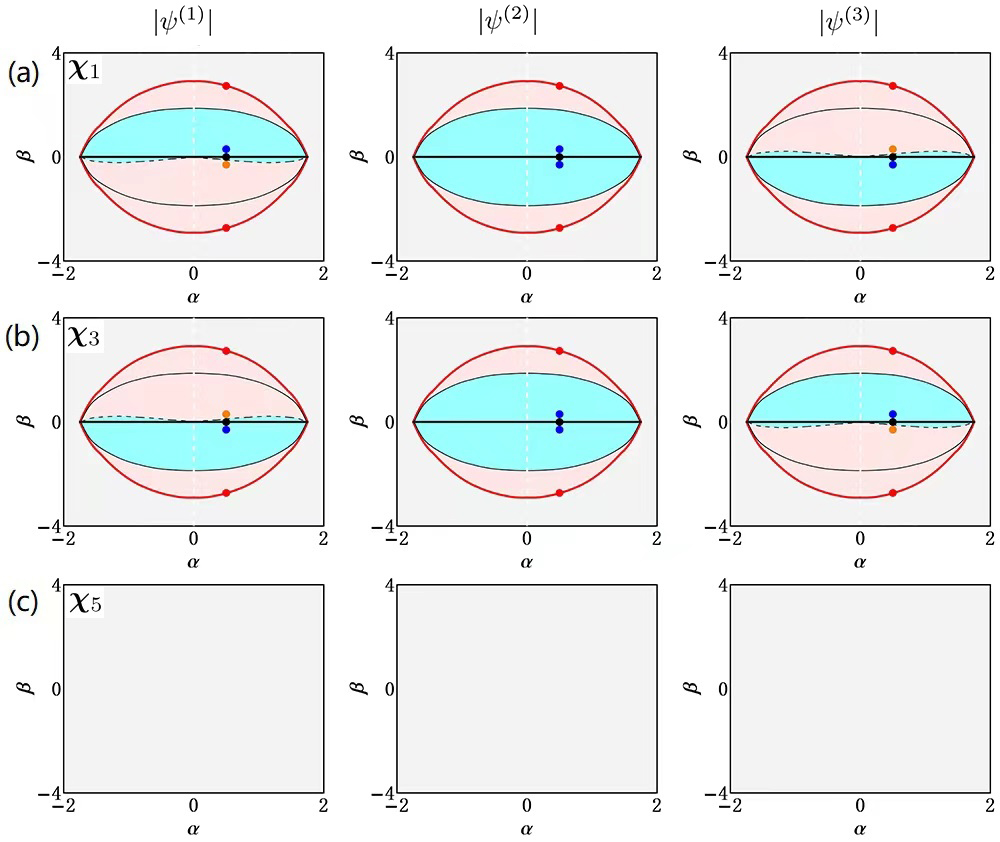}
\caption{Existence diagrams of the three-components of dark solitons  with eigenvalues $\bm{\chi}_1$, $\bm{\chi}_3$, and $\bm{\chi}_5$, given by Eqs. (\ref{Eqchi1-6}) on the ($\alpha$, $\beta$) plane. Cyan areas correspond to the components with the four-petal patterns in each period and saddle point at the centre. The pink areas correspond to the patterns with the minimum at the centre. The red solid lines correspond to the plain dark solitons (no oscillations). The black solid lines correspond to the dark solitons with the oscillating components but not oscillating total amplitude.
The blue and orange solid circles correspond to dark solitons shown in Figs. \ref{fig4-f-kms3}(a) and  \ref{fig4-f-kms3}(b) respectively.
The red and black solid circles correspond to dark solitons shown below in Figs. \ref{fig6-f-rd3}(a) and
\ref{fig6-f-rd3}(b), respectively.
Parameter $a = 1$.}
\label{fig5-f-ed3}
\end{figure*}

Figure \ref{fig5-f-ed3} shows the existence diagrams of dark vector soliton components on the ($\alpha$, $\beta$)-plane for three eigenvalues $\bm\chi_1$, $\bm\chi_3$, and $\bm\chi_5$.
Dark solitons do exist only for the case of the eigenvalues $\bm\chi_1$ and $\bm\chi_3$. In these two cases, solitons are confined to the eye-shape areas bounded by the red solid curves. Due to the symmetry (\ref{eqsymm1}), the existence regions of $\psi^{(1)}_1$ and $\psi^{(3)}_1$ components of dark solitons are symmetric around the line $\beta=0$.  Further comparison of the cases $\bm\chi_1$ and $\bm\chi_3$ shows that
the existence region of $\psi^{(1)}_1(\bm\chi_1)$ [or $\psi^{(3)}_1(\bm\chi_1)$] coincides with that of $\psi^{(3)}_1(\bm\chi_3)$ [or $\psi^{(1)}_1(\bm\chi_3)$]. Dark solitons do not exist in the grey areas.

\begin{figure*}[htbp]
\centering
\includegraphics[width=160mm]{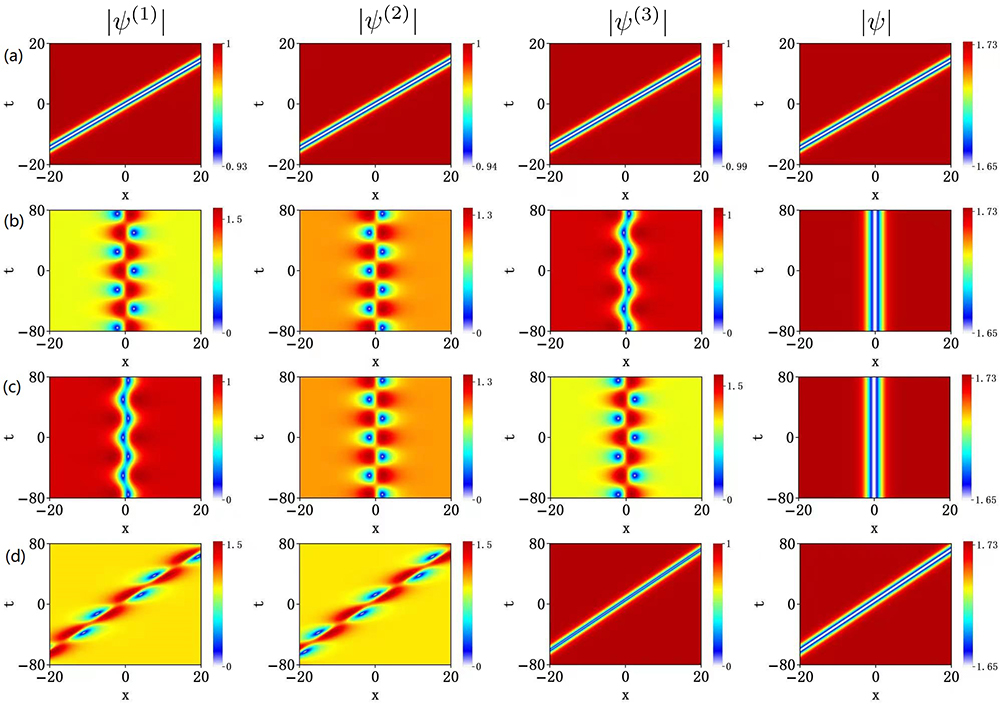}
\caption{Individual $|\psi^{(j)}_1|$ and the total amplitude distributions $|\psi|$ of the three-component dark solitons. (a) Plain dark solitons (\ref{eqds3}). (b) Dark solitons with three oscillating components (\ref{eqvbs3-1}) with $\beta=0$. (c) Dark solitons with three oscillating components (\ref{eqvbs3-2}) with $\beta=0$. (d) Moving dark solitons with two oscillating components (\ref{eqbs3-2}) with $\beta_1=\beta_2=\beta_3=0.5$. Parameters $a=1$, and $\alpha=0.5$.}
\label{fig6-f-rd3}
\end{figure*}

The regions of dark soliton existence for case $N=3$ are limited by the red solid curves  obtained from the condition:
\begin{eqnarray}
\bm\chi_{1i}=\bm\chi_{3i}=-\alpha~~\textmd{or}~~\bm\chi_{2i}=\bm\chi_{4i}=0.
\end{eqnarray}
At this boundary, the vector dark soitons have the form:
\begin{eqnarray}\label{eqds3}
\psi^{(j)}_{DS}=\psi^{(j)}_0\left\{\frac{\beta_j+\bm\chi_{r}}{\beta_j+\bm\chi}+\frac{i\alpha}{\beta_j+\bm\chi}\tanh{ \left[ \alpha(x+\bm\chi_{r} t) \right] }\right\}.
\end{eqnarray}
The difference from the dark solitons in the case $N=2$, Eq. (\ref{eqds1}), is that the group velocity $-\bm\chi_{r}$ is not zero.
The amplitude profiles for these solitons is shown in Fig. \ref{fig6-f-rd3}(a).

Inside the red solid lines, the dark soliton components are oscillating.
Taking $\beta=0$ (i.e., $\beta_j=0$, $j=1,2,3$), we obtain the soliton solution from  (\ref{eqkmb}). The explicit expressions can be represented in the following forms:
\begin{eqnarray}\label{eqvbs3-1}
\begin{split}
\psi^{(1)}_{VBS}(\bm\chi_1)&=\psi^{(1)}_{VBS}(\bm\chi_2)=\psi^{(1)}_0(\psi_{DS}+\psi^{(1)}_{BS}),\\
\psi^{(2)}_{VBS}(\bm\chi_1)&=\psi^{(2)}_{VBS}(\bm\chi_2)=\psi^{(2)}_0(\psi_{DS}+\psi^{(2)}_{BS}),\\
\psi^{(3)}_{VBS}(\bm\chi_1)&=\psi^{(3)}_{VBS}(\bm\chi_2)=\psi^{(3)}_0(\psi_{DS}+\psi^{(3)}_{BS}),
\end{split}
\end{eqnarray}
and
\begin{eqnarray}\label{eqvbs3-2}
\begin{split}
\psi^{(1)}_{VBS}(\bm\chi_3)&=\psi^{(1)}_{VBS}(\bm\chi_4)=\psi^{(1)}_0(\psi_{DS}+\psi^{(3)}_{BS}),\\
\psi^{(2)}_{VBS}(\bm\chi_3)&=\psi^{(2)}_{VBS}(\bm\chi_4)=\psi^{(2)}_0(\psi_{DS}+\psi^{(2)}_{BS}),\\
\psi^{(3)}_{VBS}(\bm\chi_3)&=\psi^{(3)}_{VBS}(\bm\chi_4)=\psi^{(3)}_0(\psi_{DS}+\psi^{(1)}_{BS}).
\end{split}
\end{eqnarray}
Here, we separated the solutions into a 'bright', $\psi^{(j)}_{BS}$, and `dark', $\psi_{DS}$, parts:
\begin{eqnarray}
&\psi_{DS}&=1+\frac{\mathcal{F}\cosh{(\alpha x)}-\mathcal{F}\sinh{(\alpha x)}}{\mathcal{I}\cosh{(\alpha x)}+\mathcal{J}\sinh{(\alpha x)}},\\
&\psi^{(1)}_{BS}&=c_1 \psi^{(2)}_{BS},~~~~\psi^{(3)}_{BS}=c_2 \psi^{(2)}_{BS},\\
&\psi^{(2)}_{BS}&=-\frac{\mathcal{F}\exp{(\frac{1}{2}i\alpha^2 t)}}{\mathcal{I}\cosh{(\alpha x)}+\mathcal{J}\sinh{(\alpha x)}},
\end{eqnarray}
where
\begin{eqnarray} \nonumber
\mathcal{F}&=&(6a^2-2\alpha^2)\alpha, ~~~~
\mathcal{I}=-a^2 \alpha(4+c_1^2+c_2^2)+\alpha^3 \\  \nonumber
\mathcal{J}&=&-\alpha(a^2(-2+c_1^2+c_2^2)+\alpha^2),
\end{eqnarray}
 with $c_1=-1.366$, $c_2=0.366$.

Oscillations in Eqs. (\ref{eqvbs3-1},\ref{eqvbs3-2}) are caused by the `bright' parts.
The corresponding wave profiles in each component together with the total soliton amplitude  are shown in Figs. \ref{fig6-f-rd3}(b) and \ref{fig6-f-rd3}(c), respectively.
These can be considered as the special cases of the solutions shown in Figs. \ref{fig4-f-kms3}(a) and \ref{fig4-f-kms3}(b) but with $\beta=0$.

Also, from Eqs. (\ref{eqvbs3-1},\ref{eqvbs3-2}), we find that
\begin{eqnarray}
|\psi^{(j)}_{VBS}(\bm\chi_1;x)|=|\psi^{(4-j)}_{VBS}(\bm\chi_3;-x)|.
\end{eqnarray}
This means that the components of $|\psi^{(j)}_{VBS}(\bm\chi_1)|$ are reversed in space components of $|\psi^{(4-j)}_{VBS}(\bm\chi_3)|$.
On the other hand, in each case, all three components are different.
Like in the case $N=2$, the total amplitude always has the shape of a dark soliton that does not change in $t$.

When $\beta_j=0$, solitons (\ref{eqvbs3-1},\ref{eqvbs3-2}) have zero velocity.
Using a Galilean transformation, we obtain the moving dark soliton solution for the case $\beta_1=\beta_2=\beta_3\neq0$. It is given by
\begin{eqnarray}\label{eqmvbs3-1}
\psi^{(j)}(t,x)&=\psi^{(j)}_{VBS}(\bm\chi_1;~t, x-\beta_1t),\\
\label{eqmvbs3-2}
\psi^{(j)}(t,x)&=\psi^{(j)}_{VBS}(\bm\chi_3;~t, x-\beta_1t).
\end{eqnarray}

In contrast to the case $N=2$, the three-component Manakov equations have additional degree of freedom influencing the dynamics of components.
For the same case $\beta_1=\beta_2=\beta_3\neq0$,
performing the Darboux transformation with a Lax spectral parameter
\begin{equation}
\label{eqbs3-sp1}
\lambda_1=\bm\chi_1+\frac{3a^2}{\bm\chi_1+\beta_1},
\end{equation}
where $\bm{\chi}_{1}=-\beta_1-i\alpha$,
we can obtain another family of dark soliton solutions given by:
\begin{eqnarray}
\label{eqbs3-2}
\begin{split}
\psi^{(1)}_{VBS}&=\psi^{(1)}_0(\psi_{DS}+\psi_{BS}),\\
\psi^{(2)}_{VBS}&=\psi^{(2)}_0(\psi_{DS}-\psi_{BS}),\\
\psi^{(3)}_{VBS}&=\psi^{(3)}_0(\psi_{DS}),
\end{split}
\end{eqnarray}
where
\begin{eqnarray}
\psi_{DS}&=&1+\frac{(\alpha^2-3a^2)\exp{[\alpha( x-\beta_1 t)-d]}}{2a^2\alpha^2\cosh{[\alpha (x-\beta_1 t)+d]}},\\
\psi_{BS}&=&\frac{i(\alpha^2-3a^2)\exp{(1/2i\alpha^2 t-d)}}{a^2\alpha\cosh{[\alpha (x-\beta_1 t)+d]}},
\end{eqnarray}
and $d=\frac{1}{2}\ln{ \left( \frac{3}{2\alpha^2}-\frac{1}{2a^2} \right) }$.

In contrast to the dark soliton solutions (\ref{eqvbs3-1},\ref{eqvbs3-2}), the oscillations in the solution (\ref{eqbs3-2}) occur only in $\psi^{(1)}$ and $\psi^{(2)}$ components. The $\psi^{(3)}$ component is a plain dark soliton. This solution is shown
in Fig. \ref{fig6-f-rd3}(d). The soliton propagates with the group velocity $v_g=\beta_1$.
Only the components $\psi^{(1)}_{VBS}$ and $\psi^{(2)}_{VBS}$ periodically exchange energy. The total amplitude profile is the same as in Fig. \ref{fig6-f-rd3}(c).

The solutions (\ref{eqbs3-2}) satisfy a simple transformation. Namely, the solution obtained by swapping the components $\psi^{(2)}_{VBS}\Leftrightarrow\psi^{(3)}_{VBS}$,
\begin{eqnarray}
\label{eqbs3-3}
\begin{split}
&&\psi^{(1)}_{VBS}&=\psi^{(1)}_0(\psi_{DS}+\psi_{BS}),\\
&&\psi^{(2)}_{VBS}&=\psi^{(2)}_0(\psi_{DS}),\\
&&\psi^{(3)}_{VBS}&=\psi^{(3)}_0(\psi_{DS}-\psi_{BS}).
\end{split}
\end{eqnarray}
is still the solution of Eqs. (\ref{eq1}).
The nonlinear superposition of (\ref{eqbs3-2}) and (\ref{eqbs3-3}) with different $\alpha$ produces the second-order dark soliton. It is presented below.

\section{Second-order dark solitons for $N=2$ with $a_j=a\neq0$}\label{Sec-N2-1}

Each of the fundamental dark solitons can be part of the nonlinear superposition of more complex structures. As in the previous works related to the scalar NLSE case \cite{JETP88,Erkintalo,SKM}, the nonlinear superposition of fundamental dark solitons in the Manakov system can be constructed using next steps in the Darboux transformation (see Appendix \ref{SecA-1}).
For the two-component Manakov system, the fundamental solution on a constant background can be obtained by using the vector eigenfunctions of the transformed Lax pair with the coefficients $\{1,1,0\}$ (see Appendix \ref{SecA-1}). A particular case is a dark soliton solution (\ref{eqkmb}) for $N=2$.

First, we consider the case with equal background amplitudes $a_j=a$. Two types of second-order dark solitons are obtained below: i) when the wavenumbers are unequal $\beta_1=-\beta_2=\beta\neq0$; ii) when the wavenumbers are equal $\beta_1=\beta_2$. In each case, the solitons have the same (zero) velocity. Then, the second-order solution is a bound state of two dark solitons.

\begin{figure}[htbp]
\centering
\includegraphics[width=84mm]{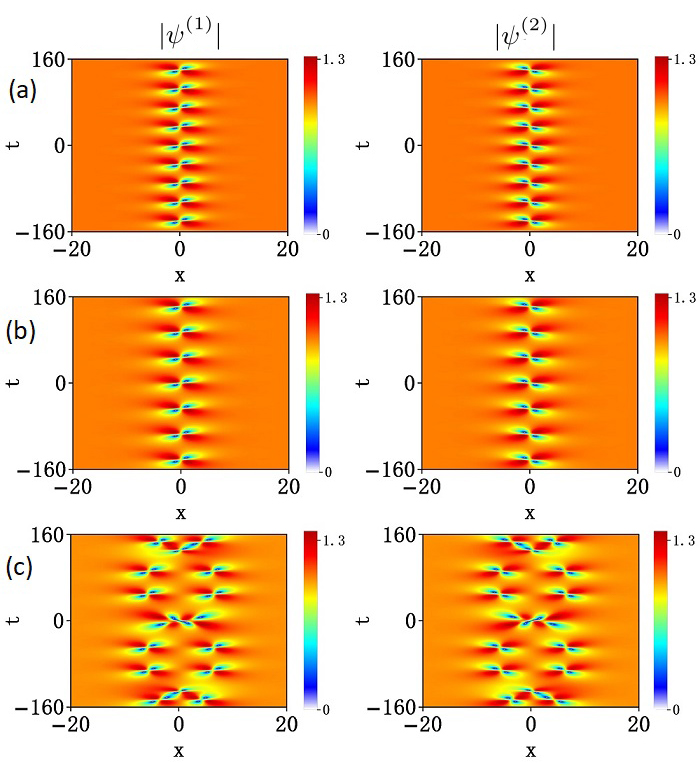}
\caption{Fundamental dark soliton components $|\psi^{(j)}_1(\bm{\chi}_1)|$ given by (\ref{eqkmb}) on the ($x$,$t$)-plane when (a) $\alpha=0.5$, and (b) $\alpha=0.4$.  (c) The two components of the nonlinear superposition $|\psi^{(j)}_2(\bm{\chi}_1)|$ of dark solitons shown in (a) and (b). Parameters $a_j=1$, and $\beta_1=-\beta_2=0.3$.}
\label{fig7-2-KMS}
\end{figure}

\subsection{Second-order dark soliton with $\beta_1=-\beta_2=\beta\neq0$}

The two components of the fundamental dark soliton (\ref{eqkmb}) for the cases $\alpha=0.5$ and $\alpha=0.4$ are shown in Figs. \ref{fig7-2-KMS}(a) and \ref{fig7-2-KMS}(b)  respectively. These components are periodic with `four-petal' type patterns in each period of oscillations. The average velocity of the dark soliton is zero.
The nonlinear superposition of these two fundamental solitons is shown in Fig. \ref{fig7-2-KMS}(c). The result of the superposition is the soliton structure oscillating with two periods. From the fundamental solution (\ref{eqkmb}), the beating period of the bound state is given by
\begin{equation}
D_t=\left|\frac{2\pi}{\Omega|_{(\alpha=\alpha_1)}-\Omega|_{(\alpha=\alpha_2)}}\right|,
\end{equation}
where $\alpha_1=0.5$ and $\alpha_2=0.4$.
The average velocity of this combined structure is also zero.

\begin{figure*}[htbp]
\centering
\includegraphics[width=130mm]{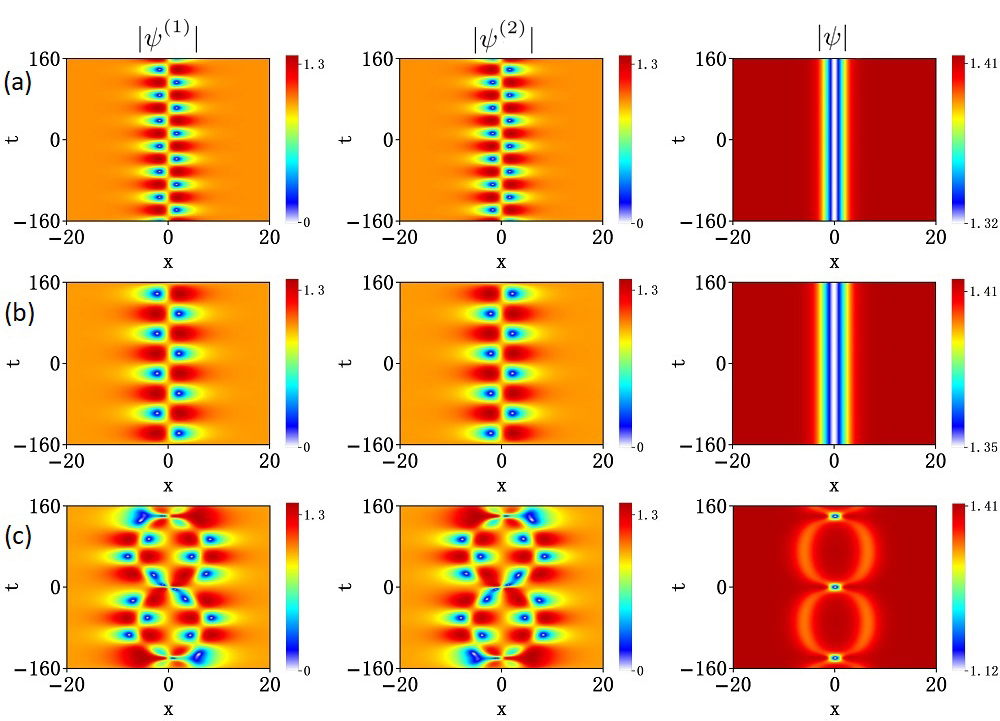}
\caption{Dark soliton with oscillating components (\ref{eqvbs1}) when (a) $\alpha=0.5$ and (b) $\alpha=0.4$.  (c) Nonlinear superposition of dark solitons shown in (a) and (b). Parameters $a_j=1$, and $\beta_1=\beta_2=0$.}
\label{fig8-2-BES}
\end{figure*}

\subsection{Second-order dark solitons with $\beta_1=\beta_2$}

When $\beta_1=\beta_2$, the exact solution is given by Eqs. (\ref{eqvbs1}),(\ref{eqvbs2}). The two components and the total amplitude for two different values of $\alpha$ are illustrated in Figs. \ref{fig8-2-BES}(a) and (b) respectively.
The nonlinear superposition of these two dark solitons again produces `double-beating' soliton pattern with two frequencies of oscillation. It is shown in Fig. \ref{fig8-2-BES}(c). However, the total amplitude profile (r.h.s. panel) shows only a single beating frequency.

\section{Second-order dark solitons for $N=2$ with $a_1\neq0$, $a_2=0$}\label{Sec-N2-2}

Let us now consider higher-order dark solitons for $N=2$ when one of the background amplitudes vanishes, e.g., $a_1\neq0$, $a_2=0$.
 For the defocusing Manakov systems, all background components cannot be simultaneously equal to zero.

We can use two approaches  to investigate the properties of these solutions.
The first one is to consider the limit $a_2\rightarrow0$ in the solution presented above.
The second one is to construct directly the new exact solution with $a_2=0$.
Here, we use the second technique and present the new exact solution although both of them lead to the same result.
We first consider the valid eigenvalues of the soliton from the general relation (\ref{eqchi}).
The associated Lax spectral parameters follow from Eq. (\ref{Eqlambda}).
Finally, the corresponding soliton solutions can be constructed by performing the Darboux transformation with these spectral parameters.

\begin{figure}[htbp]
\centering
\includegraphics[width=84mm]{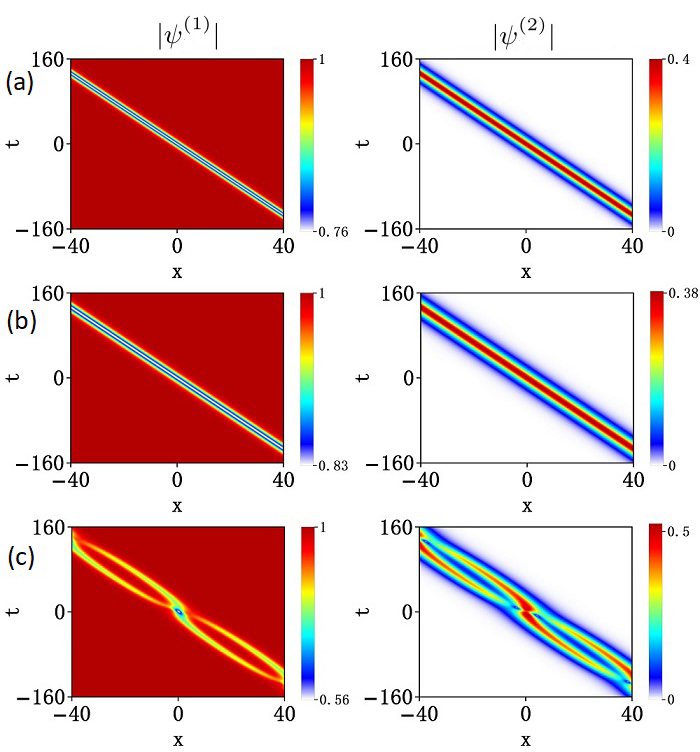}
\caption{Dark-bright soliton (\ref{eqDB}) with nonzero velocity when (a) $\alpha=0.5$ and (b) $\alpha=0.4$. (c) Nonlinear superposition of the two solitons shown in (a) and (b).
Parameters $\beta_1=0.3$, $a_1=1$, and $a_2=0$.}
\label{fig9-a2-KMS}
\end{figure}

The spectral parameter for the case $a_1\neq0$, $a_2=0$ follows from (\ref{Eqlambda}), It is given by:
\begin{equation}\label{Eqlambda-N2}
\lambda=\bm\chi+\frac{a_1^2}{\bm\chi+\beta_1},
\end{equation}
where $\bm{\chi}=\beta_1-i\alpha$ is the only valid eigenvalue obtained from (\ref{eqchi}). Using the Darboux transformation with the spectral parameter (\ref{Eqlambda-N2}), we  obtain the higher-order dark soliton. The explicit form of this solution is given by:
\begin{eqnarray}\label{eqDB}
\begin{split}
\psi_{DS}^{(1)}&=\psi_0^{(1)}+\frac{\psi_0^{(1)}(\lambda^*-\lambda)\exp{(\alpha(x+\beta_1 t)-d)}}{(2\beta_1-i\alpha)2\sinh{(\alpha(x+\beta_1 t)+d)}},\\
\psi_{BS}^{(2)}&=\exp{(i\theta_2)}\frac{(\lambda^*-\lambda)\exp{(1/2i\alpha^2 t-d)}}{2\sinh{(\alpha(x+\beta_1 t)+d)}},
\end{split}
\end{eqnarray}
where $d=\frac{1}{2}\ln{\left(\frac{4\beta_1^2+\alpha^2-a_1^2}{4\beta_1^2+\alpha^2}\right)}$.
Figures \ref{fig9-a2-KMS}(a) and \ref{fig9-a2-KMS}(b) show this soliton for two different values of $\alpha$. As we can see, when $a_1\neq0$ and $a_2=0$, the fundamental two-component solution is a dark-bright soliton pair. The velocity of the soliton pair is $-\beta_1$.
 The nonlinear superposition of these two solutions is shown in Fig. \ref{fig9-a2-KMS}(c).
 It is a dark-bright soliton solution beating with a single frequency.
These solutions can be considered as the limiting cases of the corresponding solutions shown in Fig. \ref{fig7-2-KMS}(a)-(c) when $a_2 \rightarrow 0$.

\begin{figure}[htbp]
\centering
\includegraphics[width=84mm]{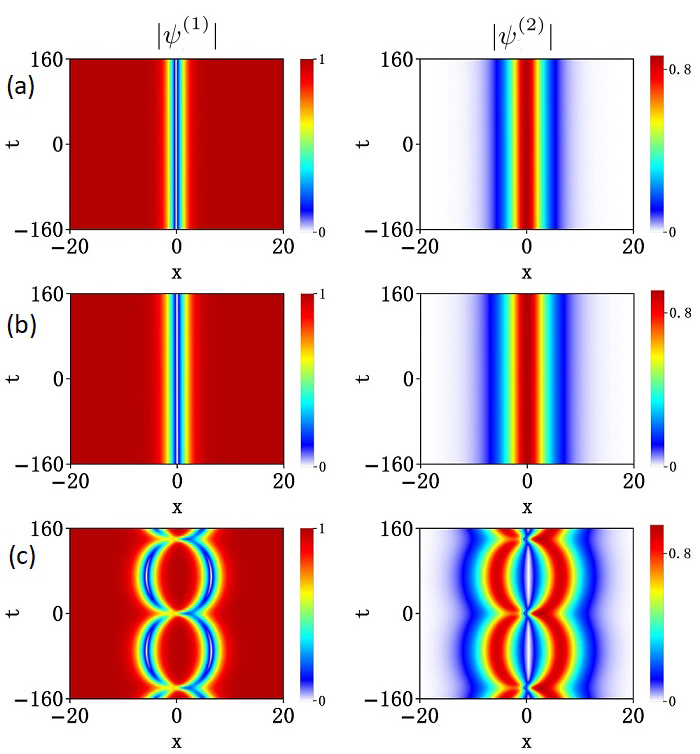}
\caption{Dark-bright soliton (\ref{eqDB}) with zero velocity when (a) $\alpha=0.5$ and (b) $\alpha=0.4$. (c) Nonlinear superposition of the two solitons shown in (a) and (b). Parameters $\beta_1=0$, $a_1=1$, and $a_2=0$.}
\label{fig10-a2-BES}
\end{figure}

\begin{figure*}[htbp]
\centering
\includegraphics[width=130mm]{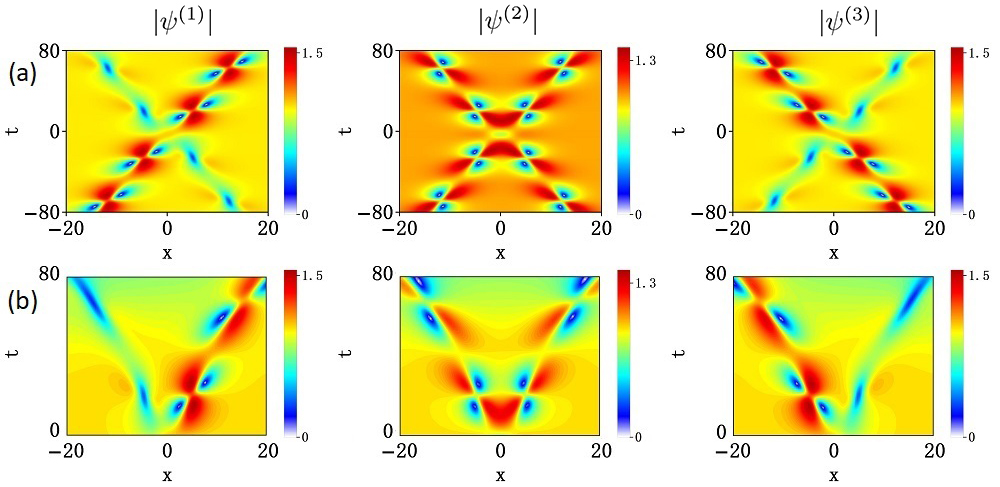}
\caption{(a) Non-degenerate soliton formed by nonlinear superposition  $|\psi^{(j)}_2(\bm{\chi}_1; \bm{\chi}_{3})|$ of two dark solitons with eigenvalues $\bm{\chi}_{1}$ and $\bm{\chi}_{3}$ shown in Fig. \ref{fig4-f-kms3}.
 (b) Numerical simulations of the solution shown in (a) starting with the initial condition generated by the exact solution at $t=0$. These simulations confirm the exact results.
Parameters $a_j=1$, $\beta=0.3$, and $\alpha=0.5$.
}
\label{fig11-2km1}
\end{figure*}

Figures \ref{fig10-a2-BES}(a) and \ref{fig10-a2-BES}(b) show the similar dark-bright soliton pairs when $\beta_1=0$. Parameters $\alpha=0.5$ and $\alpha=0.4$ are the same as before.
The soliton pairs have zero velocity since $\beta_1=0$.  The nonlinear superposition of these two solitons is shown in Fig. \ref{fig10-a2-BES}(c). The second-order soliton oscillates but its average velocity is zero.
The solutions shown in Figs. \ref{fig10-a2-BES} can be considered as the limiting cases of the solutions in Figs. \ref{fig8-2-BES} when $a_2 \rightarrow 0$.

\section{Second-order dark solitons for $N=3$ with $a_j=a\neq0$}\label{Sec5}

In contrast to the case $N=2$, the three-component Manakov system admits more eigenvalues. This has been shown in Section \ref{Sec4}. Then the number of possibilities in constructing higher-order dark solitons increases. On the other hand,
there are two combinations of the vector eigenfunctions of the transformed Lax pair to generate different fundamental solutions for $N=3$. Namely, using the combination with the coefficients of vector eigenfunctions $\{1,1,0,0\}$,
we obtain the fundamental dark soliton which coincides with (\ref{eqkmb}). The alternative combination with the coefficients $\{1,0,1,0\}$ yields the fundamental solution describing the dynamics of general breathers [see Appendix \ref{SecB-1a}].
Nonlinear superposition of these two fundamental solutions can produce new wave formation.

Like in the case $N=2$, we present below two different types of non-degenerate second-order solitons for the case $N=3$ with the equal background amplitudes $a_j=a$. i) Non-degenerate second-order solitons with unequal wavenumbers $\beta_1=-\beta_3=\beta\neq0$, $\beta_2=0$. ii) Non-degenerate second-order solitons with equal wavenumbers $\beta_1=\beta_2=\beta_3$.

\subsection{Second-order solutions with $\beta_1=-\beta_3=\beta\neq0$, $\beta_2=0$}\label{Sec5-1}

We first consider the second-order solutions formed by the nonlinear superposition of two fundamental dark solitons corresponding to two different eigenvalues $\bm{\chi}_{1}$, and $\bm{\chi}_{3}$. These superpositions also depend on the set of initial parameters $a$, $\beta$, and $\alpha$. The details of derivation of exact solutions are given in Appendix (\ref{SecB-1a}).

Figure \ref{fig11-2km1}(a) shows the nonlinear superposition of two fundamental dark solitons shown in Fig. \ref{fig4-f-kms3}. As the two original dark solitons have velocities of opposite sign ($\bm\chi_{1r}=-\bm\chi_{3r}$), the two dark solitons cross each other at $t = 0$. This superposition exhibits a typical X shape.
The first, $\psi^{(1)}$, and the third, $\psi^{(3)}$, wave components are mirror images of each other. The second wave component, $\psi^{(2)}$, is symmetric relative to the $t$ and $x$ axes. Moreover, the superposition of two dark solitons shows a typical elastic collision.
This can be proved strictly by the asymptotic analysis shown in Appendix \ref{SecC}.

In order to confirm the accuracy of exact solutions, we used direct numerical simulations of Manakov equations. Figure \ref{fig11-2km1}(b) shows the results of numerical simulations with the initial conditions extracted from the exact solution at $t=0$, namely, $\psi^{(j)}(x,t=0)$.
Comparison of the upper half of the solution in Fig. \ref{fig11-2km1}(a) with the results of numerical simulations in Fig. \ref{fig11-2km1}(b) shows that the exact solutions are indeed correct.

\begin{figure*}[htbp]
\centering
\includegraphics[width=130mm]{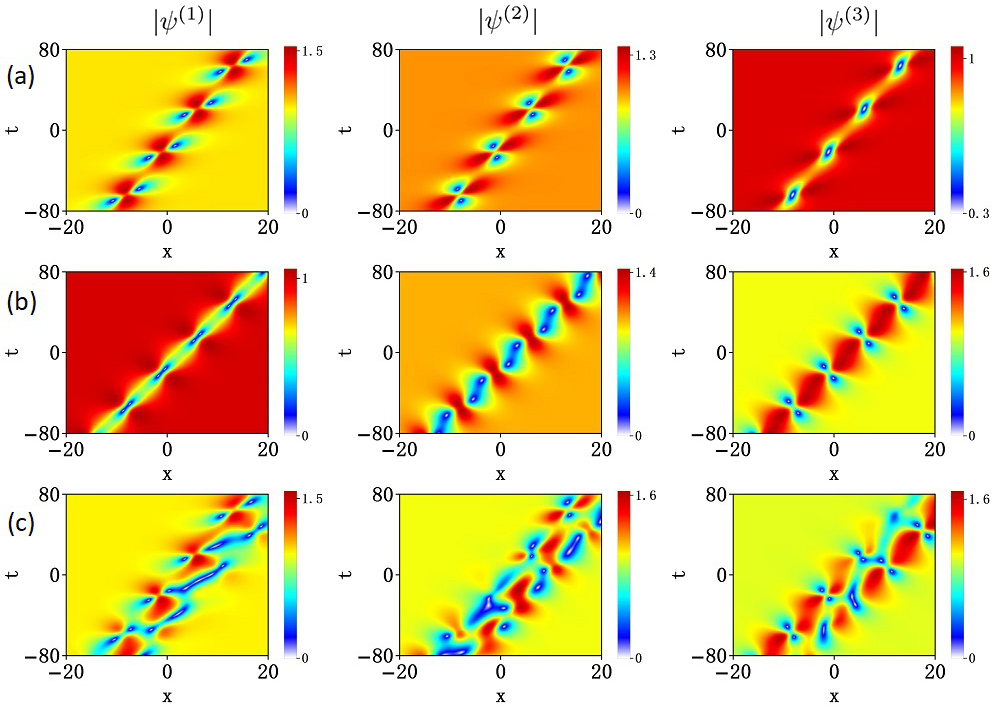}
\caption{(a) Fundamental dark soliton (\ref{eqkmb}) with $\alpha=0.5$. (b) Fundamental general breather (\ref{eqgb}) with $\alpha=0.4$. (c) Nonlinear superposition of the solutions shown in (a) and (b). Parameters $a_j=1$, and $\beta=0.3$.}
\label{fig12-2km2}
\end{figure*}

Let us now consider the nonlinear superposition of two fundamental solutions corresponding to different combinations of the vector eigenfunctions of the transformed Lax pair, i.e., $\{1,1,0,0\}$ and $\{1,0,1,0\}$. The first combination produces fundamental dark soliton while the latter combination $\{1,0,1,0\}$ produces fundamental general breather (GB). It is given by:
\begin{equation}
\psi_1^{(j)}(t,x)=\psi_{0}^{(j)}(t,x) \psi_{gb}^{(j)}(t,x),\label{eqgb}
\end{equation}
where
\begin{eqnarray}
\psi_{gb}^{(j)}&=&\frac{G_j\cosh({\bm{\Lambda}+\upsilon_j})
+H_j\cos{(\bm{\Pi}+\vartheta_j)}}{M_j\cosh{\bm{\Lambda}}+U_j\cos{\bm{\Pi}}},\label{eqgbp}
\end{eqnarray}
with
\begin{eqnarray}
\bm{\Lambda}&=&\left(\widehat{\bm\chi}_{i}-\bm\chi_{i}\right)x+
\left(\widehat{\bm\chi}_{r}\widehat{\bm\chi}_{i}-\bm\chi_{r}\bm\chi_{i}\right)t+\bm{d},~~~\label{gbcosh}\\
\bm{\Pi}&=&\left(\bm\chi_{r}-\widehat{\bm\chi}_{r}\right)x+
\frac{1}{2}\left(\bm\chi_{r}^2-\widehat{\bm\chi}_{r}^2-\bm\chi_{i}^2+\widehat{\bm\chi}_{i}^2\right)t+\bm{r}.~~~~~~
\end{eqnarray}
Here, $\bm{d}$ and $\bm{r}$ are:
\begin{eqnarray}
\bm{d}=\frac{1}{2}\ln{\left(\frac{\widehat{\bm\chi}^*-\widehat{\bm\chi}}{\bm\chi^*-\bm\chi}\right)},~~~~
\bm{r}=\frac{1}{2i}\ln{\left(\frac{\bm\chi^*-\widehat{\bm\chi}}{\widehat{\bm\chi}^*-\bm\chi}\right)},\label{gb-dr}
\end{eqnarray}
while
\begin{eqnarray}
\upsilon_j=\frac{1}{2}\ln{\left[\frac{(\bm\chi^*+\beta_j)(\widehat{\bm\chi}+\beta_j)}
{(\bm\chi+\beta_j)(\widehat{\bm\chi}^*+\beta_j)}\right]},\\
\vartheta_j=\frac{1}{2i}\ln{\left[\frac{(\widehat{\bm\chi}^*+\beta_j)(\widehat{\bm\chi}+\beta_j)}
{(\bm\chi^*+\beta_j)(\bm\chi+\beta_j)}\right]}.
\end{eqnarray}
The coefficients $G_j$, $H_j$ , $M_j$ and $U_j$ are respectively:
\begin{equation}
G_j=\frac{{(\widehat{\bm\chi}}^*+\beta_j)}{(\widehat{\bm\chi}+\beta_j)(\widehat{\bm\chi}^*-\widehat{\bm\chi})}
\exp{(\upsilon_j+\bm{d})},
\end{equation}
\begin{equation}
H_j=\frac{{(\bm\chi}^*+\beta_j)}{(\widehat{\bm\chi}+\beta_j)(\bm\chi^*-\widehat{\bm\chi})}
\exp{(i\vartheta_j+i\bm{r})},
\end{equation}
and
\begin{equation}
M_j=\frac{\exp{(\bm{d})}}{(\widehat{\bm\chi}^*-\widehat{\bm\chi})},~~~~
U_j=\frac{\exp{(i\bm{r})}}{(\bm\chi^*-\widehat{\bm\chi})}.
\end{equation}
The values $\bm\chi=\bm\chi_1~\textmd{or}~\bm\chi_3$, and $\widehat{\bm\chi}$ can be found by solving Eq.(\ref{x3}) numerically. Here we use $\widehat{\bm\chi}=\bm\chi_{1,c}$. Due to the condition $\bm\chi_{r}\neq\widehat{\bm\chi}_{r}$, this solution describes general breather rather than a dark soliton (\ref{eqkmb}).

The dark soliton (\ref{eqkmb}) and the general breather (\ref{eqgb}), each with the eigenvalue $\bm\chi=\bm\chi_1$ are shown in Figs. \ref{fig12-2km2}(a) and \ref{fig12-2km2}(b) for the cases $\alpha=0.5$ and $\alpha=0.4$ respectively.
The nonlinear superposition of these two solutions is shown in Fig. \ref{fig12-2km2}(c).

\begin{figure*}[htbp]
\centering
\includegraphics[width=160mm]{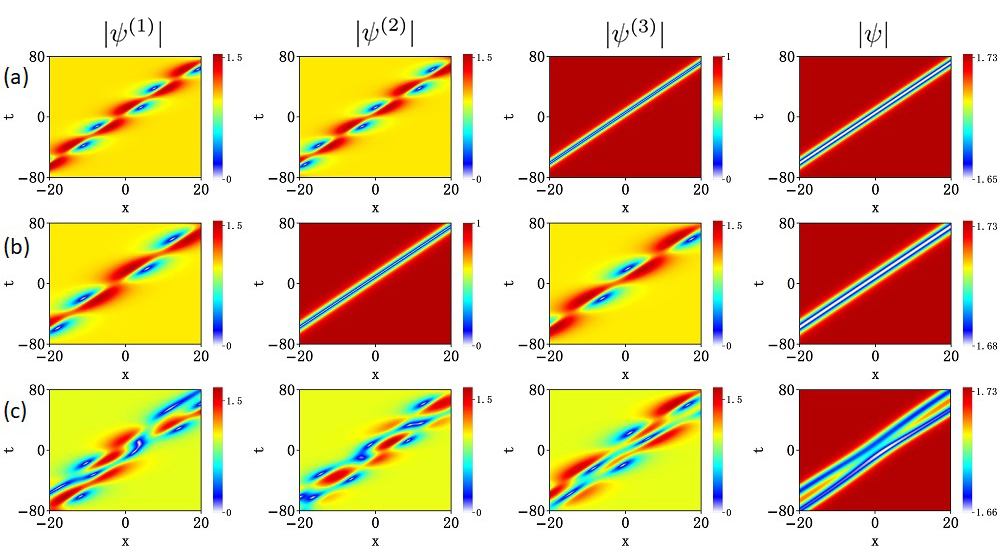}
\caption{(a) Dark soliton with two oscillating components (\ref{eqbs3-2}) with $\alpha=0.5$, (b) Dark soliton with two oscillating components (\ref{eqbs3-3}) with $\alpha=0.4$.  (c)  Second-order non-degenerate solution formed by nonlinear superposition of solutions shown in (a) and (b). Parameters $a_j=1$, and $\beta_1=\beta_2=\beta_3=0.3$.}
\label{fig13-2km3}
\end{figure*}

\begin{figure*}[htbp]
\centering
\includegraphics[width=160mm]{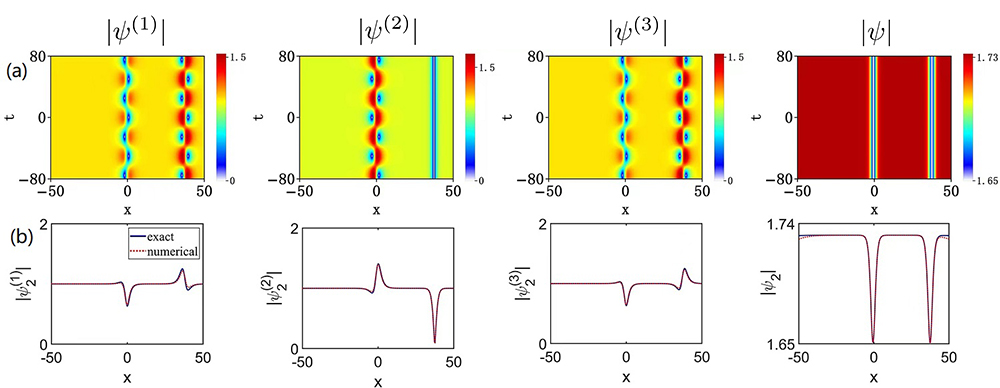}
\caption{(a) Amplitude distributions of second-order degenerate beating soliton formed by nonlinear superposition of two beating solitons shown in Figs. \ref{fig6-f-rd3}(b) and \ref{fig6-f-rd3}(c). (b) comparison of amplitude profiles between numerical simulations and exact solutions at $t=40$.
The parameters are $a_j=1$, $\beta=0$, and $\alpha=0.5$.}
\label{fig14-BE-1}
\end{figure*}

\subsection{Second-order solutions with $\beta_1=\beta_2=\beta_3$}\label{Sec5-2}

When $N=3$, the number of possible higher-order combinations is larger than in the case $N=2$.
First, we consider the nonlinear superposition of solitons with two different eigenvalues
given by Eqs. (\ref{eqbs3-2}) and (\ref{eqbs3-3}).
The details of derivation are presented in Appendix \ref{SecB-1b}.
Figure \ref{fig13-2km3}(a) shows the amplitude profiles of dark soliton (\ref{eqbs3-2}) with $\alpha=0.5$ while Fig. \ref{fig13-2km3}(b) shows the amplitude profiles of the dark soliton (\ref{eqbs3-3}) with $\alpha=0.4$. These two dark solitons have the same velocity $\beta_1$.
Their superposition is shown in Fig. \ref{fig13-2km3}(c). It shows complex beating pattern in each component. The total amplitude is a bound state of two dark solitons with weakly attractive interaction around the centre $(t,x)=(0,0)$.

Now, let us consider the nonlinear superposition of solutions given by Eqs. (\ref{eqvbs3-1}) and (\ref{eqvbs3-2}) that satisfy the condition $\beta_1=\beta_2=\beta_3=0$.  Their amplitude profiles are shown in Figs. \ref{fig6-f-rd3}(b) and \ref{fig6-f-rd3}(c) respectively. The eigenvalues $\bm\chi_1=\bm\chi_3=-i\alpha$ of these solutions are identical.
Thus, for any $\alpha$, the nonlinear superposition of these solutions is degenerate.
This solution is shown in Fig. \ref{fig14-BE-1}(a).
Due to the degeneracy, in every component, the two localised waves are well-separated in $x$ despite the zero shifts of individual solitons in the solution. The $\psi^{(1)}$ and $\psi^{(3)}$ components consist of two oscillating solitons while the $\psi^{(2)}$ component is a combination of oscillating and non-oscillating dark soliton. Moreover, the amplitudes of the r.h.s. solitons in $\psi^{(1)}$ and $\psi^{(3)}$ components are complementary. As the total amplitude should be dark solitons without oscillation, the r.h.s. soliton in the $\psi^{(2)}$ component exhibits a pure dark structure.
The total amplitude also shows two well-separated dark solitons.
 This is in sharp contrast to the non-degenerate case shown in Fig. \ref{fig13-2km3}(c).

\begin{figure*}[htbp]
\centering
\includegraphics[width=130mm]{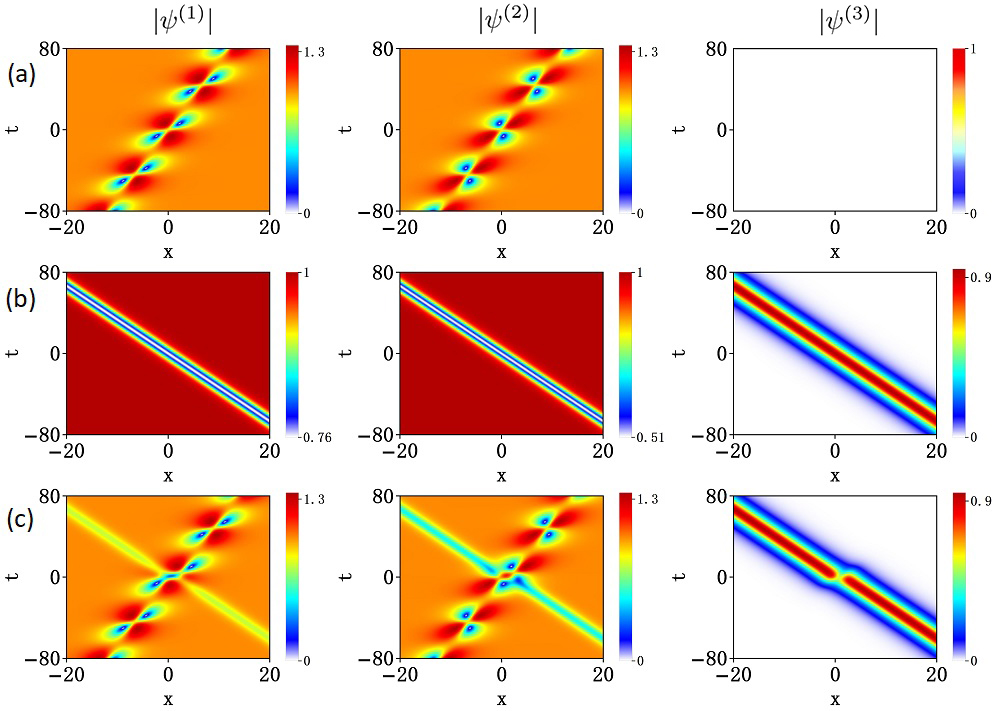}
\caption{(a) Dark soliton with two oscillating components and zero in $\psi^{(3)}$-component  when $\alpha=0.5$. (b) Dark-dark-bright soliton (\ref{gz1}) when $\alpha=0.4$. (c) Second-order non-degenerate solution formed by the nonlinear superposition of solitons shown in (a) and (b).
Parameters $\beta=0.3$, $a_1=a_2=1$, and $a_3=0$.
}
\label{fig15-gzkm1}
\end{figure*}

In order to confirm the validity of the exact solution, we performed numerical simulations starting from the initial condition which is the exact solution at $t=0$. The resulting wave profiles (dotted lines) after propagation of 40 units ($t=40$) are presented in Fig. \ref{fig14-BE-1}(b). The wave profiles according to the exact solutions are shown on the same plots by solid lines. As expected, the two profiles in each plot coincide.

\section{Second-order solutions  with $a_1=a_2\neq0$, $a_3=0$ for $N=3$}\label{Sec6}

Next, we consider the cases when one or two of the background amplitudes vanish, namely i) $a_1=a_2\neq0$, $a_3=0$; ii) $a_1\neq0$, $a_2=a_3=0$. The case when all $a_j=0$ is not allowed in the defocusing Manakov systems.
 Each of the two cases produces new nonlinear superposition.
 In this section, let us focus our attention on the case $a_1=a_2\neq0$, $a_3=0$.
The corresponding spectral parameter (\ref{Eqlambda}) reduces to
\begin{equation}\label{Eqlambda2}
\lambda=\bm\chi+\frac{a_1^2}{\bm\chi+\beta_1}+\frac{a_2^2}{\bm\chi+\beta_2},
\end{equation}
where $\bm\chi$ denotes the valid eigenvalue determined below.

\subsection{Higher-order solitons with $\beta_1=-\beta_3=\beta\neq0$, $\beta_2=0$}

The explicit expressions for the eigenvalues $\bm\chi_j$ can be obtained from Eq. (\ref{eqchi}). Namely,
\begin{eqnarray}\label{Eqchi1234-2}
\begin{split}
&\bm{\chi}_{1}=-\frac{1}{2}i\left(\alpha+\sqrt{\kappa_1-\kappa_2}\right)-\frac{1}{2}\beta,\\
&\bm{\chi}_{2}=-\frac{1}{2}i\left(\alpha-\sqrt{\kappa_1-\kappa_2}\right)-\frac{1}{2}\beta,\\
&\bm{\chi}_{3}=\beta-i\alpha,~~~~~~~~~\bm{\chi}_{4}=\beta.
\end{split}
\end{eqnarray}
where $\kappa_1=\alpha^2-4a^2-\beta^2$,
$\kappa_2=2\sqrt{4a^2+4a^2\beta^2-\alpha^2\beta^2}$. Only complex eigenvalues $\bm{\chi}_{1}$, $\bm{\chi}_{2}$, $\bm{\chi}_{3}$ are valid. They are related to each other as follows
\begin{eqnarray}
\bm\chi_{1i}+\bm\chi_{2i}=-\alpha,~~\bm\chi_{1r}=\bm\chi_{2r}.\label{eqchi13-r}
\end{eqnarray}
This means that the two eigenvalues $\bm{\chi}_{1}$, $\bm{\chi}_{2}$ in vector soliton formation play the same role. If we use $\bm{\chi}_{1}$ or $\bm{\chi}_{2}$ as the eigenvalue, we obtain vector solitons in $\psi^{(1)}$ and $\psi^{(2)}$ wave components and a zero solution in $\psi^{(3)}$ wave component. The derivation is given in Appendix \ref{SecB-2a}.
However, if we use the eigenvalue $\bm{\chi}_{3}=\beta-i\alpha$, we obtain the solution in the form of dark-dark-bright solitons. Its explicit form is given by:
\begin{eqnarray}\label{gz1}
\begin{split}
\psi^{(1)}(\bm{\chi}_{1})&=\left\{1+\frac{(\lambda^*-\lambda)\exp[\alpha (x+\beta t)-\sigma]}{(\beta_1+\bm\chi)2\sinh[\alpha (x+\beta t)+\sigma]}\right\}\psi_0^{(1)},\\
\psi^{(2)}(\bm{\chi}_{1})&=\left\{1+\frac{(\lambda^*-\lambda)\exp[\alpha (x+\beta t)-\sigma]}{2~\bm\chi\sinh[\alpha (x+\beta t)+\sigma]}\right\}\psi_0^{(2)},\\
\psi^{(3)}(\bm{\chi}_{1})&=\left\{\frac{(\lambda^*-\lambda)\exp(1/2 i \alpha^2 t-\sigma)}{2a\sinh[\alpha (x+\beta t)+\sigma]}\right\}\exp(i\theta_3).
\end{split}
\end{eqnarray}
where $\sigma=\frac{1}{2}\ln{\left(1-\frac{a^2}{(\beta_1+\bm\chi)(\beta_1+\bm\chi^*)}
-\frac{a^2}{\bm\chi\bm\chi^*}\right)}$, and $\lambda$ is given by Eq. (\ref{Eqlambda2}).

\begin{figure*}[htbp]
\centering
\includegraphics[width=130mm]{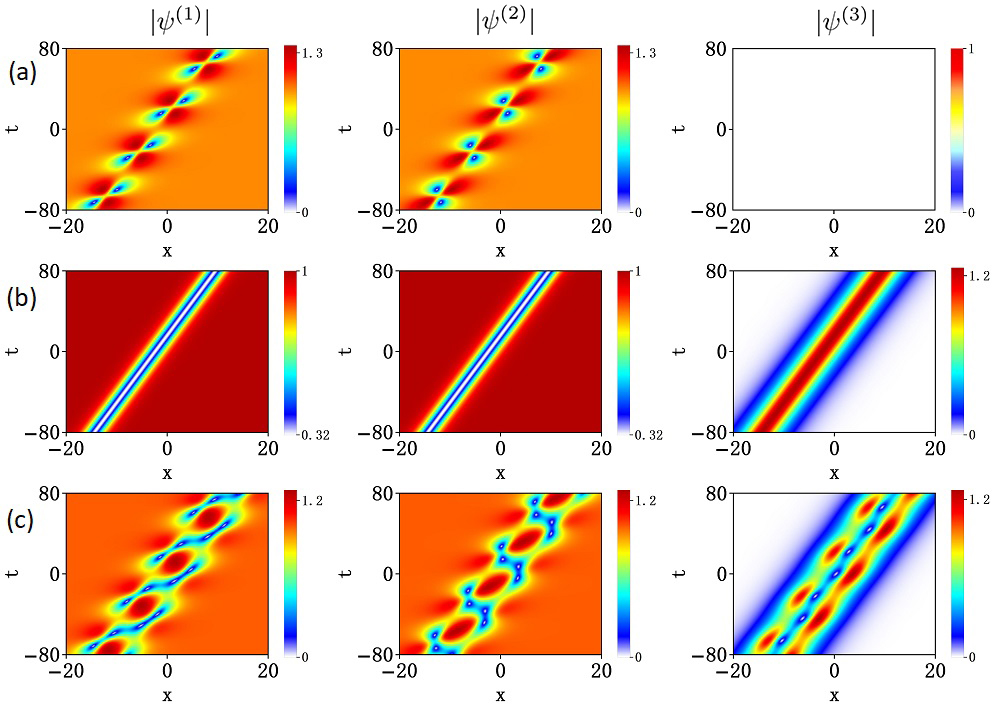}
\caption{(a) Oscillating dark soliton solution with one zero component $\psi^{(3)}$ when $\alpha=0.5$. (b) Dark-dark-bright soliton (\ref{gz1}) when $\alpha=0.4$. (c) Second-order non-degenerate solution formed by nonlinear superposition of two solitons shown in (a) and (b).
Parameters $\beta=0.3$, $a_1=a_2=1$, and $a_3=0$.
}
\label{fig16-gzkm2}
\end{figure*}

The nonlinear superposition of $\psi^{(j)}(\bm{\chi}_{1})$ and $\psi^{(j)}(\bm{\chi}_{3})$ produces new higher-order solution $\psi^{(j)}(\bm{\chi}_{1};\bm{\chi}_{3})$.
Figure \ref{fig15-gzkm1} shows the amplitude profiles of the fundamental dark solitons $|\psi^{(j)}(\bm{\chi}_{1})|$, $|\psi^{(j)}(\bm{\chi}_{3})|$, and their superposition $|\psi^{(j)}(\bm{\chi}_{1};\bm{\chi}_{3})|$.
The solution $\psi^{(j)}(\bm{\chi}_{1})$ shows a periodic four-petal pattern in $\psi^{(1)}$ and $\psi^{(2)}$ wave components while $\psi^{(3)}$ is zero. The solution $\psi^{(j)}(\bm{\chi}_{3})$ is a dark-dark-bright soliton with velocity $-\beta$.
Their superposition is the interaction of two dark solitons in $\psi^{(1)}$ and $\psi^{(2)}$ wave components. This interaction is visible in $\psi^{(3)}$-component as a phase shift of a bright soliton due to the nonlinear coupling of wave components.
The velocities of the two fundamental solitons in this example are opposite.
The solution shown in Fig. \ref{fig15-gzkm1}(c) is a particular case of that shown in Fig. \ref{fig11-2km1} when $a_3\rightarrow0$.

In analogy with the case shown in Fig. \ref{fig12-2km2}, we also consider the solitons corresponding to two different combinations of the coefficients of the vector eigenfunctions of the transformed Lax pair, i.e., $\{1,1,0,0\}$ and $\{1,0,1,0\}$. In each combination, we use $\bm{\chi}_{1}=-\frac{1}{2}i\left(\alpha+\sqrt{\kappa_1-\kappa_2}\right)-\frac{1}{2}\beta$ in Eq. (\ref{Eqchi1234-2}) as the eigenvalue.
The first combination $\{1,1,0,0\}$ produces the oscillations in the $\psi^{(1)}$ and $\psi^{(2)}$ wave components, while $\psi^{(3)}$ wave component is zero. The corresponding amplitude profiles are shown in Fig. \ref{fig16-gzkm2}(a) for $\alpha=0.5$.
The second combination $\{1,0,1,0\}$ produces dark-dark-bright soliton solution (\ref{gz1}). It is illustrated in Fig. \ref{fig16-gzkm2}(b), for $\alpha=0.4$.

\begin{figure*}[htbp]
\centering
\includegraphics[width=130mm]{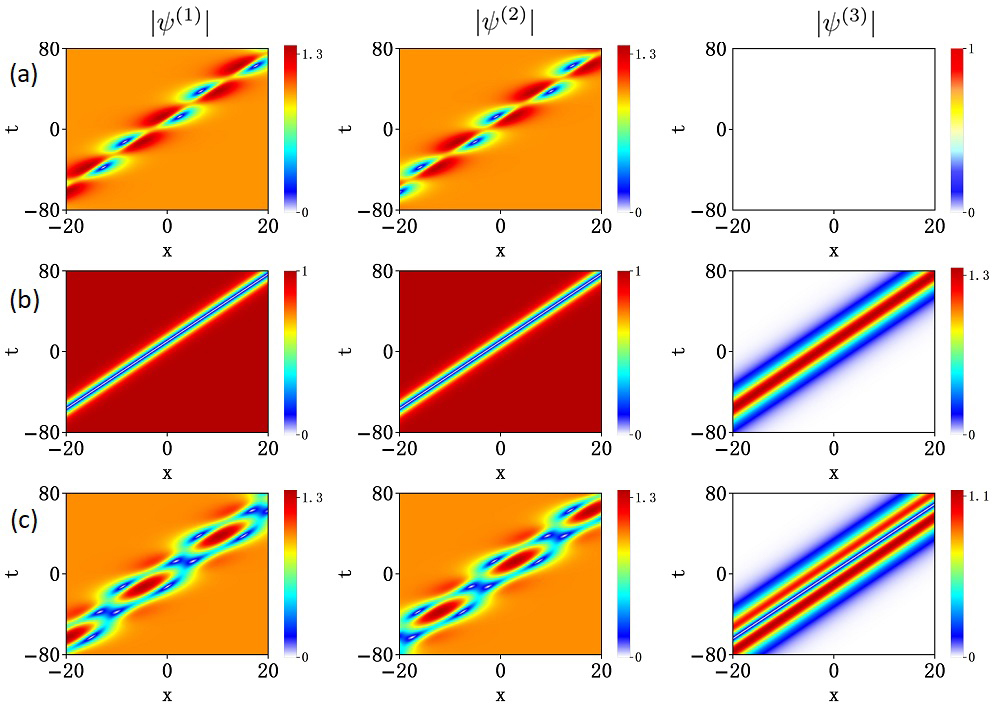}
\caption{(a) Oscillating dark soliton solution (\ref{EqVBS0-3}) with zero component in $\psi^{(3)}$ when $\alpha=0.5$. (b) Dark-dark-bright soliton (\ref{gz1}) with $\alpha=0.4$. (c) Second-order non-degenerate solution formed by nonlinear superposition of (a) and (b). Parameters $\beta_j=0.3$, $a_1=a_2=1$, and $a_3=0$.}
\label{fig17-gzkm3}
\end{figure*}

Velocities of fundamental solutions shown in Figs. \ref{fig16-gzkm2}(a) and \ref{fig16-gzkm2}(b) are unequal but very close.
As a result, their interaction shows complex oscillations of a quasi-bound state of two  solitons shown in Fig. \ref{fig16-gzkm2}(c). The first two components
$\psi^{(1)}$ and $\psi^{(2)}$ are oscillating dark solitons on the plane wave background while the component $\psi^{(3)}$ is an oscillating bright soliton.
It has a two-peak wave profile.

\subsection{Higher-order solutions with $\beta_1=\beta_2=\beta_3$}

Here, we consider the limiting cases of solutions presented in Section \ref{Sec5-2} when $a_1=a_2\neq0$, $a_3=0$. These results are analogues of the solutions shown in
Figs. \ref{fig13-2km3} and \ref{fig14-BE-1} when $a_3=0$. These are solutions (\ref{eqbs3-2}) and (\ref{eqbs3-3}) for the case $a_3=0$.
The eigenvalues (\ref{eqchi}) for equal wavenumbers $\beta_j$ when $a_1=a_2\neq0$, and $a_3=0$  are given by:
\begin{eqnarray}
\begin{split}
&&\bm{\chi}_{1}&=\bm{\chi}_{3}=-\beta_1-i\alpha,\\
&&\bm{\chi}_{2}&=\bm{\chi}_{4}=-\beta_1.
\end{split}
\end{eqnarray}
Only two of them $\bm{\chi}_{1}$ and $\bm{\chi}_{3}$ are valid eigenvalues. They are consistent with (\ref{eqbs3-2}) in the case $a_j=a$.
The Lax spectral parameter (\ref{Eqlambda2}) is
\begin{equation}\label{eqbs3-sp2}
\lambda=\bm\chi+\frac{2a_1^2}{\bm\chi+\beta_1}.
\end{equation}
This is different from Eq. (\ref{eqbs3-sp1}). Using this spectral parameter, we obtain two types of fundamental solitons for two combinations of coefficients of the eigenfunctions: $(1,1,0,0)$ and $(1,0,1,0)$.
The first combination $(1,1,0,0)$ with the eigenvalue $\bm{\chi}_{1}=-\beta_1-i\alpha$ gives:
\begin{eqnarray}\label{EqVBS0-3}
\begin{split}
\psi^{(1)}(\bm{\chi}_{1})&=\psi^{(1)}_0(\psi_{DS}+\psi_{BS})\\
\psi^{(1)}(\bm{\chi}_{1})&=\psi^{(2)}_0(\psi_{DS}-\psi_{BS})\\
\psi^{(3)}(\bm{\chi}_{1})&=0.
\end{split}
\end{eqnarray}
where
\begin{eqnarray}
\psi_{DS}&=&1+\frac{(\alpha^2-2a^2)\exp{[\alpha( x-\beta_1 t)-d]}}{2a^2\alpha^2\cosh{[\alpha (x-\beta_1 t)+d]}},\\
\psi_{BS}&=&\frac{i(\alpha^2-2a^2)\exp{(1/2i\alpha^2 t-d)}}{2a^2\alpha\cosh{[\alpha (x-\beta_1 t)+d]}},
\end{eqnarray}
and $d=\frac{1}{2}\ln{\left(\frac{1}{\alpha^2}-\frac{1}{2a^2}\right)}$. In this case, the two wave components $\psi^{(1)}$ and $\psi^{(2)}$ are oscillating while the third wave component $\psi^{(3)}$ is zero. Velocity of this soliton is $\beta_1$. The solution is shown in Fig. \ref{fig17-gzkm3}(a) for $\alpha=0.5$. The difference from the solution shown in Fig. \ref{fig13-2km3} is that, it has oscillations in $\psi^{(1)}$ and $\psi^{(2)}$ wave components. The $\psi^{(3)}$ wave component is zero.
The second combination $\{1,0,1,0\}$ yields the vector dark-dark-bright soliton solution (\ref{gz1}) where $\bm\chi_1=-\beta_1-i\alpha$. It is shown in Fig. \ref{fig17-gzkm3}(b).
Velocity of this soliton is also $\beta_1$.

The nonlinear superposition of these two fundamental solutions is shown in  Fig. \ref{fig17-gzkm3}(c). It reveals complex oscillating patterns in $\psi^{(1)}$ and $\psi^{(2)}$ wave components.
The third wave component $\psi^{(3)}$ is the two-hump bright soliton also with oscillating structure. This result is a limiting case of the solution shown in Fig. \ref{fig13-2km3} when $a_3\rightarrow 0$.

\begin{figure*}[htbp]
\centering
\includegraphics[width=130mm]{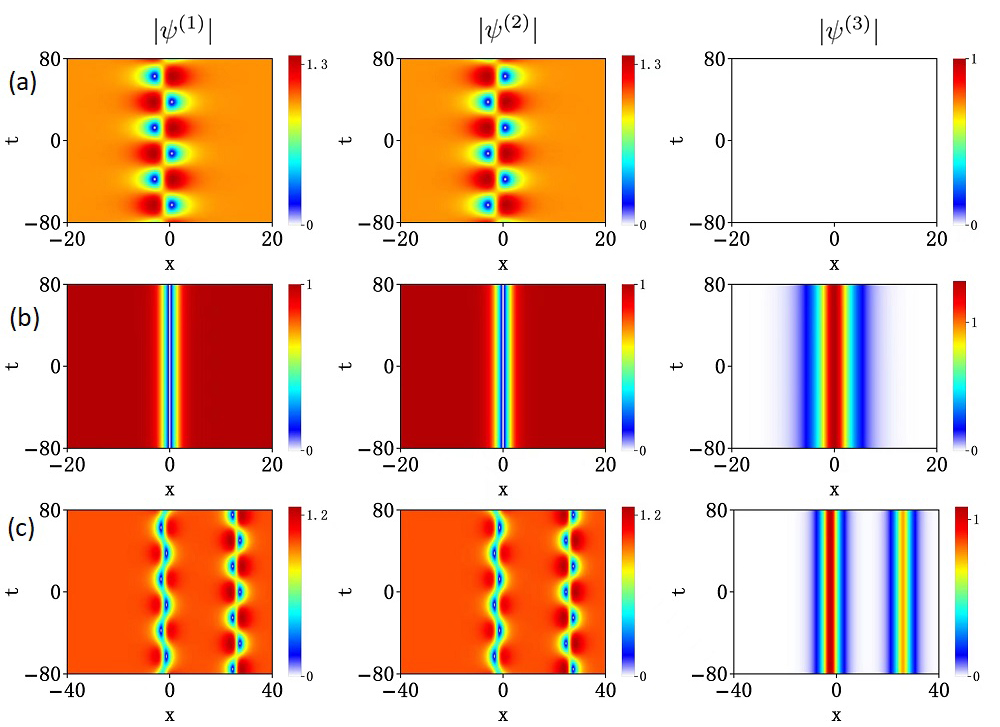}
\caption{(a) Oscillating dark soliton solution (\ref{EqVBS0-3}) with the third zero component when $\alpha=0.5$. (b) Dark-dark-bright soliton (\ref{gz1}) with $\alpha=0.5$. (c) Second-order degenerate solution formed by nonlinear superposition of solutions in (a) and (b). Parameters  $\beta=0$, and $a_3=0$. }
\label{fig18-BE-2}
\end{figure*}

Let us now consider the fundamental dark solitons (\ref{eqvbs3-1}) and (\ref{eqvbs3-2}) and their nonlinear superposition when $a_3=0$.
Similar to the case shown in Fig. \ref{fig17-gzkm3}, the solution (\ref{eqvbs3-1}) reduces to (\ref{EqVBS0-3}) with zero velocity, $\beta_j=0$. It is shown in Fig. \ref{fig18-BE-2}(a).
This solution oscillates in $\psi^{(1)}$ and $\psi^{(2)}$ wave components
 and has zero third component.
The solution (\ref{eqvbs3-2}) becomes a dark-dark-bright soliton (\ref{gz1}) with zero velocity.
It is shown in Fig. \ref{fig18-BE-2}(b).
For a given $\alpha$, these two solitons have identical eigenvalues.
Thus, their nonlinear superposition is a degenerate second-order soliton.
 It is shown in Fig. \ref{fig18-BE-2}(c). Due to the degeneracy, the two solitons are separated in space.
 They have the same period in $t$. However, their phases are shifted relative to each other.
 Oscillations are observed only in $\psi^{(1)}$ and $\psi^{(2)}$ wave components.
 The $\psi^{(3)}$ component shows two well-separated bright solitons.
These solutions are the limiting cases of those shown in Fig. \ref{fig14-BE-1} when $a_3\rightarrow 0$.

\section{Second-order solutions for $N=3$ when $a_1\neq0$, $a_2=a_3=0$}\label{Sec7}

Now, we consider the case when two background components are zero $a_2=a_3=0$, but $a_1\neq0$.
In this case, the resulting Lax spectral parameter becomes
\begin{equation}\label{Eqlambda3}
\lambda=\bm\chi+\frac{a_1^2}{\bm\chi+\beta_1},
\end{equation}
where $\bm\chi$ is determined below.
Second-order exact soliton solution is constructed at the second step of Darboux transformation using the spectral parameter (\ref{Eqlambda3}).
As in Section \ref{Sec6}, two different combinations of the coefficients of the vector eigenfunctions of the transformed Lax pair are used, i.e., $\{1,1,0,0\}$ and $\{1,0,1,0\}$.

In the first case, $\{1,1,0,0\}$, the explicit form of the soliton solution is given by:
\begin{eqnarray}\label{eq-a2a3-1}
\begin{split}
\psi^{(1)}&=\psi_0^{(1)}\left[1+\frac{(\lambda^*-\lambda)\exp(\alpha (x+\bm\chi_r t)-\xi)}{(\beta_1+\bm\chi)2\sinh(\alpha (x+\bm\chi_r t)+\xi)}\right],\\
\psi^{(2)}&=\frac{(\lambda^*-\lambda)\exp(-\frac{i}{2}(2\bm\chi_rx+(\bm\chi_r^2-\alpha^2) t)-\xi)}{2\sinh(\alpha (x+\bm\chi_r t)+\xi)\exp{(-i\theta_2)}},\\
\psi^{(3)}&=0.
\end{split}
\end{eqnarray}
where
\begin{eqnarray}
\xi=\frac{1}{2}\ln{\left(\frac{-a_1^2+(\beta_1+\bm\chi)(\beta_1+\bm\chi^*)}
{a_1^2(\beta_1+\bm\chi)(\beta_1+\bm\chi^*)}\right)}.
\end{eqnarray}
This solution describes dark and bright solitons in $\psi^{(1)}$ and $\psi^{(2)}$ wave components, respectively.  The third component is zero.

The second combination, $\{1,0,1,0\}$, provides a similar solution but with zero in the second  component:
\begin{eqnarray}\label{eq-a2a3-2}
\begin{split}
\psi^{(1)}&=\psi_0^{(1)}\left[1+\frac{(\lambda^*-\lambda)\exp(\alpha (x+\bm\chi_r t)-\xi)}{(\beta_1+\bm\chi)2\sinh(\alpha (x+\bm\chi_r t)+\xi)}\right],\\
\psi^{(2)}&=0,\\
\psi^{(3)}&=\frac{(\lambda^*-\lambda)\exp(1/2 i\bm\alpha^2 t-\xi)}{2\sinh(\alpha (x+\bm\chi_r t)+\xi)}\exp{(i\theta_3)}.
\end{split}
\end{eqnarray}

\begin{figure*}[htbp]
\centering
\includegraphics[width=130mm]{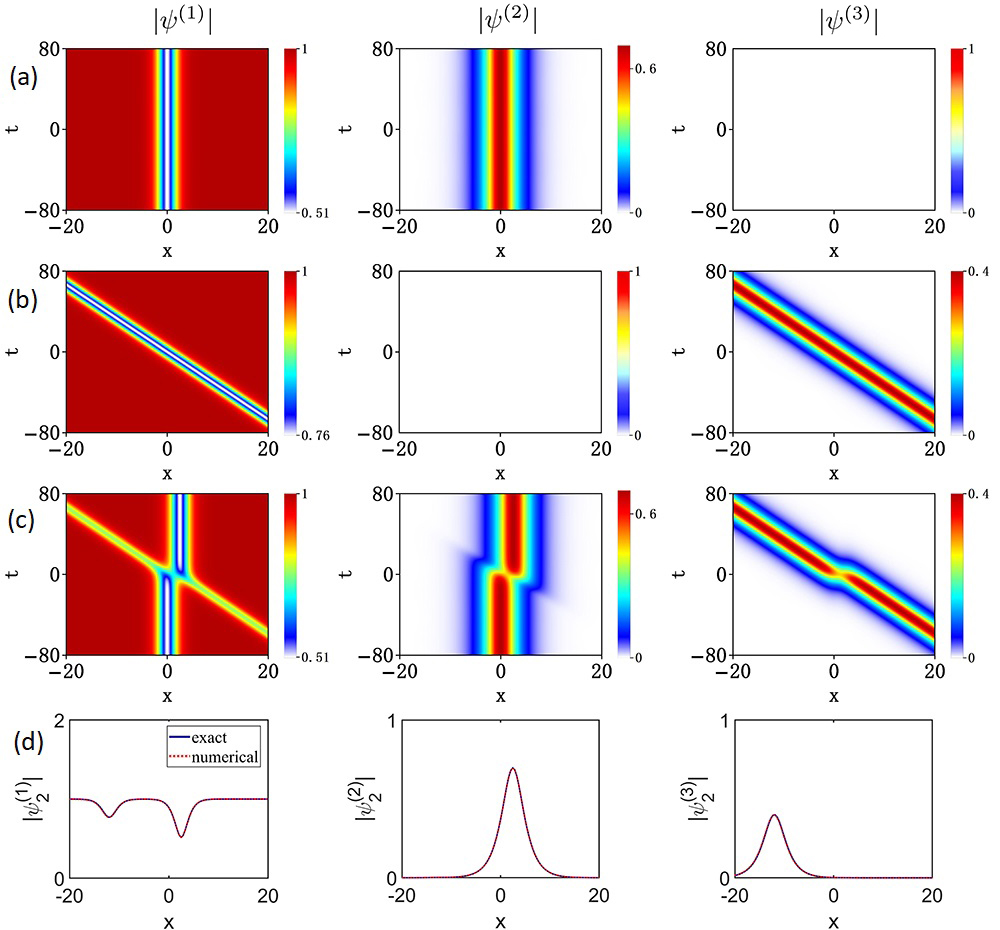}
\caption{(a) Dark-bright soliton (\ref{eq-a2a3-1}) with the third zero component when $\alpha=0.5$. (b) Dark-bright soliton (\ref{eq-a2a3-2}) with the second zero component when $\alpha=0.5$. (c) Second-order non-degenerate solution formed by nonlinear superposition of solitons shown in (a) and (b). (d) Amplitude profiles in (c) at $t=40$ obtained from the exact solutions (solid curves) and numerical simulations (dashed curve). Parameters $\beta=0.3$, $a_1=1$, and $a_2=a_3=0$.}
\label{fig19-guzi1}
\end{figure*}

Below, we consider these two solitons for specific eigenvalues.

\subsection{Second-order solutions when $\beta_1=-\beta_3=\beta\neq0$, and $\beta_2=0$}

The four eigenvalues in this case are given by:
\begin{eqnarray}\label{Eqchi1234-1}
\begin{split}
&\bm{\chi}_{1}=-i\alpha,~~~~~~~~~~~\bm{\chi}_{2}=0,\\
&\bm{\chi}_{3}=-i\alpha+\beta,~~~~~\bm{\chi}_{4}=\beta.
\end{split}
\end{eqnarray}
As discussed, only complex eigenvalues $\bm{\chi}_{1}$, $\bm{\chi}_{3}$ are valid.

Figure \ref{fig19-guzi1}(a) shows the evolution of amplitude profiles of the vector soliton (\ref{eq-a2a3-1}) corresponding to the eigenvalue $\bm{\chi}_{1}=-i\alpha$. The solution is a zero velocity dark soliton in the first component $\psi^{(1)}$ and a bright soliton in the second component $\psi^{(2)}$. The third component $\psi^{(3)}$ is zero. The amplitude profiles of the vector soliton (\ref{eq-a2a3-2}) with the eigenvalue, $\bm{\chi}_{3}=-i\alpha+\beta$, are shown in
Fig. \ref{fig19-guzi1}(b). This soliton has a non-zero velocity $-\beta$. It is a dark soliton in the first component, $\psi^{(1)}$, and bright soliton in the third component, $\psi^{(3)}$. The second component, $\psi^{(2)}$, is zero.

The nonlinear superposition of these two solitons is shown in Fig. \ref{fig19-guzi1}(c). The plot shows an elastic collision of two vector solitons with a phase shift at $t=0$. The phase shift can be clearly seen also in the second and third components of the wave field. These results are the limiting cases of those shown in Fig. \ref{fig11-2km1} when $a_2\rightarrow 0$, $a_3\rightarrow 0$.

In order to confirm the validity of the exact solutions, we performed numerical simulations of this higher-order solution starting from the initial conditions provided by the exact solution at $t=0$. The amplitude profiles of the exact solution (solid curves) and the numerical simulations (dashed curves) at $t=40$ are shown in Fig. \ref{fig19-guzi1}(d). There is an excellent agreement between them, as expected.

\begin{figure*}[htbp]
\centering
\includegraphics[width=130mm]{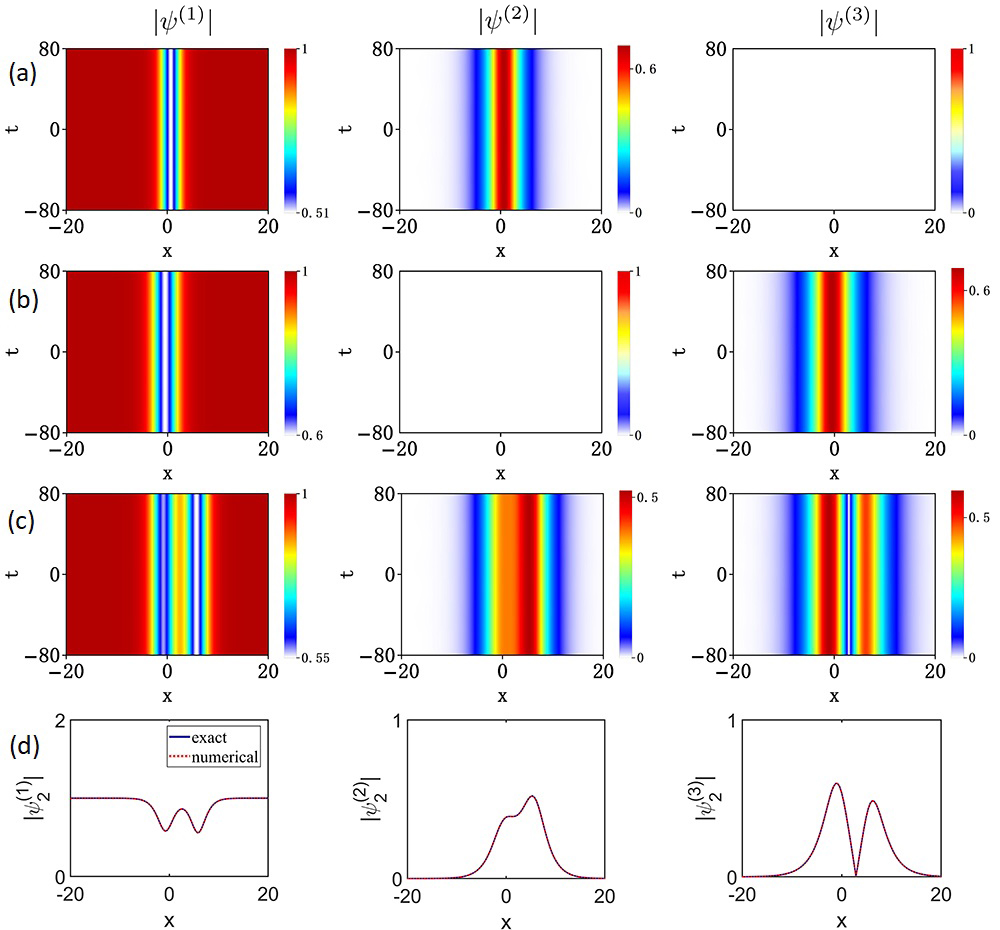}
\caption{(a) Dark-bright vector soliton (\ref{eq-a2a3-1}) with zero third component when $\alpha=0.5$. (b) Dark-bright vector soliton (\ref{eq-a2a3-2}) with the second zero component when $\alpha=0.4$. (c) Second-order non-degenerate solution formed by nonlinear superposition of vector solitons shown in (a) and (b). and (d) Amplitude profiles at $t=80$ of the second-order solution shown in (c) obtained from the exact solutions (solid curves) and from numerical simulations (dashed curves).
Parameters $\beta=0.3$, $a_1=1$, and $a_2=a_3=0$.}
\label{fig20-guzi2}
\end{figure*}

Another type of a higher-order solution is formed by the nonlinear superposition of solutions (\ref{eq-a2a3-1}) and (\ref{eq-a2a3-2}) corresponding to a single eigenvalue (either $\bm\chi_1$ or $\bm\chi_3$) but with different $\alpha$. To be specific, we chosen the eigenvalue $\bm\chi_1$. The corresponding fundamental soliton solutions are shown in Figs. \ref{fig20-guzi2}(a) and  \ref{fig20-guzi2}(b). These two solitons have zero velocity due to the condition $\bm\chi_{1r}=0$. Their superposition results in a new form of bound state of two solitons.  It is shown in Fig. \ref{fig20-guzi2}(c). The first component, $\psi^{(1)}$, is an asymmetric bound state of two dark solitons with the profile that has two unequal dips.
The $\psi^{(2)}$ and $\psi^{(3)}$ components are the asymmetric bound states of two bright  solitons with two unequal humps.

These results are limiting cases of those shown in Fig. \ref{fig12-2km2} when $a_2\rightarrow 0$,  $a_3\rightarrow 0$.
Namely, the moving vector solitons shown in Fig. \ref{fig12-2km2}(c) reduce to the ones shown in Fig. \ref{fig20-guzi2}(c) when $a_2\rightarrow 0$, and $a_3\rightarrow 0$. Validity of the results shown in Fig. \ref{fig20-guzi2}(c) are confirmed using numerical simulations starting from the initial conditions provided by the exact solution at $t=0$. Figure \ref{fig20-guzi2}(d) shows the wave profiles at $t=80$ obtained from the exact solutions (solid curves) and from the numerical simulations (dashed curves). As expected, the two profiles coincide.

\begin{figure*}[htbp]
\centering
\includegraphics[width=130mm]{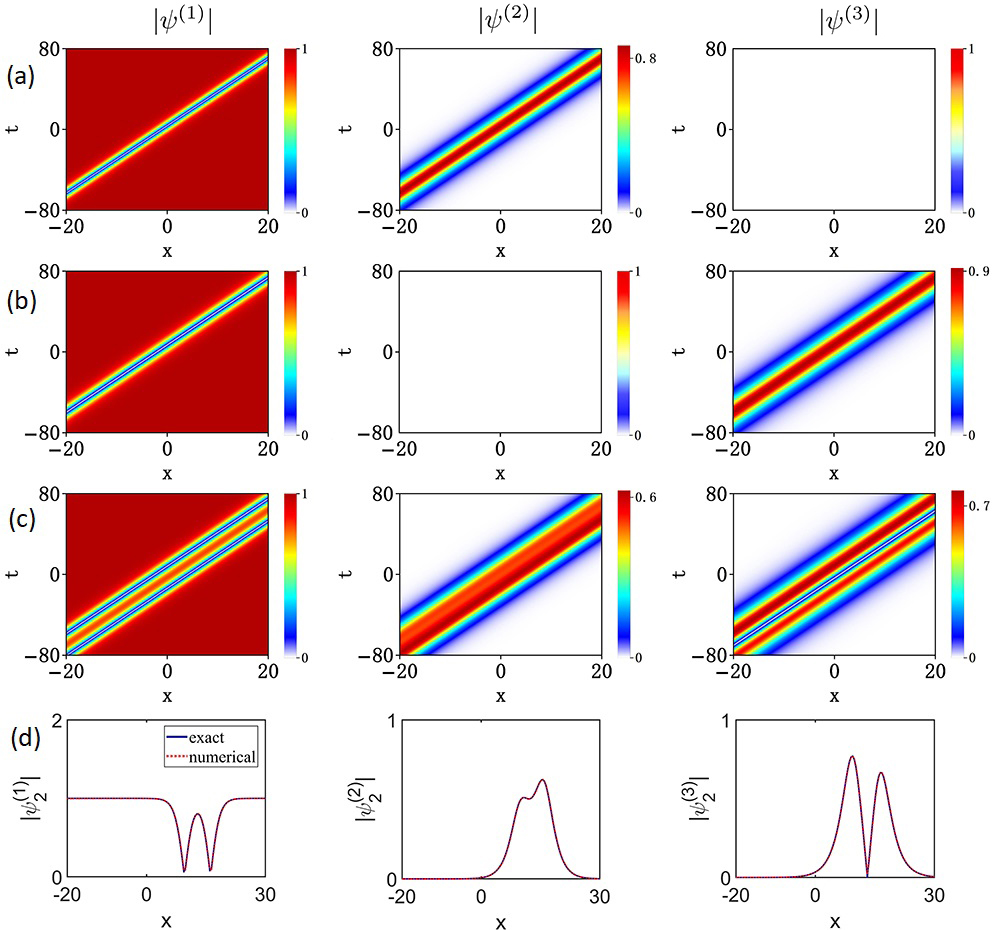}
\caption{(a) Dark-bright vector soliton (\ref{eq-a2a3-1}) with zero third component when $\alpha=0.5$. (b) Dark-bright vector soliton (\ref{eq-a2a3-2}) with zero second component when $\alpha=0.4$. (c) Second-order non-degenerate solution formed by nonlinear superposition of vector solitons shown in (a) and (b). (d) Amplitude profiles of the second-order solution shown in (c) at $t=40$ obtained from the exact solutions (solid curves) and numerical simulations (dashed curves).
Parameters $\beta_j=0.3$, $a_1=1$, and $a_2=a_3=0$.}
\label{fig21-guzi3}
\end{figure*}

\subsection{Second-order solutions when $\beta_1=\beta_2=\beta_3$}

In the case of equal wavenumbers $\beta_1=\beta_2=\beta_3$, the eigenvalues are given by
\begin{eqnarray}
\begin{split}
&&\bm{\chi}_{1}&=\bm{\chi}_{3}=-\beta_1-i\alpha,\\
&&\bm{\chi}_{2}&=\bm{\chi}_{4}=-\beta_1.
\end{split}
\end{eqnarray}
Here, only one of the complex eigenvalues (either $\bm{\chi}_{1}$ or $\bm{\chi}_{3}$) is valid. Using $\bm{\chi}_{1}$, we construct the fundamental vector soliton solutions for different combinations of the eigenfunctions. These solutions are given by (\ref{eq-a2a3-1}) and (\ref{eq-a2a3-2}). They are shown in Figs. \ref{fig21-guzi3}(a) and \ref{fig21-guzi3}(b) respectively.
These two fundamental solitons have the same velocity $\beta_1$. The soliton in Fig. \ref{fig21-guzi3}(a) has the $\psi^{(1)}$ component in the form of a dark soliton and the $\psi^{(2)}$ component in the form of a bright soliton while the third component $\psi^{(3)}$ is zero.
The soliton in Fig. \ref{fig21-guzi3}(b) has the $\psi^{(1)}$ component in the form of a dark soliton and the $\psi^{(3)}$ component in the form of a bright soliton while the second component $\psi^{(2)}$ is zero. The nonlinear superposition of these two fundamental vector solitons is shown in Fig. \ref{fig21-guzi3}(c).
Similar to the case shown in Fig. \ref{fig20-guzi2},  the superposition is a bound state of two solitons but with finite velocity $\beta_1=0.3$. The $\psi^{(1)}$ component is a bound state of two equal dark solitons. On the other hand, the $\psi^{(2)}$ ($\psi^{(3)}$) component is a bound state of two unequal in-phase (out-of-phase) bright solitons.
These results are the limiting case of those shown in Fig. \ref{fig13-2km3} when $a_2\rightarrow 0$, and $a_3\rightarrow 0$. Figure \ref{fig21-guzi3}(d) shows the comparison of the wave profiles at $t=40$ obtained from the exact solutions and numerical simulations. This way, we confirmed that the exact solutions shown here are indeed correct.

\begin{figure*}[htbp]
\centering
\includegraphics[width=130mm]{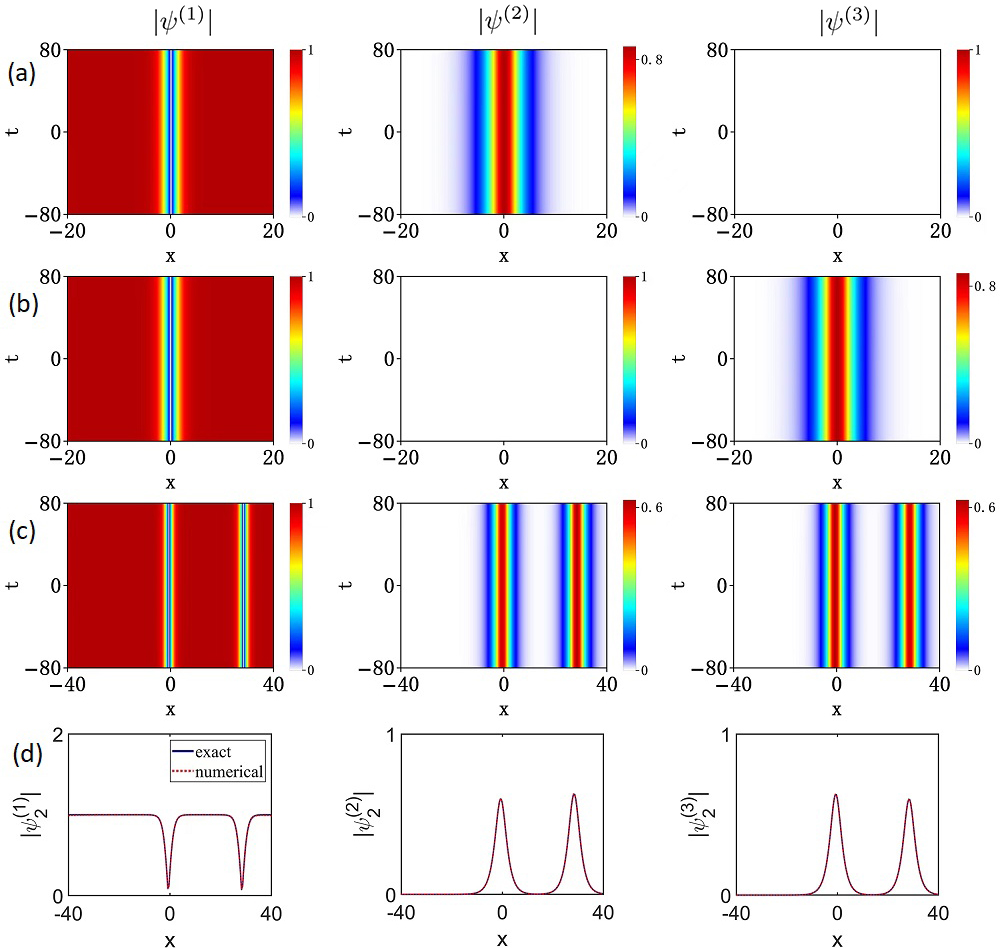}
\caption{(a) Dark-bright vector soliton (\ref{eq-a2a3-1}) with zero third component when $\alpha=0.5$. (b) Dark-bright vector soliton (\ref{eq-a2a3-2}) with zero second component when $\alpha=0.5$. (c) Second-order degenerate solution formed by nonlinear superposition of vector solitons shown in (a) and (b). (d) Amplitude profiles of the second-order solution shown in (c) at $t=40$ obtained from the exact solutions (solid curves) and numerical simulations (dashed curves).
Parameters $\beta=0$, and $a_2=a_3=0$.}
\label{fig22-BE-3}
\end{figure*}

We finally consider the second-order vector solitons corresponding to the solution shown in Fig. \ref{fig14-BE-1}(a) when $a_3\rightarrow0$ (or the solution shown in Fig. \ref{fig18-BE-2}(c) when  both $a_2\rightarrow0$, and $a_3\rightarrow0$).
Figures \ref{fig22-BE-3}(a) and \ref{fig22-BE-3}(b) show the fundamental vector solitons used for obtaining the second-order solution.
The nonlinear superposition of these two vector solitons results in a second-order soliton.
It is shown in Fig. \ref{fig22-BE-3}(c).
It is a degenerate solution because the eigenvalues of the two fundamental solutions coincide. The situation is similar to the one shown in Fig. \ref{fig18-BE-2}(c). The resulting second-order solution consists of two equal well separated dark-bright solitons propagating in parallel. Each of the wave profiles shown in Fig. \ref{fig22-BE-3}(d) is symmetric in $x$.
The parallel solitons shown in Figs. \ref{fig20-guzi2}(c), \ref{fig21-guzi3}(c), and \ref{fig22-BE-3}(c), come from the nonlinear superposition between two fundamental dark-bright solitons with the same velocity, each associated with a zero solution in different components.

\section{Conclusions}\label{Sec8}

In conclusion, we have studied fundamental vector solitons and their interaction in the \emph{defocusing} regime of Manakov equations. We derived multi-parameter family of fundamental vector soliton solutions in analytic form and presented the existence diagrams of these solitons for the two- and three-component Manakov equations.
We have found that vector solitons exist only in finite areas of the ($\alpha$, $\beta$) plane.
Within these areas, the dark soliton components oscillate.
At the boundaries of the existence diagrams, vector solitons are transformed into plain vector dark solitons.

We have also provided exact solutions for the interaction of fundamental solitons.
These are nonlinear superpositions of fundamental vector dark solitons.
We found a rich variety of interaction patterns of two solitons each with it own eigenvalue.
The two eigenvalues may differ or they can coincide.
The corresponding solutions are non-degenerate or degenerate second-order solutions respectively. We confirmed the correctness of our theoretical results using numerical simulations.

Because of the widespread fundamental and practical interest to physical systems described by the set of Manakov equations in the defocusing regime, we believe that our results may have a significant impact on experimental physics.

\section*{ACKNOWLEDGEMENTS}
The work of Liu is supported by the NSFC (Grants No. 12175178, and No. 12047502),
the Natural Science basic Research Program of Shaanxi Province (Grant No. 2022KJXX-71), and Shaanxi Fundamental
Science Research Project for Mathematics and Physics (Grant No. 22JSY016).
The work of Akhmediev is supported by the Qatar National Research Fund (grant NPRP13S-0121-200126).

\begin{appendix}
\section{Vector soliton solutions for $N=2$}\label{SecA}
\subsection{Vector solitons for $N=2$ with $a_j=a\neq0$}\label{SecA-1}
\subsubsection{Vector solitons when $\beta_1=-\beta_2=\beta\neq0$}\label{SecA-1a}

We represent Eqs. (\ref{eq1}) as the condition of compatibility of two linear equations:
\begin{eqnarray}
\bf{\Psi}_x=\textbf{U}\Psi,~~~\bf{\Psi}_t=\textbf{V}\Psi,\label{Lax-pair}
\end{eqnarray}
with the matrices
\begin{small}
\begin{eqnarray}
\begin{split}\label{UV1}
&~~~~~~~~~~~~~~~\mathbf{U}=i\left(\frac{\lambda}{2}\left(\bm\sigma+\mathbf{I}\right)+\mathbf{Q}\right),\\
&\mathbf{V}=i\left(\frac{\lambda^2}{4}\left(\bm\sigma+\mathbf{I}\right)+\frac{\lambda}{2}\mathbf{Q}-\frac{1}{2}\bm\sigma(\mathbf{Q}^2+i\mathbf{Q}_x)+\bm{a}^2\mathbf{I}\right)
\end{split}
\end{eqnarray}
\end{small}
where
\begin{eqnarray}
&\mathbf{Q}=\left(
\begin{array}{ccc}
0& \bm{-\psi^{\dagger}}\\
\bm{\psi} &\bm{0}_{N\times N}\\
\end{array}
\right),
&\bm{\sigma}=\left(
\begin{array}{ccc}
1& \bm{0}_{1\times N}\\
\bm{0}_{N\times 1} & -\mathbf{I}_{N\times N}\\
\end{array}
\right).
\end{eqnarray}
Here, the vector function $\bm{\psi}$=$\left(\psi^{(1)}, \psi^{(2)}, ... ,\psi^{(j)}\right)^\textsf{T}$, $\dagger$ denotes the matrix transpose and complex conjugate, $\mathbf{I}$ is an identity matrix, $\lambda$ is the spectral parameter, and $\bm{a}^2=\sum_{j=1}^{j=N}\left(a_j^2\right)$.
The system of Manakov equations (\ref{eq1}) follows from the compatibility condition
\begin{eqnarray}
\mathbf{U}_t-\mathbf{V}_x+[\mathbf{U}, \mathbf{V}]=0.
\end{eqnarray}
For $N=2$, using a diagonal matrix $S$=diag$(1,e^{-i\theta_1},e^{-i\theta_2})$, the Lax pair can be rewritten as:
\begin{eqnarray}\label{UV}
\begin{small}
\begin{split}
&\tilde{\mathbf{U}}=i\left(
\begin{array}{cccc}
\lambda & -a_1 & -a_2\\
a_1 & -\beta_1 & 0\\
a_2 & 0 & -\beta_2\\
\end{array}
\right),
&\tilde{\mathbf{V}}=-\frac{i}{2}\tilde{\mathbf{U}}^2+2i\bm{a}^2\mathbf{I}.
\end{split}
\end{small}
\end{eqnarray}
The linear eigenvalue problem in terms of the transformed
Lax pair (\ref{UV}) is given by
\begin{eqnarray}\label{x2}
\det(\tilde{\mathbf{U}}-i\bm{\chi})=0.
\end{eqnarray}
Eq. (\ref{x2}) admits three eigenvalues $\bm\chi_{n,l}$, ($l=a,b,c$).
To obtain the solution of Eq. (\ref{UV1}), we further diagonalise the matrices $\tilde{\mathbf{U}}$ and $\tilde{\mathbf{V}}$. Namely, we have
\begin{equation}
\varphi_{x}=\tilde{\mathbf{U}}\varphi,~~~~\varphi_{t}=\tilde{\mathbf{V}}\varphi,~~~~
\varphi=H^{-1}S\bf{\Psi}.\label{xxxx}
\end{equation}
where the transformation matrix $\mathbf{H}$ is:
\begin{eqnarray}
\begin{split}
&\mathbf{H}=\left(
\begin{array}{cccc}
1 & 1 & 1\\
\frac{a_1}{\beta_1+\bm\chi_{n,a}} & \frac{a_1}{\beta_1+\bm\chi_{n,b}} & \frac{a_1}{\beta_1+\bm\chi_{n,c}}\\
\frac{a_2}{\beta_2+\bm\chi_{n,a}} & \frac{a_2}{\beta_2+\bm\chi_{n,b}} & \frac{a_2}{\beta_2+\bm\chi_{n,c}}\\
\end{array}
\right).\\
\end{split}
\end{eqnarray}
Solving Eq. (\ref{xxxx}), we have
\begin{eqnarray}
\begin{split}
\varphi_{n,l}&=c_{n,l}\exp{\left[ i\bm\chi_{n,l}x+\frac{1}{2} \left(4\bm{a}^2+\bm\chi_{n,l}^2\right) t\right] },
\end{split}
\end{eqnarray}
where $c_{n,1}$ ($l=a,b,c$) are arbitrary constants corresponding to the vector eigenfunctions of the transformed Lax pair $\varphi_{n,l}$.
Finally, the eigenfunctions $\mathbf{\Psi}_n=(R_{n},S_{n},W_{n})$ are given by
\begin{eqnarray}
\begin{split}
R_{n}&=\varphi_{n,a}+\varphi_{n,b}+\varphi_{n,c},\\
S_{n}&=\psi_{0}^{(1)}\left(\sum_{l=a}^c\frac{\varphi_{n,l}}{\beta_1+\bm\chi_{n,l}}\right),\\
W_{n}&=\psi_{0}^{(2)}\left(\sum_{l=a}^c\frac{\varphi_{n,l}}{\beta_2+\bm\chi_{n,l}}\right).
\end{split}
\end{eqnarray}
The fundamental (first-order, $n=1$) vector solution of the two-component Manakov equations can be obtained through the Darboux transformation \cite{DT}. Namely
\begin{eqnarray}\label{eqdt2}
\begin{split}
&&\psi_{1}^{(1)}=\psi_{0}^{(1)}+\frac{(\lambda_{1}^*-\lambda_{1})R_{1}^*S_{1}}{|R_{1}|^2-|S_{1}|^2-|W_{1}|^2},\\
&&\psi_{1}^{(2)}=\psi_{0}^{(2)}+\frac{(\lambda_{1}^*-\lambda_{1})R_{1}^*W_{1}}{|R_{1}|^2-|S_{1}|^2-|W_{1}|^2}.\\
\end{split}
\end{eqnarray}
If one of the $c_{1,l}$ is zero, we obtain exact solutions that describe the dynamics of a single soliton. However, for the defocusing case, we have to set $c_{1,c}=0$ so that $\varphi_{1,c}=0$. Below we clarify this point.

Let us focus on the linear eigenvalue problem (\ref{x2}). The latter directly leads to
\begin{eqnarray}\label{lambda-appendix}
\lambda=\bm\chi+\sum_{j=1}^2\frac{a_j^2}{\bm\chi+\beta_j},
\end{eqnarray}
where the eigenvalue $\bm\chi$ is given by Eq. (\ref{eqchi}). However, not all these eigenvalues are valid, as shown in Section \ref{Sec3}.
Substituting one of the valid eigenvalues [e.g., $\bm\chi_1$ (or $\bm\chi_2$) given by (\ref{Eqchi1234})] into (\ref{lambda-appendix}),
we obtain the corresponding Lax spectrum $\lambda_1$.
Using this spectrum and solving (\ref{x2}), we obtain three eigenvalues ($\bm\chi_{1,a}$, $\bm\chi_{1,b}$, $\bm\chi_{1,c}$).
We find that $\bm\chi_{1,a}=\bm\chi_1$, $\bm\chi_{1,b}=\bm\chi_2=\bm\chi_1+i\alpha$, and $\bm\chi_{1,c}\neq\bm\chi_{1,a},\bm\chi_{1,b}$.
Clearly, $\bm\chi_{1,a}$, $\bm\chi_{1,b}$ are valid eigenvalues. However, $\bm\chi_{1,c}$ cannot be used to generate any valid solution.
Without loss of generality, we use the coefficients: $(c_{1,a},c_{1,b},c_{1,c})=(1,1,0)$.
Using this set of the coefficients, we obtain the fundamental soliton solution of the two-component Manakov equations.
The explicit form of (\ref{eqdt2}) is given by (\ref{eqkmb}).

\subsubsection{Vector solitons when $\beta_1=\beta_2$}\label{SecA-1b}

The general soliton solution derived above reduces to the soliton solution with zero velocity when $\beta=0$. The moving soliton can be obtained by employing a Galilean transformation.
Below, we show how to obtain the general vector soliton solution via the Darboux transformation.
When $\beta_1=\beta_2$, the eigenvalue $\bm\chi_{n,a}=-\beta_1-i\alpha$, $\bm\chi_{n,b}=-\beta_1$.
The transformation matrix $\mathbf{H}$ can be rewritten as:
\begin{eqnarray}
\begin{split}
&\mathbf{H}=\left(
\begin{array}{cccc}
1 & 0 & 1\\
\frac{a_1}{\beta_1+\bm\chi_{n,a}} & a_1 & \frac{a_1}{\beta_1+\bm\chi_{n,c}}\\
\frac{a_2}{\beta_2+\bm\chi_{n,a}} & -a_2 & \frac{a_2}{\beta_2+\bm\chi_{n,c}}\\
\end{array}
\right).\\
\end{split}
\end{eqnarray}
The corresponding eigenfunctions $(R_{n}, S_{n}, W_{n})$ are given by
\begin{eqnarray}\label{eq-vbs2}
\begin{split}
R_{n}&=\varphi_{n,a}+\varphi_{n,c},\\
S_{n}&=\psi_{0}^{(1)}\left(\frac{\varphi_{n,a}}{\beta_1+\bm\chi_{n,a}}+\varphi_{n,b}+
\frac{\varphi_{n,c}}{\beta_1+\bm\chi_{n,c}}\right),\\
W_{n}&=\psi_{0}^{(2)}\left(\frac{\varphi_{n,a}}{\beta_2+\bm\chi_{n,a}}-\varphi_{n,b}+
\frac{\varphi_{n,c}}{\beta_2+\bm\chi_{n,c}}\right).
\end{split}
\end{eqnarray}
Here, we still use the coefficients: $\{c_{1,a},c_{1,b},c_{1,c}\}=\{1,1,0\}$.
Substituting (\ref{eq-vbs2}) into (\ref{eqdt2}) yields the fundamental vector soliton solutions for $N=2$.
The explicit form coincides with Eq. (\ref{eqmvbs2}).

\subsection{Vector solitons for $N=2$ with $a_1\neq0$, $a_2=0$}\label{SecA-2}

When $a_2=0$, the spectral parameter is given by (\ref{Eqlambda-N2}).
The transformation matrix $\mathbf{H}$ takes the form:
\begin{eqnarray}
\begin{split}
&\mathbf{H}=\left(
\begin{array}{cccc}
1 & 0 & 1\\
\frac{a_1}{\beta_1+\bm\chi_{n,a}} & 0 & \frac{a_1}{\beta_1+\bm\chi_{n,c}}\\
0 & 1 & 0\\
\end{array}
\right).\\
\end{split}
\end{eqnarray}
Using (\ref{Eqlambda-N2}) and solving the associated Lax pair, we have the eigenfunctions:
\begin{eqnarray}\label{eq-vdbs2}
\begin{split}
R_{n}&=\varphi_{n,a}+\varphi_{n,c},\\
S_{n}&=\psi_{0}^{(1)}\left(\frac{\varphi_{n,a}}{\beta_1+\bm\chi_{n,a}}+\frac{\varphi_{n,c}}{\beta_1+\bm\chi_{n,c}}\right),\\
W_{n}&=\varphi_{n,b}\exp{\left(i\theta_2\right)}.
\end{split}
\end{eqnarray}
Substituting (\ref{eq-vdbs2}) into (\ref{eqdt2}) with the coefficients: $\{c_{1,a},c_{1,b},c_{1,c}\}=\{1,1,0\}$ yields the fundamental dark-bright soliton solution (\ref{eqDB}).

\subsection{Second-order solutions for $N=2$}\label{SecA-3}

The second-order solutions can be obtained in the next iteration of the Darboux transformation.
They are:
\begin{eqnarray}
&&\psi^{(1)}_2=\psi^{(1)}_1+(\lambda_2^*-\lambda_2)(\mathbf{P})_{12},\\
&&\psi^{(2)}_2=\psi^{(2)}_1+(\lambda_2^*-\lambda_2)(\mathbf{P})_{13}.
\end{eqnarray}
Here, $\psi^{(j)}_1$ denote the fundamental vector solution obtained above.
$(\mathbf{P})_{1i}$ represents the element of the matrix $(\mathbf{P})$ in the first row and $i$-th column, and
\begin{eqnarray}
&&\mathbf{T}=\mathbf{I}-\frac{\lambda_{1}-\lambda_{1}^*}{\lambda_2-\lambda_{1}^*}
\frac{\bf{\Psi}_{1}\bf{\Psi}_{1}^\dagger\bm\Lambda}{\bf{\Psi}_{1}^\dagger\bm\Lambda\bf{\Psi}_{1}},\\
&&\mathbf{P}=\frac{\bf{\Psi}_{2}[1]\bf{\Psi}_{2}^\dagger[1]\bm\Lambda}{\bf{\Psi}_{2}^\dagger[1]\bm\Lambda\bf{\Psi}_{2}[1]},~~~\bf{\Psi}_{2}[1]=\mathbf{T}\bf{\Psi}_{2},
\end{eqnarray}
where $\bm\Lambda=diag(1,-1,-1)$.
When the two eigenvalues are different, these solutions describe the second-order non-degenerate solitons on the same vector background.

\section{Vector soliton solutions for $N=3$}\label{SecB}
\subsection{Vector solitons for $N=3$ with $a_j=a\neq0$}\label{SecB-1}
\subsubsection{Vector solitons when $\beta_1=-\beta_3=\beta\neq0$, $\beta_2=0$}\label{SecB-1a}

For $N=3$, the Lax pair can be rewritten as:
\begin{eqnarray}\label{UV2}
\begin{small}
\begin{split}
&\tilde{\mathbf{U}}=\left(
\begin{array}{cccc}
\lambda & -a_1 & -a_2 & -a_3\\
a_1 & -\beta_1 & 0 & 0\\
a_2 & 0 & -\beta_2 & 0\\
a_3 & 0 & 0 &-\beta_3\\
\end{array}
\right),
&\tilde{\mathbf{V}}=-\frac{i}{2}\tilde{\mathbf{U}}^2+2i\bm{a}^2\mathbf{I},
\end{split}
\end{small}
\end{eqnarray}
by using a diagonal matrix $S$=diag$(1,e^{-i\theta_1},e^{-i\theta_2},e^{-i\theta_3})$.
The linear eigenvalue problem of the transformed
Lax pair (\ref{UV2}) is given by
\begin{eqnarray}\label{x3}
\det(\tilde{\mathbf{U}}-i\chi)=0.
\end{eqnarray}
There are four eigenvalues $\bm\chi_{n,l}$ ($l=a,b,c,d$).
The transformation matrix in this case is:
\begin{eqnarray}
\begin{split}\label{H3-1}
&\mathbf{H}=\left(
\begin{array}{cccc}
1 & 1 & 1 & 1\\
\frac{a_1}{\beta_1+\bm\chi_{n,a}} & \frac{a_1}{\beta_1+\bm\chi_{n,b}} & \frac{a_1}{\beta_1+\bm\chi_{n,c}} & \frac{a_1}{\beta_1+\bm\chi_{n,d}}\\
\frac{a_2}{\beta_2+\bm\chi_{n,a}} & \frac{a_2}{\beta_2+\bm\chi_{n,b}} & \frac{a_2}{\beta_2+\bm\chi_{n,c}} & \frac{a_2}{\beta_2+\bm\chi_{n,d}}\\
\frac{a_3}{\beta_3+\bm\chi_{n,a}} & \frac{a_3}{\beta_3+\bm\chi_{n,b}} & \frac{a_3}{\beta_3+\bm\chi_{n,c}} & \frac{a_3}{\beta_3+\bm\chi_{n,d}}\\
\end{array}
\right).\\
\end{split}
\end{eqnarray}
The vector solution of the transformed Lax pair is:
\begin{eqnarray}
\begin{split}
\varphi_{n,l}&=c_{n,l}\exp{\{i\bm\chi_{n,l}x+\frac{1}{2}(4\bm{a}^2+\bm\chi_{n,l}^2) t\}},
\end{split}
\end{eqnarray}
where $l=a,b,c,d$. The corresponding eigenfunctions $(R_{n},S_{n},W_{n},X_{n})$ are given by
\begin{eqnarray}
\begin{split}
&&R_{n}&=\varphi_{n,a}+\varphi_{n,b}+\varphi_{n,c}+\varphi_{n,d},\\
&&S_{n}&=\psi_{0}^{(1)}\left(\sum_{l=a}^d\frac{\varphi_{n,l}}{\beta_1+\bm\chi_{n,l}}\right),\\
&&W_{n}&=\psi_{0}^{(2)}\left(\sum_{l=a}^d\frac{\varphi_{n,l}}{\beta_2+\bm\chi_{n,l}}\right),\\
&&X_{n}&=\psi_{0}^{(3)}\left(\sum_{l=a}^d\frac{\varphi_{n,l}}{\beta_3+\bm\chi_{n,l}}\right).
\end{split}
\end{eqnarray}
The fundamental vector soliton solution ($n=1$) can be obtained at the first step of the Darboux transformation:
\begin{eqnarray}\label{eqdt3}
\begin{split}
&&\psi_{1}^{(1)}=\psi_{0}^{(1)}+\frac{(\lambda_{1}^*-\lambda_{1})R_{1}^*S_{1}}{|R_{1}|^2-|S_{1}|^2-|W_{1}|^2-|X_{1}|^2},\\
&&\psi_{1}^{(2)}=\psi_{0}^{(2)}+\frac{(\lambda_{1}^*-\lambda_{1})R_{1}^*W_{1}}{|R_{1}|^2-|S_{1}|^2-|W_{1}|^2-|X_{1}|^2}.\\
&&\psi_{1}^{(3)}=\psi_{0}^{(3)}+\frac{(\lambda_{1}^*-\lambda_{1})R_{1}^*X_{1}}{|R_{1}|^2-|S_{1}|^2-|W_{1}|^2-|X_{1}|^2}.
\end{split}
\end{eqnarray}
To obtain the solution describing a single soliton, two of the coefficients $c_{1,l}$ ($l=a,b,c,d$) should be zero.
Below, we show that three combinations of the coefficients can be used to generate valid solution.

The linear eigenvalue problem (\ref{x3}) directly leads to
\begin{eqnarray}\label{lambda-appendixB}
\lambda=\bm\chi+\sum_{j=1}^3\frac{a_j^2}{\bm\chi+\beta_j},
\end{eqnarray}
where the eigenvalue $\bm\chi$ is given by Eq. (\ref{eqchi}).
Substituting one valid eigenvalue $\bm\chi_1$ given by (\ref{Eqchi1-6}) into (\ref{lambda-appendixB}), we obtain the corresponding Lax spectrum $\lambda_1$.
Using this spectrum and solving (\ref{x3}), we have four different eigenvalues $\bm\chi_{1,a}$, $\bm\chi_{1,b}$, $\bm\chi_{1,c}$, and $\bm\chi_{1,d}$.
The valid ones are: $\bm\chi_{1,a}=\bm\chi_1$, $\bm\chi_{1,b}=\bm\chi_2=\bm\chi_1+i\alpha$.
One of the eigenvalues is invalid. Without loss of generality, we let $\bm\chi_{1,d}$ to be invalid.

There are three combinations of the coefficients that can be used to generate fundamental soliton solutions. These are:
 $\{c_{1,a},c_{1,b},c_{1,c},c_{1,d}\}=\{1,1,0,0\}$, $\{c_{1,a},c_{1,b},c_{1,c},c_{1,d}\}=\{1,0,1,0\}$, and $\{c_{1,a},c_{1,b},c_{1,c},c_{1,d}\}=\{0,1,1,0\}$.
The nontrivial finding is that the case $\{1,1,0,0\}$ leads to the three-component soliton solution.
The explicit form of (\ref{eqdt3}) is given by Eq. (\ref{eqkmb}) with $N=3$.
The two other cases $\{1,0,1,0\}$ and $\{0,1,1,0\}$ lead to the solutions describing the dynamics of general breathers.
Without loss of generality, we use the combination $\{1,0,1,0\}$.
The explicit form of the solution is given by (\ref{eqgb}).

\subsubsection{Vector solitons when $\beta_1=\beta_2=\beta_3$}\label{SecB-1b}

Here, using Darboux transformations, we derive the solutions (\ref{eqbs3-2}) and (\ref{eqbs3-3}).
When $\beta_1=\beta_2=\beta_3$, the eigenvalues (\ref{Eqchi1-6}), reduce to
\begin{eqnarray}\label{Eqchi}
\begin{split}
&&\bm{\chi}_{1}&=\bm{\chi}_{3}=-\beta_1-i\alpha,\\
&&\bm{\chi}_{2}&=\bm{\chi}_{4}=-\beta_1.
\end{split}
\end{eqnarray}
In this case, only one eigenvalue (either $\bm\chi_1$ or $\bm\chi_3$) is valid.
The transformation matrix (\ref{H3-1}) can be rewritten as:
\begin{eqnarray}
\begin{split}
&\mathbf{H}=\left(
\begin{array}{cccc}
1 & 0 & 0 & 1\\
\frac{a_1}{\beta_1+\bm\chi_{n,a}} & -a_2 & -a_3 & \frac{a_1}{\beta_1+\bm\chi_{n,d}}\\
\frac{a_2}{\beta_2+\bm\chi_{n,a}} & a_1 & 0 & \frac{a_2}{\beta_2+\bm\chi_{n,d}}\\
\frac{a_3}{\beta_3+\bm\chi_{n,a}} & 0 & a_1 & \frac{a_3}{\beta_3+\bm\chi_{n,d}}\\
\end{array}
\right).\\
\end{split}
\end{eqnarray}
The corresponding eigenfunctions are:
\begin{small}
\begin{eqnarray}
\begin{split}
R_{n}&=\varphi_{n,a}+\varphi_{n,d},\\
S_{n}&=\left(\frac{a_1\varphi_{n,a}}{\beta_1+\bm\chi_{n,a}}-a_2\varphi_{n,b}-
a_3\varphi_{n,c}+\frac{a_1\varphi_{n,d}}{\beta_1+\bm\chi_{n,d}}\right)\\
&\times\exp{\left(i\theta_1\right)},\\
W_{n}&=\left(a_2\frac{\varphi_{n,a}}{\beta_2+\bm\chi_{n,a}}+a_1\varphi_{n,b}+
\frac{a_2\varphi_{n,d}}{\beta_2+\bm\chi_{n,d}}\right)\exp{\left(i\theta_2\right)},\\
X_{n}&=\left(a_3\frac{\varphi_{n,a}}{\beta_3+\bm\chi_{n,a}}+
a_1\varphi_{n,c}+\frac{a_3\varphi_{n,d}}{\beta_3+\bm\chi_{n,d}}\right)\exp{\left(i\theta_3\right)}.
\end{split}
\end{eqnarray}
\end{small}
Substituting $(R_{n}, S_{n}, W_{n},X_{n})$ into (\ref{eqdt3}) with the coefficients $(1,1,0,0)$ or $(1,0,1,0)$ yields the fundamental soliton solutions for $N=3$.
The case $(1,1,0,0)$ yields the solution defined by Eq. (\ref{eqbs3-2}) while
the case $(1,0,1,0)$ produces the solution (\ref{eqbs3-3}). As shown above, the solutions  are connected through a simple transformation $\psi^{(2)}_{VBS}\Leftrightarrow\psi^{(3)}_{VBS}$.
The nonlinear superposition of these two solutions with different $\alpha$ is a non-degenerate  second-order solution. It is shown in Fig. \ref{fig13-2km3}(c).

The soliton solutions (\ref{eqvbs3-1}) and (\ref{eqvbs3-2}) are derived using the
condition $\beta_1=\beta_2=\beta_3=0$. These are different from the solutions (\ref{eqbs3-2}) and (\ref{eqbs3-3}).
The transformation matrix (\ref{H3-1}) in this case is:
\begin{small}
\begin{eqnarray}\label{EqH3-1}
\begin{split}
&\mathbf{H}=\left(
\begin{array}{cccc}
\frac{\beta_3+\bm\chi_{n,a}}{a_3} & \frac{\beta_2+\bm\chi_{n,b}}{a_2} & \frac{\beta_1+\bm\chi_{n,c}}{a_1}  & 1\\
\frac{a_1(\beta_3+\bm\chi_{n,a})}{a_3(\beta_1+\bm\chi_{n,a})} & \frac{a_1(\beta_2+\bm\chi_{n,b})}{a_2(\beta_1+\bm\chi_{n,b})} & 1 & \frac{a_1}{\beta_1+\bm\chi_{n,d}}\\
\frac{a_2(\beta_3+\bm\chi_{n,a})}{a_3(\beta_2+\bm\chi_{n,a})} & 1 & \frac{a_2(\beta_1+\bm\chi_{n,c})}{a_1(\beta_2+\bm\chi_{n,c})} & \frac{a_2}{\beta_2+\bm\chi_{n,d}}\\
1 & \frac{a_3(\beta_2+\bm\chi_{n,b})}{a_1(\beta_3+\bm\chi_{n,b})} & \frac{a_3(\beta_1+\bm\chi_{n,c})}{a_1(\beta_3+\bm\chi_{n,c})} & \frac{a_3}{\beta_3+\bm\chi_{n,d}}\\
\end{array}
\right).\\
\end{split}
\end{eqnarray}
\end{small}
When $a_j=a$, the eigenfunctions are given by:
\begin{eqnarray}
\begin{split}
R_{n}=&\frac{\bm\chi_{n,a}}{a}\varphi_{n,a}+\frac{\bm\chi_{n,b}}{a}\varphi_{n,b}
+\frac{\bm\chi_{n,c}}{a}\varphi_{n,c}+\varphi_{n,d},\\
S_{n}=&(\varphi_{n,a}+A_1\varphi_{n,b}+
\varphi_{n,c}+\frac{a}{\bm\chi_{n,d}}\varphi_{n,d})\exp{(i\theta_1)},\\
W_{n}=&(\varphi_{n,a}+\varphi_{n,b}+A_2\varphi_{n,b}
+\frac{a}{\bm\chi_{n,d}}\varphi_{n,d})\exp{(i\theta_2)},\\
X_{n}=&(\varphi_{n,a}+A_3\varphi_{n,b}+A_4\varphi_{n,c}+\frac{a}{\bm\chi_{n,d}}\varphi_{n,d})\exp{(i\theta_3)},
\end{split}
\end{eqnarray}
where $A_1$, $A_2$, $A_3$, $A_4$ are real constants to be determined. Here, for the case of solution (\ref{eqvbs3-1}), we have $\{A_1,A_2,A_3,A_4\}=\{-1.366, 2.732, 0.366, -3.732\}$.
For the case of solution (\ref{eqvbs3-2}), we have $\{A_1,A_2,A_3,A_4\}=\{0.366, -0.732, -1.366, -0.268\}$.

Substituting $(R_{n}, S_{n}, W_{n},X_{n})$ into (\ref{eqdt3}) with the coefficients $\{c_{1,a},c_{1,b},c_{1,c},c_{1,d}\}=\{1,1,0,0\}$
yields the fundamental soliton solutions (\ref{eqvbs3-1}) and (\ref{eqvbs3-2}).
For a given $\alpha$, solutions (\ref{eqvbs3-1}) and (\ref{eqvbs3-2}) correspond to the same eigenvalue (or Lax spectral parameter).
Therefore, their nonlinear superposition forms a degenerate second-order solution.
It is shown in Fig. \ref{fig14-BE-1}.

\subsection{Vector solitons for $N=3$ with $a_1=a_2\neq0$, $a_3=0$}\label{SecB-2}

\subsubsection{Vector solitons when $\beta_1=-\beta_3=\beta\neq0$, $\beta_2=0$}\label{SecB-2a}

The soliton solution with one zero component in $\psi^{(3)}$ and dark-dark-bright solution (\ref{gz1}) are derived below.

When $a_3=0$, the valid eigenvalues (\ref{Eqchi1-6}) reduce to
\begin{eqnarray}\label{Eqchi}
\begin{split}
&\bm{\chi}_{1}=-\frac{1}{2}i\left(\alpha+\sqrt{\kappa_1-\kappa_2}\right)-\frac{1}{2}\beta,\\
&\bm{\chi}_{2}=-\frac{1}{2}i\left(\alpha-\sqrt{\kappa_1-\kappa_2}\right)-\frac{1}{2}\beta,\\
&\bm{\chi}_{3}=\beta-i\alpha,~~~~~~~~~\bm{\chi}_{4}=\beta.
\end{split}
\end{eqnarray}
Among them, the eigenvalues $\bm\chi_1$ , $\bm\chi_2$ and $\bm\chi_3$ are valid.
The transformation matrix $\mathbf{H}$ takes the form:
\begin{eqnarray}
\begin{split}
&\mathbf{H}=\left(
\begin{array}{cccc}
1 & 1 & 0 & 1\\
\frac{a_1}{\beta_1+\bm\chi_{n,a}} & \frac{a_1}{\beta_1+\bm\chi_{n,b}} & 0 & \frac{a_1}{\beta_1+\bm\chi_{n,d}}\\
\frac{a_2}{\beta_2+\bm\chi_{n,a}} & \frac{a_2}{\beta_2+\bm\chi_{n,b}} & 0 & \frac{a_2}{\beta_2+\bm\chi_{n,d}}\\
0 & 0 & 1 & 0\\
\end{array}
\right).\\
\end{split}
\end{eqnarray}
The corresponding eigenfunctions are:
\begin{eqnarray}
\begin{split}
R_{n}&=\varphi_{n,a}+\varphi_{n,b}+\varphi_{n,d},\\
S_{n}&=\psi_{0}^{(1)}\left(\frac{\varphi_{n,a}}{\beta_1+\bm\chi_{n,a}}+\frac{\varphi_{n,b}}{\beta_1+\bm\chi_{n,b}}
+\frac{\varphi_{n,d}}{\beta_1+\bm\chi_{n,d}}\right),\\
W_{n}&=\psi_{0}^{(2)}\left(\frac{\varphi_{n,a}}{\beta_2+\bm\chi_{n,a}}+\frac{\varphi_{n,b}}{\beta_2+\bm\chi_{n,b}}
+\frac{\varphi_{n,d}}{\beta_2+\bm\chi_{n,d}}\right),\\
X_{n}&=\varphi_{n,c}\exp{\left(i\theta_3\right)}.
\end{split}
\end{eqnarray}
Substituting $(R_{n}, S_{n}, W_{n},X_{n})$ into (\ref{eqdt3}) with the coefficients $(1,1,0,0)$
yields the soliton solution with one zero component in $\psi^{(3)}$. It is illustrated in Figs. \ref{fig15-gzkm1}(a) and \ref{fig16-gzkm2}(a). The case $(1,0,1,0)$ produces the dark-dark-bright soliton solution (\ref{gz1}).

\subsubsection{Vector solitons when $\beta_1=\beta_2=\beta_3$}\label{SecB-2b}

The soliton solution (\ref{EqVBS0-3}) with one zero component in $\psi^{(3)}$)
is derived below.

When $a_3=0$, the eigenvalues (\ref{Eqchi1-6}) reduce to
\begin{eqnarray}\label{Eqchi}
\begin{split}
&&\bm{\chi}_{1}&=\bm{\chi}_{3}=-\beta_1-i\alpha,\\
&&\bm{\chi}_{2}&=\bm{\chi}_{4}=-\beta_1.
\end{split}
\end{eqnarray}
Among them, only one eigenvalue (either $\bm\chi_1$ or $\bm\chi_3$) is valid.
The transformation matrix $\mathbf{H}$ takes the form:
\begin{eqnarray}
\begin{split}
&\mathbf{H}=\left(
\begin{array}{cccc}
1 & 0 & 0 & 1\\
\frac{a_1}{\beta_1+\bm\chi_{n,a}} & -a_2 & 0 & \frac{a_1}{\beta_1+\bm\chi_{n,d}}\\
\frac{a_2}{\beta_2+\bm\chi_{n,a}} & a_1 & 0 & \frac{a_2}{\beta_2+\bm\chi_{n,d}}\\
0 & 0 & a_1 & 0\\
\end{array}
\right).\\
\end{split}
\end{eqnarray}
The corresponding eigenfunctions are:
\begin{eqnarray}
\begin{split}
R_{n}&=\varphi_{n,a}+\varphi_{n,d},\\
S_{n}&=\left(\frac{a_1\varphi_{n,a}}{\beta_1+\bm\chi_{n,a}}-a_2\varphi_{n,b}+\frac{a_1\varphi_{n,d}}{\beta_1+\bm\chi_{n,d}}\right)\exp{\left(i\theta_1\right)},\\
W_{n}&=\left(a_2\frac{\varphi_{n,a}}{\beta_2+\bm\chi_{n,a}}+a_1\varphi_{n,b}+
\frac{a_2\varphi_{n,d}}{\beta_2+\bm\chi_{n,d}}\right)\exp{\left(i\theta_2\right)},\\
X_{n}&=\left(a_1\varphi_{n,c}\right)\exp{\left(i\theta_3\right)}.
\end{split}
\end{eqnarray}
Substituting $(R_{n}, S_{n}, W_{n},X_{n})$ into (\ref{eqdt3}) with the coefficients $(1,1,0,0)$
yields the soliton solution with one zero component in $\psi^{(3)}$ (\ref{EqVBS0-3}).
The case $(1,0,1,0)$ produces the dark-dark-bright solution (\ref{gz1}).
For a given $\alpha$, the two solutions (\ref{EqVBS0-3}) and (\ref{gz1}) have identical eigenvalues (or Lax spectral parameter).
Therefore, their nonlinear superposition forms a degenerate second-order solution. It is shown in
Fig. \ref{fig18-BE-2}.

\subsection{Vector solitons for $N=3$ with $a_1\neq0$, $a_2=a_3=0$}\label{SecB-3}

The solutions (\ref{eq-a2a3-1}) and (\ref{eq-a2a3-2}) with $a_2=a_3=0$ are derived below.
The spectral parameter, $\lambda$, in this case is given by
\begin{eqnarray}\label{lambda-appendix-a2}
\lambda=\bm\chi+\frac{a_1^2}{\bm\chi+\beta_1},
\end{eqnarray}
while the transformation matrix, $\mathbf{H}$, is
\begin{small}
\begin{eqnarray}
\begin{split}
&\mathbf{H}=\left(
\begin{array}{cccc}
1 & 0 & 0  & 1\\
\frac{a_1}{\beta_1+\bm\chi_{n,a}} & 0 & 0 & \frac{a_1}{\beta_1+\bm\chi_{n,d}}\\
0 & 1 & 0 & 0\\
0 & 0 & 1 & 0\\
\end{array}
\right).\\
\end{split}
\end{eqnarray}
\end{small}
The components of the eigenfunction are:
\begin{eqnarray}
\begin{split}
R_{n}&=\varphi_{n,a}+\varphi_{n,d},\\
S_{n}&=\psi_{0}^{(1)}\left(\frac{\varphi_{n,a}}{\beta_1+\bm\chi_{n,a}}
+\frac{\varphi_{n,d}}{\beta_1+\bm\chi_{n,d}}\right),\\
W_{n}&=\varphi_{n,b}\exp{\left(i\theta_2\right)},\\
X_{n}&=\varphi_{n,c}\exp{\left(i\theta_3\right)}.
\end{split}
\end{eqnarray}
Substituting $(R_{n}, S_{n}, W_{n},X_{n})$ into (\ref{eqdt3}) with the coefficients $\{c_{1,a},c_{1,b},c_{1,c},c_{1,d}\}=\{1,1,0,0\}$
yields the dark-bright-zero soliton solution (\ref{eq-a2a3-1}). The case $\{1,0,1,0\}$ produces the dark-zero-bright soliton solution (\ref{eq-a2a3-2}).
For a given $\alpha$, the solutions (\ref{eq-a2a3-1}) and (\ref{eq-a2a3-2}) have identical eigenvalues (or Lax spectral parameter).
Thus, their nonlinear superposition forms a degenerate second-order solution. It is shown in
Fig. \ref{fig22-BE-3}.

\subsection{Second-order solutions for $N=3$}\label{SecB-4}

The second-order solutions of the three-component Manakov equations are given by:
\begin{eqnarray}
&&\psi^{(1)}_2=\psi^{(1)}_1+(\lambda_2^*-\lambda_2)(\mathbf{P})_{12},\\
&&\psi^{(2)}_2=\psi^{(2)}_1+(\lambda_2^*-\lambda_2)(\mathbf{P})_{13},\\
&&\psi^{(3)}_2=\psi^{(3)}_1+(\lambda_2^*-\lambda_2)(\mathbf{P})_{14}.
\end{eqnarray}
Here $(\mathbf{P})_{1i}$ is the matrix element of $\mathbf{P}$ in the first row and $i$-th column, and
\begin{eqnarray}
&&\mathbf{T}=\mathbf{I}-\frac{\lambda_{1}-\lambda_{1}^*}{\lambda_2-\lambda_{1}^*}
\frac{\bf{\Psi}_{1}\bf{\Psi}_{1}^\dagger\bm\Lambda}{\bf{\Psi}_{1}^\dagger\bm\Lambda\bf{\Psi}_{1}},\\
&&\mathbf{P}=\frac{\bf{\Psi}_{2}[1]\bf{\Psi}_{2}^\dagger[1]\bm\Lambda}{\bf{\Psi}_{2}^\dagger[1]
\bm\Lambda\bf{\Psi}_{2}[1]},~~~\bf{\Psi}_{2}[1]=\mathbf{T}\bf{\Psi}_{2},
\end{eqnarray}
where $\bm\Lambda=diag(1,-1,-1,-1)$ and $\dagger$ denotes the matrix transpose and complex conjugate.

\section{Asymptotic analysis}\label{SecC}

We show here the asymptotic behavior of two solitons described by the second-order soliton solution $\psi_2^{(j)}(t,x)$. 
Namely, when $t\rightarrow\pm\infty$, we have
\begin{eqnarray}\label{Ceq1}
\left\{
\begin{array}{ccc}
\psi^{(j)}_{2l}&=&\psi_0^{(j)}\frac{\mathcal{M}_{l}^-}{\mathcal{N}_{l}^-},~~t\rightarrow-\infty,\\
\psi^{(j)}_{2l}&=&\psi_0^{(j)}\frac{\mathcal{M}_{l}^+}{\mathcal{N}_{l}^+},~~t\rightarrow+\infty,
\end{array}
\right.
\end{eqnarray}
where the subscript $l(=1,2)$ represents the single soliton $1$ and $2$. Moreover,
\begin{eqnarray}
\begin{split}
\mathcal{M}_{1}^-&=z_{03}^j\exp{(\omega_{1r})}+z_{02}^j\exp{(-\omega_{1r})}+z_{05}^j\exp{(\Delta_1)}\\
&+z_{06}^j\exp{(-\Delta_1)},\\
\mathcal{M}_{1}^+&=z_{01}^j\exp{(\omega_{1r})}+z_{04}^j\exp{(-\omega_{1r})}+z_{07}^j\exp{(\Delta_1)}\\
&+z_{08}^j\exp{(-\Delta_1)},\\
\mathcal{M}_{2}^-&=z_{04}^j\exp{(\omega_{2r})}+z_{02}^j\exp{(-\omega_{2r})}+z_{12}^j\exp{(\Delta_2)}\\
&+z_{10}^j\exp{(-\Delta_2)},\\
\mathcal{M}_{2}^+&=z_{01}^j\exp{(\omega_{2r})}+z_{03}^j\exp{(-\omega_{2r})}+z_{09}^j\exp{(\Delta_2)}\\
&+z_{11}^j\exp{(-\Delta_2)},\\
\mathcal{N}_{1}^-&=m_{01}\exp{(\omega_{1r})}+m_{02}\exp{(-\omega_{1r})}+m_{03}\exp{(\Delta_1)}\\
&+m_{04}\exp{(-\Delta_1)},\\
\mathcal{N}_{1}^+&=m_{05}\exp{(\omega_{1r})}+m_{06}\exp{(-\omega_{1r})}+m_{07}\exp{(\Delta_1)}\\
&+m_{08}\exp{(-\Delta_1)},\\
\mathcal{N}_{2}^-&=m_{06}\exp{(\omega_{2r})}+m_{02}\exp{(-\omega_{2r})}+m_{09}\exp{(\Delta_2)}\\
&+m_{10}\exp{(-\Delta_2)},\\
\mathcal{N}_{2}^+&=m_{05}\exp{(\omega_{2r})}+m_{01}\exp{(-\omega_{2r})}+m_{11}\exp{(\Delta_2)}\\
&+m_{12}\exp{(-\Delta_2)}.\nonumber
\end{split}
\end{eqnarray}
Here, $\omega_{lr}=(\bm\chi_{li}^{n}-\bm\chi_{li}^{m})x+(\bm\chi_{lr}^{n}\bm\chi_{li}^{n}-\bm\chi_{lr}^{m}\bm\chi_{li}^{m})t$,
$\Delta_{l}=i[(\bm\chi_{lr}^{m}-\bm\chi_{lr}^{n})x+((\bm\chi_{lr}^{m})^2-(\bm\chi_{li}^{m})^2+(\bm\chi_{li}^{n})^2-(\bm\chi_{lr}^{n})^2)/2 t]$. Note that
$\bm\chi_{l}^{k}$ $(k=m,n)$ are the valid eigenvalues of the soliton solutions we have analyzed. For the case $N=2$, we let $\bm\chi_{l}^{m}=\bm\chi_a$, $\bm\chi_{l}^{m}=\bm\chi_b$, where $\bm\chi_{a,b}$ are the complex roots of Eq. (\ref{x2}).
For the case $N=3$, we must have $\bm\chi_{l}^{m}=\bm\chi_a$, $\bm\chi_{l}^{m}=\bm\chi_b$ or $\bm\chi_c$, with $\bm\chi_{a,b,c}$ being the complex roots of Eq. (\ref{x3}).
Complex parameters ($z_{01}^j \sim z_{12}^j$, $m_{01} \sim m_{12}$) are given at the end of Appendix.

On the other hand, we shall rewrite the single soliton solution as
\begin{eqnarray}\label{Ceq2}
\psi_{1l}^{(j)}=\psi_0^{(j)}\frac{\mathcal{M}_{l}}{\mathcal{N}_{l}},
\end{eqnarray}
where
\begin{eqnarray}
\begin{split}
\mathcal{M}_{l}&=\frac{(\bm\chi_{l}^{m*}+\beta_j)\exp{(\omega_{lr})}}{(\bm\chi_{l}^{m}+\beta_j)(\bm\chi_{l}^{m*}-\bm\chi_{l}^{m})}+
\frac{(\bm\chi_{l}^{n*}+\beta_j)\exp{(\Delta_l})}{(\bm\chi_{l}^{m}+\beta_j)(\bm\chi_{l}^{n*}-\bm\chi_{l}^{m})}\\&
+\frac{(\bm\chi_{l}^{m*}+\beta_j)\exp{(-\Delta_l})}{(\bm\chi_{l}^{n}+\beta_j)(\bm\chi_{l}^{m*}-\bm\chi_{l}^{n})}+
\frac{(\bm\chi_{l}^{n*}+\beta_j)\exp{(-\omega_{lr})}}{(\bm\chi_{l}^{n}+\beta_j)(\bm\chi_{l}^{n*}-\bm\chi_{l}^{n})},\\
\mathcal{N}_{l}&=\frac{\exp{(\omega_{lr})}}{\bm{\chi}_{l}^{m*}-\bm{\chi}_{l}^{m}}+\frac{\exp{(\Delta_l)}}{\bm{\chi}_{l}^{n*}-\bm{\chi}_{l}^{m}}+
\frac{\exp{(-\Delta_l)}}{\bm{\chi}_{l}^{m*}-\bm{\chi}_{l}^{n}}+\frac{\exp{(-\omega_{lr})}}{\bm{\chi}_{l}^{n*}-\bm{\chi}_{l}^{n}}\nonumber
\end{split}
\end{eqnarray}
By comparing (\ref{Ceq1}) with (\ref{Ceq2}), we have
\begin{eqnarray}\label{Ceq4}
\left\{
\begin{array}{ccc}
\psi^{(j)}_{2l}=&\exp{(\phi_{l}^{(n)})}\psi^{(j)}_{1l}\{\pm\omega_{lr}\pm s_{l}^{(n)},\pm\Delta_l\pm \theta_{l}^{(n)}\},\\ &t\rightarrow-\infty,\\
\psi^{(j)}_{2l}=&\exp{(\phi_{l}^{(m)})}\psi^{(j)}_{1l}\{\pm\omega_{lr}\pm s_{l}^{(m)},\pm\Delta_l\pm \theta_{l}^{(m)}\},\\ &t\rightarrow+\infty.
\end{array}
\right.
\end{eqnarray}
Here $\phi_{l}^{(k)}$ $(k=m,n)$  denote the additional phase of vector plane wave, $s_{l}^{(k)}$ and $\theta_{l}^{(k)}$ are the position and phase shifts, respectively. Their expressions are given by
\begin{eqnarray}
\begin{small}
\begin{split}
&s_{1}^{(k)}=\frac{1}{2}\ln{\left[\frac{(\bm\chi_{1}^{m}-\bm\chi_{2}^{k})(\bm\chi_{2}^{k}-\bm\chi_{1}^{n*})(\bm\chi_{1}^{m*}-\bm\chi_{2}^{k*})(\bm\chi_{2}^{k*}-\bm\chi_{1}^{n})}
{(\bm\chi_{1}^{m*}-\bm\chi_{2}^{k})(\bm\chi_{2}^{k}-\bm\chi_{1}^{n})(\bm\chi_{1}^{m}-\bm\chi_{2}^{k*})(\bm\chi_{2}^{k*}-\bm\chi_{1}^{n*})}\right]},\\
&s_{2}^{(k)}=\frac{1}{2}\ln{\left[\frac{(\bm\chi_{1}^{k}-\bm\chi_{2}^{m})(\bm\chi_{1}^{k*}-\bm\chi_{2}^{m*})(\bm\chi_{1}^{k*}-\bm\chi_{2}^{n})(\bm\chi_{1}^{k}-\bm\chi_{2}^{n*})}
{(\bm\chi_{1}^{k*}-\bm\chi_{2}^{m})(\bm\chi_{1}^{k}-\bm\chi_{2}^{m*})(\bm\chi_{1}^{k}-\bm\chi_{2}^{n})(\bm\chi_{1}^{k*}-\bm\chi_{2}^{n*})}\right]},\\
&\theta_{1}^{(k)}=i\arg{\left[\frac{(\bm\chi_{1}^{m}-\bm\chi_{2}^{k})(\bm\chi_{1}^{n}-\bm\chi_{2}^{k*})}{(\bm\chi_{1}^{n}-\bm\chi_{2}^{k})(\bm\chi_{1}^{m}-\bm\chi_{2}^{k*})}\right]},\\
&\theta_{2}^{(k)}=i\arg{\left[\frac{(\bm\chi_{2}^{m}-\bm\chi_{1}^{k})(\bm\chi_{2}^{n}-\bm\chi_{1}^{k*})}{(\bm\chi_{1}^{k*}-\bm\chi_{2}^{m})(\bm\chi_{2}^{k}-\bm\chi_{2}^{n})}\right]},\\
&\phi_{1}^{(k)}=i\arg{\left[\frac{\beta_j+\bm\chi_{2}^{k*}}{\beta_j+\bm\chi_{2}^{k}}\right]},~~~
\phi_{2}^{(k)}=i\arg{\left[\frac{\beta_j+\bm\chi_{1}^{k*}}{\beta_j+\bm\chi_{1}^{k}}\right]}.\nonumber
\end{split}
\end{small}
\end{eqnarray}
Equation (\ref{Ceq4}) clearly shows that
the second-order solution $\psi_2^{(j)}(t,x)$ at $t\rightarrow\pm\infty$ turns out to be the elastic collision between two single solitons with position shift $s_{l}^{(k)}$, phase shift $\theta_{l}^{(k)}$, and additional phase of vector plane wave $\phi_{l}^{(k)}$. As an example, we have shown the elastic collision for $N=3$ in Fig. \ref{fig11-2km1} (a).
Finally, the remaining complex parameters are
\begin{widetext}
\begin{eqnarray}
\begin{split}
&z_{01}^j=\frac{(\beta_j+\bm\chi_{1}^{m*})(\bm\chi_{1}^{m}-\bm\chi_{2}^{m})(\bm\chi_{1}^{m*}-\bm\chi_{2}^{m*})(\beta_j+\bm\chi_{2}^{m*})}
{(\beta_j+\bm\chi_{1}^{m})(\bm\chi_{1}^{m}-\bm\chi_{1}^{m*})(\bm\chi_{1}^{m*}-\bm\chi_{2}^{m})(\beta_j+\bm\chi_{2}^{m})(\bm\chi_{1}^{m}-\bm\chi_{2}^{m*})(\bm\chi_{2}^{m}-\bm\chi_{2}^{m*})},\\
&z_{02}^j=\frac{(\beta_j+\bm\chi_{1}^{n*})(\bm\chi_{1}^{n}-\bm\chi_{2}^{n})(\bm\chi_{1}^{n*}-\bm\chi_{2}^{n*})(\beta_j+\bm\chi_{2}^{n*})}
{(\beta_j+\bm\chi_{1}^{n})(\bm\chi_{1}^{n}-\bm\chi_{1}^{n*})(\bm\chi_{1}^{n*}-\bm\chi_{2}^{n})(\beta_j+\bm\chi_{2}^{n})(\bm\chi_{1}^{n}-\bm\chi_{2}^{n*})(\bm\chi_{2}^{n}-\bm\chi_{2}^{n*})},\\
&z_{03}^j=\frac{(\beta_j+\bm\chi_{1}^{m*})(\bm\chi_{1}^{m}-\bm\chi_{2}^{n})(\bm\chi_{1}^{m*}-\bm\chi_{2}^{n*})(\beta_j+\bm\chi_{2}^{n*})}
{(\beta_j+\bm\chi_{1}^{m})(\bm\chi_{1}^{m}-\bm\chi_{1}^{m*})(\bm\chi_{1}^{m*}-\bm\chi_{2}^{n})(\beta_j+\bm\chi_{2}^{n})(\bm\chi_{1}^{m}-\bm\chi_{2}^{n*})(\bm\chi_{2}^{n}-\bm\chi_{2}^{n*})},\\
&z_{04}^j=\frac{(\beta_j+\bm\chi_{1}^{n*})(\bm\chi_{1}^{n}-\bm\chi_{2}^{m})(\bm\chi_{1}^{n*}-\bm\chi_{2}^{m*})(\beta_j+\bm\chi_{2}^{m*})}
{(\beta_j+\bm\chi_{1}^{n})(\bm\chi_{1}^{n}-\bm\chi_{1}^{n*})(\bm\chi_{1}^{n*}-\bm\chi_{2}^{m})(\beta_j+\bm\chi_{2}^{m})(\bm\chi_{1}^{n}-\bm\chi_{2}^{m*})(\bm\chi_{2}^{m}-\bm\chi_{2}^{m*})},\\
&z_{05}^j=\frac{(\beta_j+\bm\chi_{1}^{n*})(\bm\chi_{1}^{m}-\bm\chi_{2}^{n})(\bm\chi_{1}^{n*}-\bm\chi_{2}^{n*})(\beta_j+\bm\chi_{2}^{n*})}
{(\beta_j+\bm\chi_{1}^{m})(\bm\chi_{1}^{m}-\bm\chi_{1}^{n*})(\bm\chi_{1}^{n*}-\bm\chi_{2}^{n})(\beta_j+\bm\chi_{2}^{n})(\bm\chi_{1}^{m}-\bm\chi_{2}^{n*})(\bm\chi_{2}^{n}-\bm\chi_{2}^{n*})},\\
&z_{06}^j=\frac{(\beta_j+\bm\chi_{1}^{m*})(\bm\chi_{1}^{n}-\bm\chi_{2}^{n})(\bm\chi_{1}^{m*}-\bm\chi_{2}^{n*})(\beta_j+\bm\chi_{2}^{n*})}
{(\beta_j+\bm\chi_{1}^{n})(\bm\chi_{1}^{n}-\bm\chi_{1}^{m*})(\bm\chi_{1}^{m*}-\bm\chi_{2}^{n})(\beta_j+\bm\chi_{2}^{n})(\bm\chi_{1}^{n}-\bm\chi_{2}^{n*})(\bm\chi_{2}^{n}-\bm\chi_{2}^{n*})},\\
&z_{07}^j=\frac{(\beta_j+\bm\chi_{1}^{n*})(\bm\chi_{1}^{m}-\bm\chi_{2}^{m})(\bm\chi_{1}^{n*}-\bm\chi_{2}^{m*})(\beta_j+\bm\chi_{2}^{m*})}
{(\beta_j+\bm\chi_{1}^{m})(\bm\chi_{1}^{m}-\bm\chi_{1}^{n*})(\bm\chi_{1}^{n*}-\bm\chi_{2}^{m})(\beta_j+\bm\chi_{2}^{m})(\bm\chi_{1}^{m}-\bm\chi_{2}^{m*})(\bm\chi_{2}^{m}-\bm\chi_{2}^{m*})},\\
&z_{08}^j=\frac{(\beta_j+\bm\chi_{1}^{m*})(\bm\chi_{1}^{n}-\bm\chi_{2}^{m})(\bm\chi_{1}^{m*}-\bm\chi_{2}^{m*})(\beta_j+\bm\chi_{2}^{m*})}
{(\beta_j+\bm\chi_{1}^{n})(\bm\chi_{1}^{n}-\bm\chi_{1}^{m*})(\bm\chi_{1}^{m*}-\bm\chi_{2}^{m})(\beta_j+\bm\chi_{2}^{m})(\bm\chi_{1}^{n}-\bm\chi_{2}^{m*})(\bm\chi_{2}^{m}-\bm\chi_{2}^{m*})},\\
&z_{09}^j=\frac{(\beta_j+\bm\chi_{1}^{m*})(\bm\chi_{1}^{m}-\bm\chi_{2}^{m})(\bm\chi_{1}^{m*}-\bm\chi_{2}^{n*})(\beta_j+\bm\chi_{2}^{n*})}
{(\beta_j+\bm\chi_{1}^{m})(\bm\chi_{1}^{m}-\bm\chi_{1}^{m*})(\bm\chi_{1}^{m*}-\bm\chi_{2}^{m})(\bm\beta_j+\bm\chi_{2}^{m})(\bm\chi_{1}^{m}-\bm\chi_{2}^{n*})(\bm\chi_{2}^{m}-\bm\chi_{2}^{n*})},\\
&z_{10}^j=\frac{(\beta_j+\bm\chi_{1}^{n*})(\bm\chi_{1}^{n*}-\bm\chi_{2}^{m*})(\beta_j+\bm\chi_{2}^{m*})(\bm\chi_{1}^{n}-\bm\chi_{2}^{n})}
{(\beta_j+\bm\chi_{1}^{n})(\bm\chi_{1}^{n}-\bm\chi_{1}^{n*})(\bm\chi_{2}^{m*}-\bm\chi_{1}^{n})(\bm\chi_{1}^{n*}-\bm\chi_{2}^{n})(\bm\chi_{2}^{m*}-\bm\chi_{2}^{n})(\beta_j+\bm\chi_{2}^{n})},\\
&z_{11}^j=\frac{(\beta_j+\bm\chi_{1}^{m*})(\bm\chi_{1}^{m*}-\bm\chi_{2}^{m*})(\beta_j+\bm\chi_{2}^{m*})(\bm\chi_{1}^{m}-\bm\chi_{2}^{n})}
{(\beta_j+\bm\chi_{1}^{m})(\bm\chi_{1}^{m}-\bm\chi_{1}^{m*})(\bm\chi_{2}^{m*}-\bm\chi_{1}^{m})(\bm\chi_{1}^{m*}-\bm\chi_{2}^{n})(\bm\chi_{2}^{m}-\bm\chi_{2}^{n})(\beta_j+\bm\chi_{2}^{n})},\\
&z_{12}^j=\frac{(\beta_j+\bm\chi_{1}^{n*})(\bm\chi_{1}^{n}-\bm\chi_{2}^{m})(\bm\chi_{1}^{n*}-\bm\chi_{2}^{n*})(\beta_j+\bm\chi_{2}^{n*})}
{(\beta_j+\bm\chi_{1}^{n})(\bm\chi_{1}^{n}-\bm\chi_{1}^{n*})(\bm\chi_{1}^{n*}-\bm\chi_{2}^{m})(\beta_j+\bm\chi_{2}^{m})(\bm\chi_{1}^{n}-\bm\chi_{2}^{n*})(\bm\chi_{2}^{m}-\bm\chi_{2}^{n*})}.\nonumber
\end{split}
\end{eqnarray}
\end{widetext}
\begin{widetext}
\begin{eqnarray}
\begin{split}
&m_{01}=\frac{1}{(\bm\chi_{1}^{m*}-\bm\chi_{1}^{m})(\bm\chi_{2}^{n*}-\bm\chi_{2}^{n})}-
\frac{1}{(\bm\chi_{1}^{m*}-\bm\chi_{2}^{n})(\bm\chi_{2}^{n*}-\bm\chi_{1}^{m})},&&
m_{02}=\frac{1}{(\bm\chi_{1}^{n*}-\bm\chi_{1}^{n})(\bm\chi_{2}^{n*}-\bm\chi_{2}^{n})}-
\frac{1}{(\bm\chi_{1}^{n*}-\bm\chi_{2}^{n})(\bm\chi_{2}^{n*}-\bm\chi_{1}^{n})},\\
&m_{03}=\frac{1}{(\bm\chi_{1}^{n*}-\bm\chi_{1}^{m})(\bm\chi_{2}^{n*}-\bm\chi_{2}^{n})}-
\frac{1}{(\bm\chi_{1}^{n*}-\bm\chi_{2}^{n})(\bm\chi_{2}^{n*}-\bm\chi_{1}^{m})},&&
m_{04}=\frac{1}{(\bm\chi_{1}^{m*}-\bm\chi_{1}^{n})(\bm\chi_{2}^{n*}-\bm\chi_{2}^{n})}-
\frac{1}{(\bm\chi_{1}^{m*}-\bm\chi_{2}^{n})(\bm\chi_{2}^{n*}-\bm\chi_{1}^{n})},\\
&m_{05}=\frac{1}{(\bm\chi_{1}^{m*}-\bm\chi_{1}^{m})(\bm\chi_{2}^{m*}-\bm\chi_{2}^{m})}-
\frac{1}{(\bm\chi_{1}^{m*}-\bm\chi_{2}^{m})(\bm\chi_{2}^{m*}-\bm\chi_{1}^{m})},&&
m_{06}=\frac{1}{(\bm\chi_{1}^{n*}-\bm\chi_{1}^{n})(\bm\chi_{2}^{m*}-\bm\chi_{2}^{m})}-
\frac{1}{(\bm\chi_{1}^{n*}-\bm\chi_{2}^{m})(\bm\chi_{2}^{m*}-\bm\chi_{1}^{n})},\\
&m_{07}=\frac{1}{(\bm\chi_{1}^{n*}-\bm\chi_{1}^{m})(\bm\chi_{2}^{m*}-\bm\chi_{2}^{m})}-
\frac{1}{(\bm\chi_{1}^{n*}-\bm\chi_{2}^{m})(\bm\chi_{2}^{m*}-\bm\chi_{1}^{m})},&&
m_{08}=\frac{1}{(\bm\chi_{1}^{m*}-\bm\chi_{1}^{n})(\bm\chi_{2}^{m*}-\bm\chi_{2}^{m})}-
\frac{1}{(\bm\chi_{1}^{m*}-\bm\chi_{2}^{m})(\bm\chi_{2}^{m*}-\bm\chi_{1}^{n})},\\
&m_{09}=\frac{1}{(\bm\chi_{1}^{n*}-\bm\chi_{1}^{n})(\bm\chi_{2}^{n*}-\bm\chi_{2}^{m})}-
\frac{1}{(\bm\chi_{1}^{n*}-\bm\chi_{2}^{m})(\bm\chi_{2}^{n*}-\bm\chi_{1}^{n})},&&
m_{10}=\frac{1}{(\bm\chi_{1}^{n*}-\bm\chi_{1}^{n})(\bm\chi_{2}^{m*}\bm-\chi_{2}^{n})}-
\frac{1}{(\bm\chi_{2}^{m*}-\bm\chi_{1}^{n})(\bm\chi_{1}^{n*}-\bm\chi_{2}^{n})},\\
&m_{11}=\frac{1}{(\bm\chi_{1}^{m*}-\bm\chi_{1}^{m})(\bm\chi_{2}^{n*}-\bm\chi_{2}^{m})}-
\frac{1}{(\bm\chi_{1}^{m*}-\bm\chi_{2}^{m})(\bm\chi_{2}^{n*}-\bm\chi_{1}^{m})},&&
m_{12}=\frac{1}{(\bm\chi_{1}^{m*}-\bm\chi_{1}^{m})(\bm\chi_{2}^{m*}-\bm\chi_{2}^{n})}-
\frac{1}{(\bm\chi_{2}^{m*}-\bm\chi_{1}^{m})(\bm\chi_{1}^{m*}-\bm\chi_{2}^{n})}.\nonumber
\end{split}
\end{eqnarray}
\end{widetext}

\end{appendix}

\end{document}